%% file: templateArxiv.tex
\documentclass{article}

\usepackage{arxiv}

\usepackage{tikz}

\usepackage{filecontents}

\usepackage[utf8]{inputenc} 
\usepackage[T1]{fontenc}    
\usepackage{hyperref}       
\usepackage{xurl}            
\usepackage{booktabs}       
\usepackage{amsfonts}       
\usepackage{nicefrac}       
\usepackage{microtype}      
\usepackage{lipsum}
\usepackage{fancyhdr}       
\usepackage{graphicx}       
\graphicspath{{media/}}     
\usepackage{amsthm}
\newtheorem{definition}{Definition}
\newtheorem{remark}{Remark}
\usepackage{amsmath}
\usepackage{amssymb}
\usepackage{booktabs}
\usepackage{subcaption}
\usepackage[numbers]{natbib}
\NewCommandCopy{\origciteauthor}{\citeauthor}
\makeatletter
\RenewDocumentCommand{\citeauthor}{s m !o}{%
  {\let\hyper@natlinkstart\@gobble
   \let\hyper@natlinkend\relax
   \let\hyper@natlinkbreak\@firstoftwo
   \IfBooleanTF{#1}{\origciteauthor*{#2}}{\origciteauthor{#2}}}%
  \IfValueT{#3}{#3}~\cite{#2}}
\makeatother

\usepackage{multirow}
\usepackage{makecell}
\usepackage{threeparttable}
\usepackage{tabularx}
\usepackage{adjustbox}
\usepackage{bm}
\usepackage{xcolor}
\usepackage{dsfont}
\usepackage[table]{xcolor}  
\usepackage{booktabs,multirow,makecell,threeparttable}
\usepackage{enumitem}
\usepackage{amsthm}         
\usepackage{pifont}
\usepackage{rotating}

\newcounter{rqcounter}
\renewcommand{\therqcounter}{RQ\arabic{rqcounter}}
\newcommand{\rqitem}[1]{\refstepcounter{rqcounter}\label{#1}\textbf{\therqcounter.}\ }

\usepackage[table]{xcolor}\usepackage{booktabs,multirow,threeparttable,bm,graphicx}
\definecolor{lexcol}{HTML}{1F5FA9}\definecolor{embcol}{HTML}{A8431F}\definecolor{best}{HTML}{2A78D6}
\newcommand{\win}[3]{\cellcolor{best!18}\bfseries #1\textsuperscript{\textcolor{#2}{#3}}}
\newcommand{\dset}[1]{\multirow{3}{*}{\rotatebox[origin=c]{90}{#1}}}

\usepackage{fancyhdr}
\fancyheadoffset{0pt}
\title{Subgroup Membership Inference Audits of Differentially Private Synthetic Text} 

\def\tsc#1{\csdef{#1}{\textsc{\lowercase{#1}}\xspace}}
\tsc{WGM}
\tsc{QE}
\tsc{EP}
\tsc{PMS}
\tsc{BEC}
\tsc{DE}

\usepackage{bigfoot}

\DeclareNewFootnote{AAffil}[arabic]
\DeclareNewFootnote{ANote}[fnsymbol]

\usepackage{etoolbox}
\makeatletter
\patchcmd\maketitle{\def\@makefnmark{\rlap{\@textsuperscript{\normalfont\@thefnmark}}}}{}{}{}
\makeatother

\makeatletter
\def\thanksAAffil#1{
  \footnotemarkAAffil\protected@xdef\@thanks{\@thanks%
        \protect\footnotetextAAffil[\the \c@footnoteAAffil]{#1}}%
}
\def\thanksANote#1{%
  \footnotemarkANote%
  \protected@xdef\@thanks{\@thanks%
        \protect\footnotetextANote[\the \c@footnoteANote]{#1}}%
}
\makeatother

\author{
Yidan Sun%
\footnotemarkAAffil[1] \hspace{0.03cm}$^{,}$\thanksANote{Corresponding author: \texttt{y.sun1@imperial.ac.uk}}\\
\And
Viktor Schlegel%
\footnotemarkAAffil[1] \\
\And
Srinivasan Nandakumar%
\thanksAAffil{Imperial College London, Imperial Global Singapore}\\
\And
Siew Kei Lam%
\thanksAAffil{Nanyang Technological University, Singapore}\\
\And
Anil Anthony Bharath%
\footnotemarkAAffil[1] \hspace{0.03cm}$^{,}$\thanksAAffil{Imperial College London, United Kingdom}\\
}

\begin{document}
\maketitle

\begin{abstract}
Synthetic data releases are increasingly proposed in the literature as a means
of sharing realistic data replicas in lieu of sensitive private datasets. Even
when the worst-case privacy leakage of such releases is bounded by means of
differential privacy (DP), in practice a residual risk remains. Membership
inference attack (MIA) audits are conducted to empirically quantify this risk.
However, existing methods only measure average-case risk for randomly drawn
records, which might conceal the risk to vulnerable subgroups. To highlight
this issue, we define a subgroup-targeted membership inference game in which
the target pool is an explicit parameter, and instantiate it with an audit of
32 proxies under three scenarios with different levels of attacker knowledge,
across four datasets, three generators (DP-SGD fine-tuning, API-based
prompting, and activation steering), and five privacy budgets. The audit shows
that synthetic releases leak subgroup membership and that prior attacks
systematically underestimate this leakage. DP is effective at the aggregate
level: it substantially reduces average leakage at every budget we test. Three
observations temper this picture. First, the remaining leakage is concentrated rather than spread out: under DP, a tenth of the records carries roughly 40\% of it. 
Second, the protection DP delivers in practice is uneven: within its
worst-case guarantee, the noise removes more of the measured leakage from
random records than from high-risk ones---and a merged-pool audit that
scores both record types against shared negatives confirms this at the
record level.
Third, \emph{which} records leak
proves to be a property of the release mechanism rather than of the record
alone, so record-level risk cannot be assessed independently of the release.
\end{abstract}


\input{1_introduction}
\input{2_related_works}
\input{3_method}
\input{4_experiment_setup}

\input{5_results}

\input{6_conclusion_limitations}

\section*{Acknowledgments}
This research is part of the IN-CYPHER programme and is supported by the National Research Foundation, Prime Minister’s Office, Singapore under its Campus for Research Excellence and Technological Enterprise (CREATE) programme.

\bibliographystyle{unsrtnat}  
\bibliography{references}  

\newpage
\input{Appendix}

\end{document}

%% file: 1_introduction.tex
\section{Introduction}

\begin{figure*}[t]
  \centering
  \includegraphics[width=0.95\linewidth]{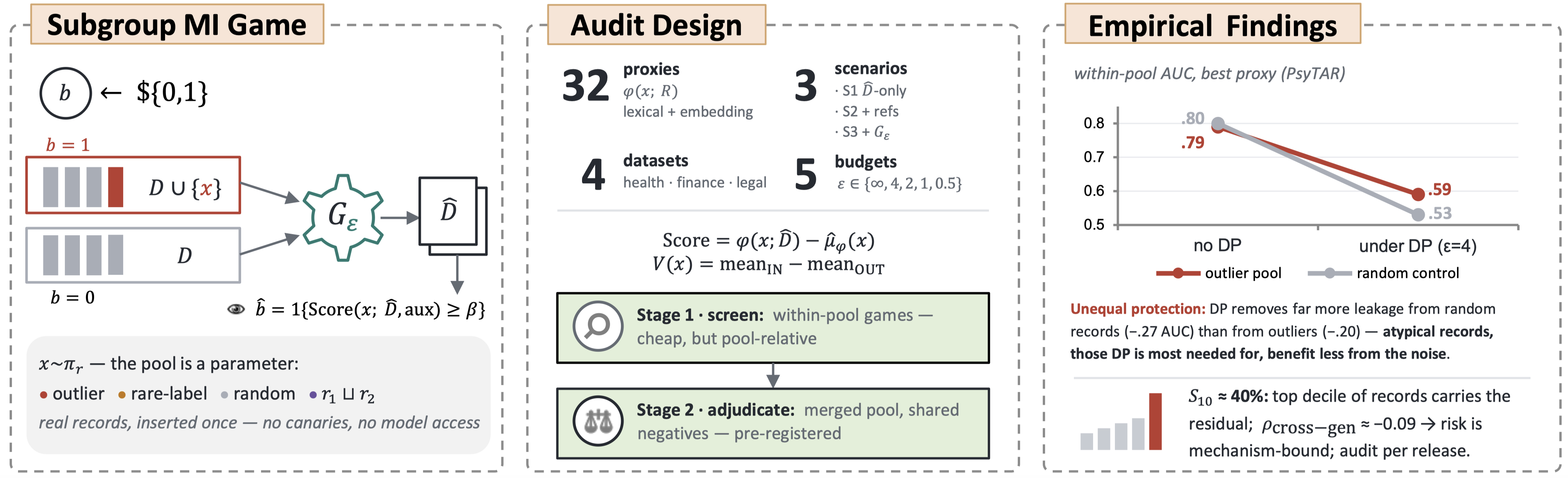}
  \caption{Overview: a pool-parameterized membership game (left), the audit instantiating it (middle), and the central finding (right) — DP removes less leakage from atypical records than from random ones, and the residual concentrates on a top decile that no record attribute predicts.}
  \label{fig:overview}
\end{figure*}

Synthetic data bear the potential to be used as a privacy-preserving substitute for direct data sharing in high-stakes domains such as healthcare~\citep{wu2025term2note}, finance~\citep{assefa2020generating,fca2024synthetic}, and law~\citep{zhao2025controlled,bellovin2019privacy}
By releasing generated records instead of the original dataset, practitioners hope to retain utility while reducing disclosure risk~\cite{stadler2022synthetic}.
However, synthetic data are not inherently safe: when the generation pipeline overfits or memorizes training examples, the released synthetic dataset can still reveal whether particular records were used in training~\cite{meeus2025canary,sun2025synbench}. This risk is most fundamentally quantified empirically via Membership Inference Attacks (MIA)~\cite{shokri2017membership}: an adversary is tasked to determine, given a target record and the output of a computation (i.e., synthetic dataset) whether the target was used in the computation (i.e. membership). Repeating this over an evaluation set of target records indicates the average ability of an adversary to distinguish members from non-members~\cite{carlini2022membership,zarifzadeh2023low}. Reducing (or bounding) the adversary's MIA capability gives a notion of \emph{plausible deniability}: if it is impossible to tell whether an individual's data was used in the computation, all subsequent inferences regarding that data (e.g., attribute inference) might not hold. 

Conducting MIA audits on synthetic texts is hard: an attacker needs to extract information from a whole dataset (i.e. hundred of individual records) and for unstructured data like text, no obvious aggregate statistics, as is the case for e.g., tabular data~\cite{stadler2022synthetic}, exist. 
Therefore existing privacy studies for synthetic data tend to omit membership inference audits~\cite{wu2025term2note,yue2023synthetic}. A tempting alternative is to evaluate membership inference of random private records on the \emph{model} that was trained to generate synthetic data, such as the state-of-the-art RMIA~\cite{zarifzadeh2023low} attack. These attacks assume stronger adversary knowledge: a release-based adversary lacks access to the generator, and cannot be applied to training-free synthesis methods~\cite{Xie2024DifferentiallyText,banayeeanzade2026epsvec}.
Canary-based attacks~\cite{meeus2025canary}, meanwhile, insert crafted worst-case sequences (potentially multiple times), which, while yielding upper bounds on memorization, evaluate the privacy leakage risk of synthetic examples instead of naturally occurring ones.
However, sampling random (or artificial) target records reports a generic \emph{average-case} risk and can miss an important detail: privacy leakage may be \emph{concentrated} on a small, high-risk subset of records: samples that are unusually distinctive relative to the background distribution~\cite{feldman2020longtail,carlini2019secret,bagdasaryan2019disparate}, rather than distributed uniformly across the population. Low empirical MIA scores in average-case scenarios may therefore underestimate privacy harms for such outlier samples.

To investigate this issue, we study \emph{subgroup-targeted} membership inference for synthetic data releases. We formalize a membership inference game parameterized by the \emph{target pool}: target records are drawn from a designated pool, chosen based on a desired characteristic. The rest of the private data to synthesise from are drawn from its complement. Because the pool can be arbitrarily defined, the same formulation can be used to conduct the high-risk subgroups audits, random control audits, and---by taking unions of pools---a design in which different kinds of records are scored against a common reference distribution. This formulation isolates whether pool membership creates disproportionate inference risk, and is particularly relevant for privacy auditing, where concentrated-risk failures matter more than average-case results.

To conduct a rigorous membership inference audit within this framework, we adapt normalized-evidence membership inference attacks~\cite{meeus2025canary,zarifzadeh2023low} to synthetic data \emph{releases}:  The exact likelihood-ratio test---how much more probable the observed release dataset is, when the target is a member ---is intractable to calculate for a whole dataset (rather than model weights): therefore we replace it with  easily measurable proxy scores, such as n-gram overlap and embedding distance, all normalized by scenario-dependent calibration.
This allows us to cleanly separate the evidence of \emph{proxy}, which is purely extracted from releases (lexical or embedding-based), for a target record, from \emph{scenario-specific calibration}, which normalizes that evidence according to the attacker's knowledge, and the same proxy family can be calibrated under progressively stronger attacker capabilities.

This allows us to model attacker capability as an explicit axis of the audit:
We study three release-based attacker scenarios---a \emph{release-only} attacker, a
\emph{raw-reference} attacker, and a \emph{generator-assisted release} attacker---to
measure how much membership leakage is inferable from the release alone and what auxiliary
calibration and generator access can add.

Finally, because harm may be concentrated \emph{within} a pool, we complement aggregate
metrics with per-record vulnerability scores obtained from repeated attack episodes,
and ask whether a data holder could predict, from intrinsic record attributes, for which
records membership is more likely to be inferred.

Our experiments across four datasets, three DP text generators, and five privacy budgets show that this risk is real and structured: it concentrates on few records, differs across pools, and depends on the mechanism.
On high-risk records, the best attacker reaches AUC $0.79$--$0.83$, identifying 34–61\% of true members at a false-positive rate below 5\% when the generator is non-private, 39--52\% of them exposed to an attacker that sees nothing but the release, and leakage persists above chance at every DP budget (AUC $0.55$--$0.65$, TPR up to $\approx$18\% against a 5\% chance rate).
DP is effective in aggregate---average AUC falls substantially at every
budget. But its protection is uneven in two ways. Across pools, DP removes
less leakage from high-risk records than from random ones: without DP,
targeting a high-risk pool buys the attacker little or nothing, yet under
every DP budget we test, the high-risk pools remain more distinguishable
than random controls. And within a pool, the residual
leakage is concentrated: a tenth of the records carries roughly 40\% of
the per-record vulnerability---up to 72\%---at every \(\varepsilon\),
though which records occupy that top decile changes from release to
release.

\paragraph{Contributions.}
\begin{itemize}[leftmargin=*,itemsep=2pt,topsep=2pt]
  \item \textbf{A pool-parameterized membership game} whose target pool is
  an explicit parameter, instantiating high-risk audits, random
  controls, and union pools within one formulation (\S\ref{sec:game}).
  \item \textbf{A comprehensive release-based audit}: 32 proxies, three
  attacker scenarios, four datasets, three generators, five budgets.
  Leakage persists above chance at every budget, added attacker knowledge
  reshuffles which proxy wins rather than how much leaks, and no prior
  fixed attack matches the audit (best AUC in 69/78 cells)
  (\S\ref{sec:results-leakage}--\ref{sec:results-baselines}).
  \item \textbf{DP's protection is uneven}: it removes less leakage from
  high-risk pools than from random controls, and a tenth of the records
  carries ${\sim}40\%$ of what remains at every $\varepsilon$
  (\S\ref{sec:results-concentration}, \S\ref{sec:control-gap}).
  \item \textbf{Which records leak is a property of the mechanism, not the record}: 
  vulnerability does not transfer across generators and no
  intrinsic attribute predicts it, so risk cannot be triaged from the data
  alone and must be audited per release (\S\ref{sec:attributes}).
  \item \textbf{A two-stage audit protocol.} Within-pool scores are only
  meaningful relative to their own pool, so we screen with full within-pool games and adjudicate flagged cells with a merged-pool audit that scores all record types against shared negatives (\S\ref{sec:two-stage}).
\end{itemize}

%% file: 2_related_works.tex

\section{Related Work}
\label{sec:related}

\paragraph{Membership inference against models.}
Membership inference attacks (MIA) ask whether a specific record was part of a model's training set, and have become a standard empirical measure of privacy leakage since the shadow-model formulation of~\citeauthor{shokri2017membership}.
\citeauthor{carlini2022membership} recast MIA from first principles,
showing that per-example calibration and evaluation at low false-positive rates
(rather than average accuracy) are essential to expose high-confidence leakage;
LiRA and its successors operationalize this through likelihood-ratio tests against
shadow models.
RMIA~\cite{zarifzadeh2023low} reduces the cost of this approach by normalizing a
target's membership evidence against a model-averaged baseline that approximates the
marginal probability of the record, yielding strong attacks with far fewer reference
models.

Our work inherits RMIA's normalized-evidence view and \citeauthor{carlini2022membership}['s] 
emphasis on
low-FPR evaluation, but departs from the model-access assumption common to this line:
our adversary never queries a model or observes its likelihoods, and instead reasons
only from a released synthetic dataset.

\paragraph{Privacy of synthetic data.}
A parallel line studies whether releasing synthetic records, rather than a trained
model, protects the underlying data.
\citeauthor{stadler2022synthetic} showed that synthetic tabular data does not
provide privacy for free and can remain vulnerable to inference attacks, and
subsequent frameworks such as TAPAS~\cite{houssiau2022tapas} and
density-based attacks like DOMIAS~\cite{vanbreugel2023domias}
formalized membership inference directly against synthetic releases, typically by
comparing a target's local density under the synthetic distribution to that under a
reference distribution.
These works focus on tabular data and average-case risk.
We adopt the same release-based threat model---the adversary sees only synthetic
outputs and auxiliary reference data---but target text generation, high-risk
subgroups, and the low-FPR regime rather than aggregate tabular attackability.

\paragraph{Auditing synthetic text and canaries.}
For text, memorization is commonly audited by inserting crafted \emph{canaries} and measuring their exposure~\cite{carlini2019secret}.
\textsc{Canary}~\cite{meeus2025canary} adapts normalized-evidence MIA to LLM-generated synthetic text: because release likelihoods are unavailable, it replaces them with tractable proxy signals computed from synthetic outputs, and amplifies detectability by repeating each canary $n_{\mathrm{rep}}>1$ times.
This yields a valuable upper bound on worst-case memorization, but measures the exposure of artificial sequences designed to be memorized rather than of real records.
Our audit differs in target and intent: we insert naturally occurring subgroup records \emph{once}, so the reported leakage reflects the exposure of genuine members of a high-risk population, and we study a family of heterogeneous proxies together with scenario-dependent calibration rather than a fixed signal.
Closest to our setting, SYNBENCH~\cite{sun2025synbench} benchmarks DP text generators with a release-based membership attack on natural records, sharing our IN/OUT reference construction, but audits average-case risk with a single dataset-level n-gram signal and aggregate AUC. 
We instead make concentrated risk the object of the audit: a subgroup-targeted
lone-record game with matched random controls, heterogeneous proxies under explicit
attacker-capability scenarios, and evaluation at low-FPR operating points and the
level of individual records.

\paragraph{Differentially private text generation.}
Differential privacy~\cite{dwork2006calibrating} is the standard defense for private data synthesis.
Two design points are representative of current DP text generators.
DP-SGD--based methods~\cite{dp-transformers} fine-tune language models with clipped,
noised gradients and then sample synthetic text, enforcing privacy during training.
Inference-based methods such as Aug-PE~\cite{Xie2024DifferentiallyText} instead privatize
corpus statistics and prompt a foundation model to match them, enforcing privacy
through the released statistics rather than model training.
We audit both, across a range of budgets, and connect our game to DP through its
add-one-record adjacency relation, which upper-bounds any adversary's achievable ROC
at a given budget.

\paragraph{Subgroup, outlier, and disparate memorization.}
A growing body of work shows that memorization and privacy risk are not uniform across
a dataset.
Outliers and atypical examples are memorized more
readily~\cite{feldman2020longtail,carlini2019secret},
and privacy harms and DP's utility costs can fall unevenly across
subgroups~\cite{bagdasaryan2019disparate}.
Most closely related, prior MIA analyses have observed that average-case
metrics can understate the risk to the most vulnerable records~\cite{carlini2022membership,long2018understanding}.
We build on this observation but make the subgroup the explicit object of the audit:
we define a high-risk subgroup, formalize a subgroup-targeted release-based game with a matched random-target control, and characterize both whether the subgroup is disproportionately exposed and how that exposure is distributed \emph{within} the
subgroup.
To our knowledge, this concentrated-risk, release-based formulation for synthetic text
has not been studied under an explicit range of attacker capabilities.

%% file: 3_method.tex
\section{Method}

\paragraph{Overview.}
We formalize \emph{subgroup-targeted membership inference} for synthetic data release as a release-based hypothesis test \emph{parameterized by the target pool}: a selection rule is used to decide which records the adversary must infer membership for, and running the same game with different selection rules ---high-outlierness, rare-label, uniformly random, and composites of two---separates the effect of targeting from the leakage a release exhibits regardless of targeting (Sections~\ref{sec:game}--\ref{sec:pool-rules}). 
The game asks the adversary to distinguish two worlds: under $H_1$ the
training set contains the target (one record drawn from the pool) while
under $H_0$ (the null world) it contains no record of the pool at all.
Because extracting likelihoods from synthetic releases is intractable, the adversary scores membership through a family of lexical and embedding-space proxies
(Section~\ref{sec:proxy-framework}), calibrated under three progressively stronger
attacker scenarios (Sections~\ref{sec:scenarios}).
Repeated episodes yield a per-record vulnerability score \(V(x)\) for the concentration analysis of Section~\ref{sec:results-concentration}.

\subsection{Problem Setting}
\label{sec:setting}
Let \(\pi\) be a distribution over records \(x\in\mathcal X\), and \(C=(x_1,\dots,x_M)\) a finite candidate corpus drawn i.i.d.\ from \(\pi\). Let further \mbox{\(\hat D \leftarrow G_\varepsilon(D)\)} a mechanism to generate synthetic data, applied to a private training set \(D \subseteq C\) of size \(N\).
The adversary observes a target \(x\), the release \(\hat D\), and scenario-dependent side information \(\mathrm{aux}\) (Section~\ref{sec:scenarios}) and has to predict whether \(x \in D\).

\paragraph{Target pools.}
We treat the targeted records as a parameter of the audit.
A \emph{selection rule} \(r\) maps the corpus to a pool
\(\mathcal G_r = \sigma_r(C) \subseteq C\) with \(|\mathcal G_r| = n\);
the target is then drawn uniformly from the pool,
\mbox{\(\pi_r \sim \mathrm{Uniform}(\mathcal G_r)\)}, and the private training records
uniformly from its complement,
\mbox{\(\bar\pi_r \sim \mathrm{Uniform}(C \setminus \mathcal G_r)\)}.
Pools are defined relative to the population (as estimated by the corpus): whether a record is an \textsc{Outlier} (a ranking over $C$), \textsc{Random} (a uniform draw from $C$) or features \textsc{Rare} labels, depends on the population.

\subsection{Pool-Parameterized Membership Inference Game}
\label{sec:game}
\begin{definition}[Pool-Parameterized MI on Synthetic Release]
\label{def:game}
Fix a selection rule \(r\), \(\mathcal G_r=\sigma_r(C)\).
\begin{enumerate}[leftmargin=*, itemsep=1pt, topsep=2pt]
    \item Sample a target \(x \sim \pi_r\) and a fair bit
    \(b \overset{\$}{\leftarrow}\{0,1\}\).
    \item If \(b=1\), set \(D := D_0\cup\{x\}\) with \(D_0 \sim (\bar\pi_r)^{N-1}\);
    if \(b=0\), draw \(D \sim (\bar\pi_r)^{N}\).
    \item Release \(\hat D \leftarrow G_\varepsilon(D)\) and give
    \((x,\hat D,\mathrm{aux})\) to the adversary.
    \item The adversary predicts
    \(\hat b = \mathbf{1}\{\mathrm{Score}(x;\hat D,\mathrm{aux})\ge \beta\}\).
\end{enumerate}
We write $\mathrm{Adv}_r$ for the adversary's advantage in the game under
rule $r$, reported as ROC AUC and, where relevant, TPR at low FPR.
\end{definition}

Equivalently the attack distinguishes \(H_1: D = D_0\cup\{x\}\) from
\(H_0: D \sim (\bar\pi_r)^{N}\) (null world), with \(x\sim\pi_r\) in both worlds: only membership
status differs, so the game isolates the evidence a release carries about pool records.
Taking \(r=\textsc{Random}\) gives \(\pi_r = \bar\pi_r = \mathrm{Unif}(C)\), recovering
the average-case game, so \(\mathrm{Adv}_r - \mathrm{Adv}_{\textsc{Random}}\) measures
what targeting buys the adversary. 
Unlike RMIA and \textsc{Canary}, which fix the target set implicitly, 
the pool is an explicit parameter, which is what makes the selected-vs-control comparison possible.

\begin{remark}[Composite pools]
\label{rem:composite}
For \(r = r_1 \sqcup r_2\), step~1 also records the constituent the target came from and
\(\bar\pi_r\) is uniform over \(C\setminus(\mathcal G_{r_1}\cup\mathcal G_{r_2})\).
Both constituents are then scored against one reference distribution, so their
advantages are directly comparable and per-record risk can be regressed on attributes
over the union of the two ranges---neither of which holds when \(\mathrm{Adv}_{r_1}\)
and \(\mathrm{Adv}_{r_2}\) come from separate within-pool games
(\S\ref{sec:gap-within}).
\end{remark}

\subsection{Target-Pool Selection Rules}
\label{sec:pool-rules}
We study four rules:
\begin{itemize}[leftmargin=*, itemsep=1pt, topsep=2pt]
  \item \textsc{Outlier}: the \(n\) records of highest outlierness according to a specified measure, representing atypical records.
  \item \textsc{Rare}: records with a low-frequency label.
  \item \textsc{Random}: \(n\) records drawn uniformly from \(C\): a \emph{control} that removes selection while preserving everything else.
  \item $r_1 \sqcup r_2$: a \emph{composite} pool joining the two constituent pools (Remark~\ref{rem:composite}).
\end{itemize}
Under every rule the remaining records are partitioned into pools of private candidates auxiliary reference data by the same procedure (\S\ref{sec:subgroup-construction}). The generation pipeline, \(|C|\), \(N\), pool construction, proxy family, and how negatives are sampled is fixed. 
Thus, instantiations differ only in \(r\), so any observable difference is attributable to the choice of the pool. 
A target pool generated by one single rule supports a \emph{within-pool} game whose scores are calibrated against that pool's own complement---cheap (one release each) but pool-relative and not comparable across rules (\S\ref{sec:gap-within}).
Composite pools remove this at the cost of one extra release per pair.
We therefore screen with single rule pools and use composite pools for the cross-rule and per-record comparisons that screening flags.

\subsection{Release-Based Proxy Auditing}
\label{sec:proxy-framework}
An ideal attack would directly compare release likelihoods,
\(\Lambda_r(x;\hat D) = P(\hat D\mid H_1)/P(\hat D\mid H_0)\) with
\(P(\hat D\mid H_1)=\mathbb{E}_{D_0\sim(\bar\pi_r)^{N-1}}[P(\hat D\mid D_0\cup\{x\})]\)
and \(P(\hat D\mid H_0)=\mathbb{E}_{D\sim(\bar\pi_r)^{N}}[P(\hat D\mid D)]\).
As these are intractable to compute, the adversary instead measures a set of proxy metrics
\(\phi(x;R)\), where the observation object \(R\) is the target release \(\hat D\), a
raw reference set \(D_{\mathrm{ref}}^{(j)}\), or a synthetic reference release
\(\hat D_{\mathrm{ref}}^{(j)}\), depending on the scenario.
The same set of proxies is used for all setups (\S\ref{sec:scoring}); only the reference objects
available for calibration change, and two game instantiations differ only through
\(\bar\pi_r\).
Some proxies are based on probability (\(n\)-gram log-probability, Gaussian
log-likelihood); others (similarity, overlap, distance) are treated as empirical
membership-evidence scores compared against a scenario-dependent baseline.

\subsection{Attacker Scenarios and Reference-Based Calibration}
\label{sec:scenarios}\label{sec:calibration}

For each selection rule, we study three attackers, all with access to \(\hat D\) and \(x\), but with differing access to auxiliary reference information. 

\emph{Scenario~1 (release-only)} has no reference data and no generator access, so it uses unnormalized evidence \(\mathrm{Score}^{(1)}_\phi=\phi(x;\hat D)\)---the weakest form of attacker with the lowest assumptions.
\emph{Scenario~2 (raw-reference)} cannot query the generator but has access to auxiliary data from \(\bar\pi_r\).
\emph{Scenario~3 (generator-assisted)} additionally has release-level query access to the same pipeline.

Raw evidence \(\phi(x;\hat D)\) conflates membership signal with the target's intrinsic difficulty.
Following RMIA~\cite{zarifzadeh2023low} and \textsc{Canary}~\cite{meeus2025canary},
we calibrate against the support \(x\) receives when its membership is marginalized,
\(\mu^\phi(x) = \tfrac{1}{2}\bigl(\mathbb{E}[\phi(x;\hat D)\mid H_1]
+ \mathbb{E}[\phi(x;\hat D)\mid H_0]\bigr)\).
the release-based analogue of RMIA's \(\Pr(x)\), estimated without bias by averaging
over references that contain \(x\) in exactly half of the cases (RMIA Eq.~9;
\textsc{Canary} Eq.~3).
We estimate \(\mu^\phi(x)\) with a common IN/OUT construction shared by the two
reference-based scenarios: build \(M_{\mathrm{ref}}\) reference sets from an auxiliary
pool drawn from \(\bar\pi_r\), insert the target into exactly \(M_{\mathrm{ref}}/2\) of
them (IN) and leave the rest (OUT), and average proxy evidence,
\[
\widehat{\mu}^{\phi}(x)
= \frac{1}{M_{\mathrm{ref}}}\sum_{j=1}^{M_{\mathrm{ref}}} \phi\bigl(x;R^{(j)}\bigr)
= \tfrac12\bigl(\widehat{\mu}^{\phi}_{\mathrm{IN}}(x)+\widehat{\mu}^{\phi}_{\mathrm{OUT}}(x)\bigr),
\]
where \(R^{(j)}\) is the raw set \(D_{\mathrm{ref}}^{(j)}\) in Scenario~2 and the
generated release \(\hat D_{\mathrm{ref}}^{(j)}\leftarrow G_\varepsilon(D_{\mathrm{ref}}^{(j)})\)
in Scenario~3.
Only Scenario~3 estimates \(\mu^\phi(x)\) in the release space; Scenario~2 applies the same construction to raw data and thus gives a generator-free calibration of target difficulty.

Writing \(\widehat{\mu}^{\phi}_{(s)}(x)\) for the scenario-\(s\) baseline
(\(\mathrm{raw}\) for \(s{=}2\), \(\mathrm{syn}\) for \(s{=}3\)), the score is
\[
\mathrm{Score}^{(s)}_{\phi}(x;\hat D) =
\begin{cases}
  \phi(x;\hat D), & s = 1,\\[6pt]
  \phi(x;\hat D) - \widehat\mu^{\phi}_{(s)}(x),
    & \begin{array}[t]{@{}l@{}}
        s \in \{2,3\},\\
        \phi \text{ in log-space},
      \end{array}\\[10pt]
  \dfrac{\phi(x;\hat D)}{\widehat\mu^{\phi}_{(s)}(x) + \delta},
    & s \in \{2,3\} \text{ otherwise.}
\end{cases}
\]
i.e.\ difference normalization for log-space proxies and ratio normalization for non-logarithmic (nonnegative) ones, matching each proxy's scale.
Comparing the three scenarios characterizes how targeted risk scales with available auxiliary information, and specifically whether release-level generator access adds power beyond reference calibration alone.

\subsection{Proxy-Based Scoring}
\label{sec:scoring}
We use \(m=32\) proxies \(\Phi=\{\phi_1,\dots,\phi_m\}\), each a distinct form of release
support or memorization (Appendix~\ref{app:proxies}).
We report attacks per individual proxy, as a preliminary study found that a learned meta-classifier does not outperform individual proxies (see Appendix~\ref{app:meta}).

\subsection{Attack Episodes and Sample-Level Vulnerability}
\label{sec:vulnerability}
Each attack is run over \(T\) \emph{episodes}---repeated plays of
Definition~\ref{def:game}, each drawing fresh membership bits, assembling \(D\),
generating a release, and scoring every target.
Let \(s_t(x)\) be the score of \(x\) in episode \(t\), and \(T_1(x),T_0(x)\) the member
and non-member episodes for \(x\) (we keep targets with at least one of each and require
a minimum count per condition).
The sample-level vulnerability is
\begin{equation}
  V(x)
= \frac{1}{|T_1(x)|}\sum_{t\in T_1(x)} s_t(x)
- \frac{1}{|T_0(x)|}\sum_{t\in T_0(x)} s_t(x).
\label{eq_sample_vul}
\end{equation}
Since raw scores are not comparable across proxies, \(V(x)\) is computed within a fixed
proxy and scenario, and cross-proxy/cross-scenario analyses use rank statistics.
For the same reason \(V(x)\) is not comparable across rules run as separate within-pool
games---they normalize against different complements---which is why composite pools
(Remark~\ref{rem:composite}) are needed to make that comparison well posed.
We then study the distribution of \(V(x)\), rank-based concentration, cross-scenario
consistency, and its correlation with attributes such as outlierness, lexical rarity,
duplication, and length, to determine whether leakage is broad or concentrated on a few
records.

%% file: 4_experiment_setup.tex
\section{Experiment Setup}
Our experiments broadly aim to investigate the following five research questions: 
\rqitem{rq:leak} Can synthetic text releases leak membership of high-risk records
under DP? 
\rqitem{rq:side} How does this risk scale with the attacker's knowledge? 
\rqitem{rq:baselines} Do existing release-based attacks suffice for subgroup
auditing? 
\rqitem{rq:uniform} Does DP protect the records within a target pool uniformly? 
\rqitem{rq:exposed} Can a data holder predict which records are at risk, and is the high-risk pool more exposed than ordinary records?

This section describes the setup of the experiments needed to pursue these questions.
\subsection{Datasets and Domains}
\label{sec:datasets}

We evaluate on four datasets across three domains, spanning two orders of
magnitude in length and label density (Table~\ref{tab:datasets}), to test
whether subgroup-targeted leakage persists across data sources with
different linguistic and label structure.
In \textbf{healthcare} we use \textsc{N2C2'08}~\cite{uzuner2009recognizing}, clinical
discharge summaries annotated for obesity and co-morbidities, and
\textsc{PsyTAR}~\cite{Zolnoori2019}, social-media posts labeled for adverse drug
effects---formal clinical documentation versus informal user-generated discourse.
In \textbf{finance} we use \textsc{DMSAFN}~\cite{dmsafn},
financial-news text.
In \textbf{legal} we use \textsc{EurLex}~\cite{chalkidis2019large}, EU legislative
documents annotated with EuroVoc concepts, reduced to the eight most frequent.

\begin{table}[t]
\centering
\footnotesize
\setlength{\tabcolsep}{3.5pt}
\begin{tabular}{lrrrr}
\toprule
 & \textsc{N2C2} & \textsc{PsyTAR} & \textsc{DMSAFN} & \textsc{EurLex} \\
\midrule
Domain            & health & health & fin.\ & legal \\
$N$               & 604 & 5102 & 3876 & 4942 \\
$|\mathcal{Y}|$   & 16 & 7 & 3 & 8 \\
Comb.             & 429 & 38 & 3 & 95 \\
Card.\ (mn/mx)    & 4.8/11 & 1.3/5 & 1.0/1 & 1.6/6 \\
Words (mn/md)     & 1124/1070 & 14/12 & 23/21 & 191/119 \\
Tok.\ md/p95/mx   & 1.8k/3.2k/4.0k & 15/40/194 & 26/57/134 & 166/852/2.0k \\
\bottomrule
\end{tabular}
\caption{Corpus-level statistics. \emph{Comb.}: distinct label combinations;
\emph{Card.}: positive labels/record; mn/md/mx = mean/median/max. Tokens use the
Llama-3.2-1B tokenizer. \textsc{DMSAFN} is single-label, the rest multi-label.}
\label{tab:datasets}
\end{table}

\subsection{Target-Pool Construction}
\label{sec:subgroup-construction}

The game of Section~\ref{sec:game} is parameterized by a target pool
\(P \subset \mathcal D_{\mathrm{full}}\).
We construct three pools---two notions of ``high risk'' and a
control---and their compositions.

\paragraph{Outlier pool (geometric).}
We encode all records with a per-dataset sentence encoder (see \ref{app:targetpool}) 
and take records flagged by a high-percentile distance-to-mean
threshold combined with an optional Local Outlier Factor criterion for locally isolated records.

\paragraph{Rare-label pool (semantic).}
As an embedding-free notion of high risk, we select records whose label combinations
fall below a frequency threshold (see \ref{app:targetpool}).
Notably, the rare-label pool shares \emph{zero}
records with the outlier pool \(O\) where applicable, even though by design, they are not mutually exclusive.

\paragraph{Random pools (control).}
For each dataset we draw three independent uniform samples
(\texttt{rand0/1/2}), size-matched to the \emph{outlier} pool, that indicate how much outlier subgroup MIA signal stand out against a random-population control for the dataset. 
The rare-label pools are compared against the same controls.

\paragraph{Induced partition.}
Each pool \(P\) induces its own partition
\(\mathcal D_{\mathrm{full}} = P \cup C \cup R\) (disjoint), where \(C\) builds the
private training sets and \(R\) supplies attacker-side iid auxiliary data.
The partition is shared across the three attacker scenarios, giving a consistent comparison of attack
strength across both scenarios and pools.

\subsection{Synthetic Text Generators}
\label{sec:generators}

We use three baselines at distinct design points for DP text synthesis.
\textbf{Aug-PE}~\cite{Xie2024DifferentiallyText} is API-based: it estimates privatized
feature statistics from the corpus and iteratively queries an LLM to generate and evolve
matching text, enforcing privacy through the statistics rather than through DP training.
\textbf{DP-Transformers} (\textbf{DP-Gen})~\cite{dp-transformers} is end-to-end neural:
it fine-tunes with DP-SGD (per-example clipping, additive noise) and samples by standard
decoding, enforcing privacy during training.
\textbf{EPSVec}~\cite{banayeeanzade2026epsvec} steers a frozen LLM with a privatized
\emph{dataset vector}---per-layer hidden-state differences between sensitive examples and
model-generated references, clipped and Gaussian-noised, then added at every decoding
step---enforcing privacy once at vector release and decoupling the privacy cost from the
number of samples.

\paragraph{Backbones.}
All three share the same backbone family and scale (Llama-3.2-1B), isolating
the privacy mechanism from generator capacity; DP-Gen and EPSVec use the
base checkpoint, while Aug-PE's prompt-driven generation requires the
instruction-tuned one.

\subsection{Baselines}
\label{sec:setup-baselines}

We compare against the two release-based MIA families applicable to our threat model,
re-instantiated in our game under the identical protocol (same pools, releases, trial
plan).
\textbf{Canary's Echo}~\cite{meeus2025canary} scores membership from the
release through two signal families---$n$-gram support under a model fit
on the synthetic text, and embedding similarity ($\mathrm{SIM}_{\mathrm{emb}}$, $k{=}25$)---optionally calibrated against
reference models via RMIA~\cite{zarifzadeh2023low}; we port all four
published arms unchanged, fitting references on raw text in Scenario~2 and
shadow-generator releases in Scenario~3.
Their targets are canaries injected 12--16 times while ours are real records appearing
once, so their published AUCs are not head-to-head comparable.
\textbf{DOMIAS}~\cite{vanbreugel2023domias} scores \(\log p_G(x)-\log p_R(x)\); its
published settings have \(n/d\geq 78\) whereas ours has \(n/d\approx 0.26\)--\(1\), where
a full-covariance Gaussian is rank-deficient, so we sweep four estimators
(Gaussian/KDE \(\times\) raw/PCA, fit only on attacker-accessible data) and report the
best.
For fairness, each family is collapsed to its best arm per
(dataset, generator, \(\varepsilon\)) cell, mirroring our own worst-case selection.

\subsection{Proxy Families}
\label{sec:proxies}

Proxies are scores \(\phi(x;R)\) as in Section~\ref{sec:scenarios}, with the
observation object \(R\) set by the scenario; calibration is per feature, and
Table~\ref{appd:tab:proxy_families} lists all proxies and hyperparameters.
\emph{Lexical.} \quad A retrieve-then-compare design: \(x\) is a BM25 query retrieving its
top-\(k\) candidates from \(R\) (\(k=50\)), over which we compute
surface-overlap statistics ranging from character \(n\)-gram containment
(\(n=5\))\footnote{Unlike prior audits that fit the \(n\)-gram model on the
whole release~\cite{meeus2025canary,sun2025synbench}, we fit it only on the
top-\(k\) retrieved candidates, localizing evidence to where single-record
memorization surfaces.} to an entity/number F1, each aggregated by max and
mean (Table~\ref{appd:tab:proxy_families}); an add-1 bigram model fit on
the candidates scores verbatim reuse via \(\log p(x)\).
\emph{Embedding-space.} \quad Records are embedded by the per-dataset scoring encoder; relative to
\(R\) we collect neighborhood similarity (cosine, dot-product, CSLS),
distance and density statistics (Euclidean, Mahalanobis), and a Gaussian
log-likelihood \(\ell_R(x)\) (Table~\ref{appd:tab:proxy_families}).

\subsection{Evaluation Protocol}
\label{subsec:eval_protocol}

Fix a target pool \(P\) (Section~\ref{sec:subgroup-construction}).
We run the game of Definition~\ref{def:game} over 100 \emph{game instances}, each
producing a dedicated release \(\hat D_i \leftarrow G_\varepsilon(D_i)\) per
(dataset, generator, \(\varepsilon\)) setting; the private set \(D_i\) is built from
the candidate pool \(C\) induced by \(P\).
For \(b_i=1\), the positive target \(x_i^+ \in P\) is fixed across rounds and scored
once.
For \(b_i=0\), \(D_i\) contains no record of \(P\), so any \(x\in P\) is a valid
non-member; we pre-select a fixed pool \(\mathcal P_i \subset P\) of \(K=20\) candidates
drawn without replacement, distinct across negative instances.
Positives and negatives are thus drawn from the \emph{same} pool \(P\), so the
resulting ROC measures membership signal within \(P\) rather than any difference
between pools.
Across \(T=50\) \emph{rounds}, one candidate is drawn uniformly from \(\mathcal P_i\)
per negative instance; the pools \(\{\mathcal P_i\}\) and per-round assignments
\(\{k(t)\}\) are fixed across all generators and \(\varepsilon\), so cross-method
differences reflect the generator's effect rather than negative-sampling variance.
The round-\(t\) evaluation set is
\begin{equation}
    \mathcal{E}^{(t)}
    = \bigl\{(x_i^+,\hat{D}_i) : b_i{=}1\bigr\}
    \cup \bigl\{(x_{i,k(t)}^-,\hat{D}_i) : b_i{=}0\bigr\}.
    \label{eq:eval_set}
\end{equation}
The protocol is identical for every pool instantiation; only \(P\) (and hence the
induced \(C\) and \(R\)) changes.

\paragraph{Setup and intervals.}
Scenarios follow \S\ref{sec:scenarios} with $M_{\mathrm{ref}}=4$ (two IN, two OUT)
per target. All $T=50$ rounds share the same 100 releases and differ only in sampled
negatives, so the reported CIs capture negative-sampling variability alone: they are
narrow by construction and do not reflect uncertainty over independent pipeline
re-runs.

\subsection{Evaluation Metrics}
\label{subsec:metrics}

\paragraph{AUC-ROC.}
Our primary metric is AUC-ROC, the probability that a random member scores higher
than a random non-member.
We fit one ROC per round over \(\mathcal{E}^{(t)}\) and report the mean of the
resulting \(T=50\) values with a 95\% CI, \(\hat\mu \pm 1.96\,\hat\sigma/\sqrt{T}\).
Unless noted, each setting reports the proxy with the highest mean AUC.
Since that proxy is chosen on the same episodes it is scored on, these numbers are an
\emph{empirical worst-case audit}: what the best proxy would have revealed in
hindsight, not what a deployable attack achieves.
The meta-classifier (Appendix~\ref{app:meta}) selects on training episodes only and is
the deployable counterpart.

\paragraph{TPR at low FPR.}
A useful membership attack must be confident, not just better than chance, so we also
report TPR at FPR\,\(\leq 0.05\): the fraction of members recovered while falsely
flagging at most 5\% of non-members.
By default this uses the AUC-selected proxy.

%% file: 5_results.tex
\section{Experimental Results}

\input{tables/table_main_per_generator}

\subsection{Subgroup Membership Leakage Across Generators and Budgets}
\label{sec:results-leakage}

Answering \textbf{RQ1}, Tables~\ref{tab:main_dpgen}, \ref{tab:main_augpe}, and~\ref{tab:main_epsvec} report best-feature AUC and TPR at FPR\,$\leq$\,5\% for the outlier pools under each generator, scenario and $\varepsilon$ combination. Table~\ref{tab:main_rare} repeats the audit on the rare-label pools. Per-combination confidence intervals are given in Tables~\ref{tab:overall_main_dptransformers}--\ref{tab:overall_main_EPSVec} and~\ref{tab:overall_main_rare} (Appendix~\ref{app:more-results}).

\paragraph{Non-privately trained generators expose high-risk records at practically
exploitable rates.} When the generator is fine-tuned on the private data without DP,
releasing synthetic text is close to releasing a membership oracle for the high-risk
pool.
Under DP-SGD fine-tuning at $\varepsilon=\infty$, the best proxy reaches AUC
0.79--0.83 on \textsc{PsyTAR}, \textsc{DMSAFN}, and \textsc{EurLex}, identifying
34--61\% of true members at a false-positive rate below 5\%
(Table~\ref{tab:main_dpgen}).
A release-only attacker (S1) is already within $0.01$--$0.04$ AUC of the best calibrated
attacker for these combinations: with strong memorization, no auxiliary data is
needed.
This cannot be attributed to a serendipitous choice of proxy: even the
deployable fusion attack (Appendix~\ref{app:meta}, Tables~\ref{tab:meta-outlier}
and~\ref{tab:meta-rare}), which selects and weights proxies by nested cross-validation
without seeing the evaluation data, still reaches AUC $0.76$--$0.79$.
The spike is specific to fine-tuning: MIA scores on data generated by training-free generators 
is far below the reported \textsc{DPGen} numbers, but it also doesn't drop to chance levels, with scores largely unaffected by the application of DP (Tables~\ref{tab:main_augpe}
and~\ref{tab:main_epsvec}). 

\paragraph{DP suppresses most of this leakage, but not all of it.}
\textbf{At every finite budget we test, subgroup membership remains inferable above
chance.}
Across all datasets and scenarios under DP-SGD fine-tuning, the AUC for the best-performing proxy is no more than
$0.52$--$0.65$, and TPR@FPR\,$\leq$\,5\% reaches 18\%---3.6$\times$ the chance
rate---at $\varepsilon=4$ on \textsc{N2C2'08} and 16\% at $\varepsilon=0.5$ on
\textsc{DMSAFN}.
The drop from the non-private ceiling is large (from $0.80$ to $0.59$ on average),
confirming that DP-SGD effectively removes a large proportion of the membership signal.

\input{tables/table_main_rare_compact}

\paragraph{Rare-label pools confirm the pattern and expose leakage the outlier
audit misses.}
When the pool is selected by label statistics alone, similar observations apply.
Under non-private fine-tuning, the rare pools leak at least as strongly as the
outlier pools on the same corpora---best AUC 0.86 on \textsc{PsyTAR} (TPR 59\%)
versus 0.80 for the outlier pool---and under DP they sit on the same floor
(0.53--0.66 across generators and budgets; Table~\ref{tab:main_rare}).
Notably, for \textsc{N2C2'08} the
non-private audit stays relatively weak ($0.64$ AUC), while its \emph{rare-label} pool---records with at least eight co-morbidities---reaches $0.77$ with TPR 28\% (Table.~\ref{tab:main_rare}), suggesting that high-risk subgroups that have a disproportionately high membership signal differ across datasets. 

\subsection{Scaling with Attacker Side Information}
\label{sec:results-scenarios}

\begin{figure}[!htbp]
\centering
\includegraphics[width=0.6\linewidth]{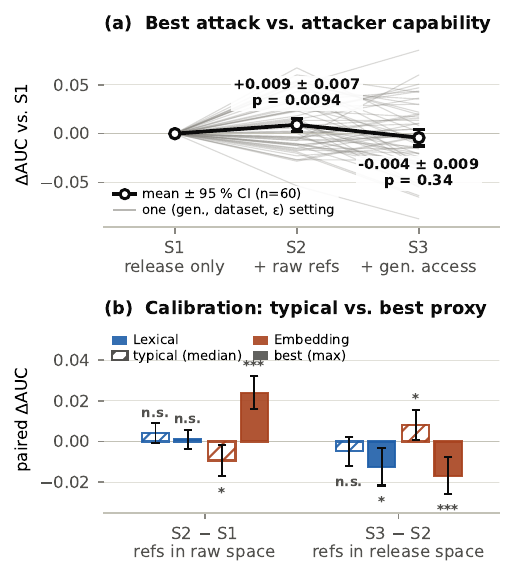}
\caption{Effect of attacker side information over all 60
(generator, dataset, $\varepsilon$) settings.
\textbf{(a)}~Change in best-feature AUC relative to the release-only attacker S1
(grey: individual settings; black: mean with 95\,\% CI).
\textbf{(b)}~Paired change of the median-performing and the best-performing proxy of
each view at each step of the scenario ladder.}
\label{fig:scenario_scaling}
\end{figure}

Answering \textbf{RQ2}: our three scenarios add capability monotonically (\S\ref{sec:scenarios}), and a natural expectation is that more capability yields a stronger attack. Our results indicate that this is not the case across all 60 combination of generator, dataset and $\varepsilon$, raw-reference calibration (S2) shifts the best-feature AUC by $+0.009$ on average ($95\,\%$ CI $\pm0.007$) relative to S1, and generator access (S3) by $-0.004$ ($\pm0.009$). 
S2 yields a statistically detectable but practically negligible gain ($+0.009$, $95\%$ CI $\pm0.007$); S3 is indistinguishable from zero ($-0.004 \pm 0.009$).
(Figure~\ref{fig:scenario_scaling}a). 
Side information changes \emph{which} proxy wins, not \emph{how much} leaks: under S2 the \emph{median} embedding proxy loses $0.013$ AUC while the \emph{best} gains $0.02$ (Figure~\ref{fig:scenario_scaling}b), so the attack improves only if the attacker re-selects its proxy afterward---and the winning scenario and channel (\textcolor{lexcol}{\textbf{L}} vs.\ \textcolor{embcol}{\textbf{E}}) indeed change from cell to cell across Tables~\ref{tab:main_dpgen}--\ref{tab:main_epsvec}, with no majority. 
These findings underline the importance of considering (and optimizing for) the best proxy for each setting (i.e, dataset, background knowledge and DP protection) separately, to not under-report MIA scores due to sub-par proxy choice.

\subsection{Comparison with Prior Release-Based Attacks}
\label{sec:results-baselines}

\input{tables/table_baseline_outlier}

\input{tables/table_baselines_rare}
Answering \textbf{RQ3}: Section~\ref{sec:results-scenarios} suggests that the most informative proxy moves from setting to setting, so any attack that commits to one signal would under-measure leakage. 
We test this directly by re-instantiating the two release-based MIA families applicable to our threat model under the same protocol: \textsc{Canary}~\cite{meeus2025canary} and \textsc{DOMIAS}~\cite{vanbreugel2023domias} (see details in ~\ref{sec:setup-baselines}).
The two baselines fall short for different reasons. DOMIAS is
structurally confined to one kind of evidence: all four of its estimators are density models over embeddings, so leakage that surfaces as verbatim wording cannot be captured more explicitly than via embedding similarity. \textsc{Canary} can be instantiated with both $n$-gram and embedding similarity proxies but only one fixed signal of each kind, whereas the informative signal varies more finely: within each family (\textcolor{lexcol}{\textbf{L}} or.\ \textcolor{embcol}{\textbf{E}}), the best proxy is different based on the specific setup (indicated by superscripts of the \textsc{Ours} columns in Table~\ref{tab:baselines-main}). This variance holds for both baselines as well (Table~\ref{tab:baselines-main}, superscripts in \textsc{Canary} and DOMIAS columns): For \textsc{Canary}, the best proxy alternates between the choices of
$n$-gram, RMIA-combined, and similarity forms, and for DOMIAS's between its Gaussian, KDE, and PCA estimators---so even within a single family there is no configuration that is guaranteed to perform best across different settings.
Table~\ref{tab:baselines-main} also shows that our MIA setup significantly outperforms the baselines, on 30 of 36 settings (across both outlier and rare label pools). Notably, our methods recovers AUC up to 0.83 on real records that were inserted only once, which is above the 0.62--0.77 reported for \textsc{Canary}, when canaries were injected 12--16 times.

\subsection{Aggregate vs. Individual Protection: Leakage Concentration under DP}
\label{sec:results-concentration}

\begin{figure*}[t]
  \centering
  \includegraphics[width=\textwidth]{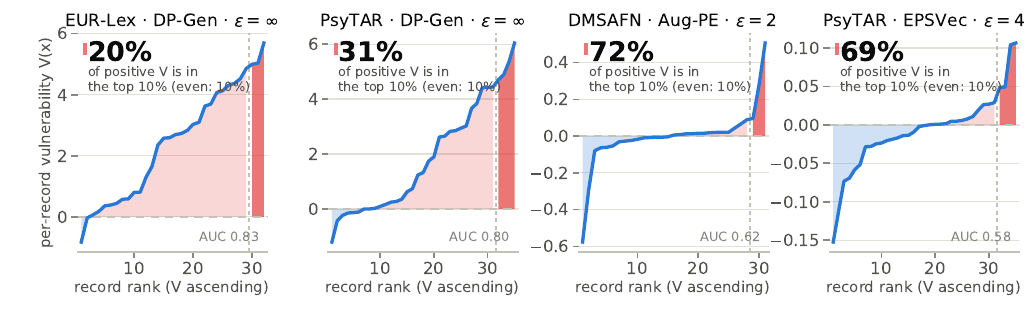}
  \caption{Sorted per-record vulnerability $V(x)$ for four representative cells
            (note the differing $y$-scales). The dark slice is the positive mass held by
            the top decile of records (dashed line: decile boundary); the printed
            percentage is the top-decile leakage share $S_{10}$
            (Eq.~\ref{eq:top-decile-share}; ${\approx}10\%$ under even leakage).
            Left: non-private fine-tuning leaks broadly; right: under suppressed
            memorization the residual is top-heavy.}
  \label{fig:v-distribution}
\end{figure*}

Answering \textbf{RQ4}, this section asks how the leakage of
\S\ref{sec:results-leakage} is distributed across records, measured by the
per-record vulnerability \(V(x)\) (Eq.~\ref{eq_sample_vul}), computed as
follows. For each \emph{combination} of dataset, generator and \(\varepsilon\), the scores of its best proxy---the same
worst-case selection as the main tables---are sign-aligned and
\(z\)-normalized, so magnitudes are comparable directly. \(V(x)\) is
then the record's mean score over the combinations's episodes in which it was the
inserted member, minus its mean over those in which it served as a
non-member candidate. A record receives a score only if it was both member and non-member at least once. Of the roughly 60 records in a pool,
19--36 qualify.

\paragraph{Top-decile leakage share.}
Order a combination's \(n\) qualifying records by decreasing \(V\), write
\(V^{+}_{(i)}=\max\{V_{(i)},0\}\), and let
\(k=\max\{1,\lfloor 0.1\,n+0.5\rfloor\}\).
The \emph{top-decile leakage share} is
\begin{equation}
  S_{10} \;=\; \frac{\sum_{i=1}^{k} V^{+}_{(i)}}{\sum_{i=1}^{n} V^{+}_{(i)}},
  \label{eq:top-decile-share}
\end{equation}
the share of the combination's total positive vulnerability that falls on its top tenth of records; records with \(V\le 0\) contribute to neither sum.
Even leakage gives \(S_{10}\approx 10\%\); values approaching 100\% mean the release's residual risk is disproportionately concentrated in a small subset records.
\(S_{10}\) is the percentage printed in each panel of
Figure~\ref{fig:v-distribution}. It is reported in full together with the share of records scoring \(V>0\) for all 55 configurations in Appendix~\ref{app:vdist}
(Table~\ref{tab:v-distribution-appendix}), and the medians and ranges below are
computed based on that table.

\paragraph{Without DP, fine-tuned releases leak broadly, not selectively.}
Under non-private DP-SGD fine-tuning, \textbf{most subgroup records (not a vulnerable few) carry positive membership signal}.
In the four \(\varepsilon=\infty\) fine-tuning combinations, 77--94\% of qualifying records (median 82\%) score \(V>0\), per-record magnitudes reach $50$--$100$ times those under any DP budget, and concentration is modest: \(S_{10}\) is 20--37\% (median
30\%), the lowest of any regime we audit.

\paragraph{Under fine-tuning, DP thins the leakage and shifts it toward the top
decile.}
In addition to changing overall membership inference risk, \textbf{DP changes how this risk is distributed: it is concentrated in a small subset of records}.
Across DP-Gen's 16 DP combinations, the share of records scoring positive falls from a
median of 82\% to 61\%, while \(S_{10}\) rises from a median of 30\% to 40\% (range
28--69\%) and exceeds the even-spread value in 15 of 16 combinations threefold, against
2 of 4 without DP (Appendix~\ref{app:vdist}).
While the mean observations suggest a trend, individual configurations deviate in both directions. For example, \textsc{DMSAFN} at \(\varepsilon=4\)
is strongly concentrated (29\% positive, \(S_{10}=69\%\)) while the same dataset at
\(\varepsilon=0.5\) remains broad (71\%, \(S_{10}=28\%\)). Therefore, we report this as suggestive rather than conclusive evidence for membership signal  concentration under DP.

\paragraph{Concentration is the default wherever bulk memorization is absent---with
or without DP.}
For the two mechanisms that don't exhibit the no-DP leakage spike, the shape
of the vulnerability distribution is already top-heavy at zero noise and barely moves
with the budget.
EPSVec's \(\varepsilon=\infty\) combinations have \(S_{10}\) of 36--82\% (median 47\%;
\textsc{PsyTAR} reaches 82\% with no noise at all), statistically indistinguishable
from its DP combinations (median 41\%); Aug-PE moves from a median of 31\% at its zero-noise
setting to 40\% under noise.
Pooling all 60 combinations, \(S_{10}\) exceeds three times the even-spread value in 50
(median 39\%, range 20--82\%).
Concentration therefore tracks the \emph{absence of bulk memorization}---whether
achieved by DP noise or by mechanism design---rather than DP itself, the record-level
counterpart of the mechanism-relative picture of \S\ref{sec:results-leakage}.

\paragraph{Reporting only average AUC scores misses the concentration of membership inference risk.}
Releases \textbf{with the same AUC can have very different top-decile shapes}, and
the \textbf{concentration of risk in the top decile is a per-release property} and not tied to a specific set of records.
For DP, a median of 40\% (up to max 72\%) of the residual positive vulnerability is concentrated in the top decile and, yet an aggregate
audit that certifies a release by mean AUC alone would report only the near-floor
we reported in Section~\ref{sec:results-leakage}.
Whether a record inhabits the top decile changes across different configurations, in line with prior findings~\cite{carlini2022privacy}. 
A high \(S_{10}\) suggests to audit each release rather
than maintaining a list of high-risk records; the next section shows that the top
decile's membership is in fact hard to predict from record attributes alone.


\subsection{Intrinsic Attributes Do Not Predict Which Records Leak}
\label{sec:attributes}

If a data holder could identify at-risk records from the private data
alone, triage would be possible without a costly audit. We test five
release-independent attributes---text length, embedding outlierness (the
selection statistic of \S\ref{sec:subgroup-construction}),
near-duplication, rare-$n$-gram density, and entity/numeral
density---against the within-pool rank of $V(x)$ in all 42 pools the
audit covers. Scoring $V(x)$ requires choosing one scenario, proxy pair
per pool; we report a \emph{conservative} rule (chosen on an 80\%
training partition) as primary and the main tables' \emph{worst-case}
rule as a sensitivity analysis. Computation details, two selection-free
rules under which every conclusion is unchanged, and the full per-pool
matrix are in Appendices~\ref{app:attributes}
and~\ref{app:selection-robustness}.

\paragraph{Under the conservative rule, no attribute predicts
vulnerability}
\label{sec:attr-null}

\begin{figure}[t]
  \centering
  \includegraphics[width=0.7\columnwidth]{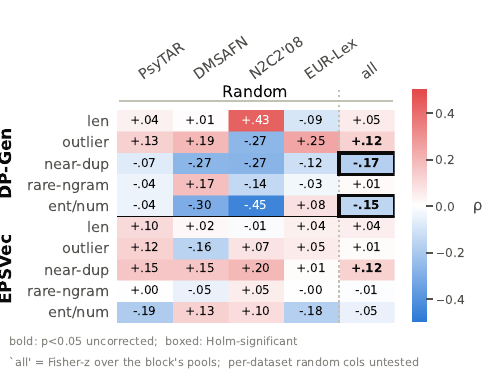}
  \caption{Spearman $\rho$ between five release-independent record attributes and per-record vulnerability rank in the random pools, per generator.
  Per-dataset columns are Fisher-$z$ combinations of that dataset's three disjoint draws and carry no test; \emph{all} pools the twelve draws (337 records) with Holm correction over the five attributes within a generator.
  Bold: nominal $p<0.05$; boxed: Holm-significant. The outlier and rare-label pools are in Figure~\ref{fig:attributes-heatmap-full}, Appendix~\ref{app:attributes}.}
  \label{fig:attributes-heatmap}
\end{figure}

\input{tables/table_control_dpgen_with_rare}

Of the 210 attribute--pool tests, none survives Holm correction.
Pooling the twelve random pools by Fisher's~$z$
(Figure~\ref{fig:attributes-heatmap}) surfaces only weak tendencies, all
under DP-SGD fine-tuning: near-duplication $-0.17$ (Holm $p=0.01$),
entity density $-0.15$ ($p=0.03$), outlierness $+0.12$ ($p=0.09$); for \textsc{EPSVec} every combined $\rho$ lies within $\pm0.12$.
Correlations of this size cannot be meaningfully used to predict membership inference risk from records.

\paragraph{Worst-case selection sharpens the tendencies but cannot
change their meaning}
\label{sec:attr-worstcase}

Re-selecting the proxy on all episodes strengthens the tendencies described above: outlierness $+0.20$, near-duplication $-0.20$, entity density $-0.15$, all clearing correction. Correlations for \textsc{EPSVec} remain essentially zero (Table~\ref{tab:selection-robustness}).  
Our findings suggest that which records a release exposes is dictated by the choice of release mechanism, DP budget, and other audit configurations and is not easily predictable from the records. Making triaging a priori infeasible.

\subsection{High-Risk vs.\ Random Pools}
\label{sec:control-gap}

In this section we answer the second part of \textbf{RQ5}, i.e., whether high-risk records are easier to infer membership comparing with ordinary records.
We proceed in two steps. 
First (\S\ref{sec:gap-within}), we run the
within-pool game on each outlier and rare-label pool and compare its
separability against a random pool under an identical protocol.
These gaps measure how distinguishable a pool's records are as members versus
as non-members, but the non-members are still selected within the pool, it cannot say whether an individual high-risk \emph{record} is more exposed (Remark~\ref{rem:composite}).
Second (\S\ref{sec:two-stage}), we answer the record-level question with
an experiment with composite pools: we merge an outlier pool with a
random pool, which allows us to score both record types against the same negatives, making the comparison well posed---run on two configurations (\textsc{EurLex} $\times$ DP-Gen $\times$ $\epsilon=\infty$; \textsc{PsyTAR} $\times$ DP-Gen $\times$ $\epsilon=4$), which in the first step are observed having statistically significant difference in AUC between outlier and control pools.

\paragraph{Within-pool distinguishability: outlier and rare-label pools
vs.\ random pools}
\label{sec:gap-within}

Table~\ref{tab:control-dpgen-or} reports the
gap in within-pool AUC between high-risk pools with random
controls, for \textsc{DP-Gen} and \textsc{EPSVec}, respectively. 
The two generators behave differently without DP and exhibit similar behavior once DP is applied. 
First, under \emph{non-private fine-tuning} (DP-Gen, $\varepsilon=\infty$), outlier pools are \emph{not} easier to attack: on two of three datasets they are significantly \emph{less} distinguishable than the controls (gaps $-0.01$ to $-0.06$;
\textsc{EUR-Lex} is the exception, $+0.14$/$+0.18$), while the rare-label
gaps are small and mixed in sign ($+0.02$ to $+0.03$ on \textsc{PsyTAR} and \textsc{N2C2'08}, $-0.04/0.05$ on \textsc{EUR-Lex}). 
Second, \emph{adding DP to fine-tuning} inverts this trend: 
at $\varepsilon=4$ all six of DP-Gen's
outlier gaps are positive ($+0.02$ to $+0.06$, all significantly), and its
rare-label gaps turn mixed ($-0.03$ to $+0.06$). 
Third, the training-free \textsc{EPSVec} shows inverted trends regardless of noise:
11 of 12 outlier gaps are positive ($+0.01$ to $+0.06$), its rare-label gaps
peak on \textsc{EUR-Lex} ($+0.05$ to $+0.09$), and its control AUC never
exceeds $0.56$.
Thus, when the generator memorizes broadly, as is the case  for non-private \textsc{DP-Gen},
\textbf{ordinary records are as easy do infer membership from as much as atypical ones}.
However, without broad memorisation, e.g., because of the use of DP or training-free methods, membership signal from random pools are closer to chance than while target from high-risk pools are left as the more distinguishable targets.

\paragraph{Membership inference game with unified pools}
\label{sec:two-stage}

\begin{table}[t]
\centering\footnotesize
\setlength{\tabcolsep}{3pt}
\caption{Merged-pool audits of the two flagged cells. Both record types sit in one pool and face the same negative candidates, so $\Delta$ (outlier-origin $-$ random-origin members) directly measures which records are easier to attack; absolute values are against the merged pool's negatives and are not comparable to the within-pool tables. Brackets: 95\% bootstrap over member records; \textbf{bold}: interval excludes zero.}
\label{tab:union-scoreboard}
\begin{tabular}{@{}llccl@{}}
\toprule
Cell & & Outl.\ & Rand.\ & $\Delta$ \\
\midrule
EUR-Lex, $\varepsilon=\infty$ & AUC & 0.73 & 0.67 & $+0.066$ {\scriptsize[-0.03,+0.16]} \\
 & TPR & 0.50 & 0.12 & $\mathbf{+0.378}$ {\scriptsize[+0.25,+0.51]} \\
\addlinespace[2pt]
PsyTAR, $\varepsilon=4$ & AUC & 0.58 & 0.46 & $\mathbf{+0.127}$ {\scriptsize[+0.02,+0.24]} \\
 & TPR & 0.07 & 0.03 & $+0.040$ {\scriptsize[-0.02,+0.11]} \\
\bottomrule
\end{tabular}
\par\vspace{1pt}\scriptsize TPR at FPR$\leq$5\%.
\end{table}

So far, the statements regarding membership signal differences between random and outlier targets were made at population level. 
This leaves  open one final question: is an outlier \emph{record}
easier to attack than a randomly drawn one? To answer it directly, we
merge the outlier pool with a random pool (the pools
share no record), rerun the membership game on that merged pool, schedule every record as the inserted member equally often, and draw negative candidates from the whole merged pool. With this setup,  both outliers and random examples are attacked under identical conditions and their scores can be compared directly.

Table~\ref{tab:union-scoreboard} shows the results: without DP the outlier
records' extra exposure shows up at the low-FPR end of the ROC; under DP it
survives only in the ranking, i.e., outlier records still rank
systematically higher than ordinary records in the attack's score ordering.
Without DP (\textsc{EUR-Lex}, $\varepsilon=\infty$), outlier records are far
more vulnerable: at 5\% false positives, 50\% of outlier members are
identified against 12\% of random members ($\Delta=+0.38$), while the
overall AUC difference is modest and not significant ($+0.07$). Under DP
(\textsc{PsyTAR}, $\varepsilon=4$), the advantage changes character: at 5\%
false positives neither type can be identified with confidence (TPR $0.07$
vs.\ $0.03$; the $+0.04$ difference has an interval that spans zero), yet
outlier members still rank above random members across the full score
range, with an AUC difference of $+0.127$---the one gap in the table whose
interval excludes zero.

%% file: tables/table_main_per_generator.tex
\definecolor{lexcol}{HTML}{1F5FA9}
\definecolor{embcol}{HTML}{A8431F}
\definecolor{best}{HTML}{2A78D6}

\newcommand{\ce}[2]{#1$_{#2}$}
\newcommand{\cL}[2]{\cellcolor{best!18}\bfseries
  #1$^{\textcolor{lexcol}{\mathrm{L}}}_{\mathbf{#2}}$}
\newcommand{\cE}[2]{\cellcolor{best!18}\bfseries
  #1$^{\textcolor{embcol}{\mathrm{E}}}_{\mathbf{#2}}$}
\newcommand{\epshead}{ & & $\bm{\infty}$ & $\bm{4}$ & $\bm{2}$ & $\bm{1}$ & $\bm{.5}$\\}
\begin{table*}[t]
\caption{Subgroup-targeted MI across privacy budgets $\varepsilon$ (columns) and
attacker scenarios S1--S3 (rows), for the three generators.
Each cell reports the best-proxy AUC with TPR\%@FPR$\le$5\% as a subscript (worst-case audit, \S\ref{subsec:metrics});
Shaded cells mark the best scenario per dataset and
$\varepsilon$, AUC ties broken by TPR; the superscript indicates the winning proxy
family, \textcolor{lexcol}{\textbf{L}}exical or \textcolor{embcol}{\textbf{E}}mbedding, respectively.
Subcaptions indicate the generator and distribution of best proxies per best scenario.
}
\label{tab:main}
\centering\footnotesize\setlength{\tabcolsep}{2pt}
\renewcommand{\arraystretch}{1.2}

\begin{subtable}[t]{0.32\textwidth}
\centering
\caption{\textsc{DP-Gen} (11\,\textcolor{lexcol}{\textbf{L}} vs.\ 9\,\textcolor{embcol}{\textbf{E}}).}
\label{tab:main_dpgen}
\begin{tabular}{@{}ll ccccc@{}}
\toprule
\epshead
\midrule
\dset{\textsc{PsyTAR}}
 & S1 & \ce{.79}{46} & \ce{.59}{12} & \ce{.59}{13} & \cE{.60}{11} & \ce{.61}{4}\\
 & S2 & \cL{.80}{53} & \cL{.59}{10} & \ce{.57}{6} & \ce{.57}{9} & \cL{.64}{6}\\
 & S3 & \ce{.76}{45} & \ce{.57}{15} & \cL{.59}{11} & \ce{.57}{8} & \ce{.63}{7}\\
\cmidrule(l){2-7}
\dset{\textsc{DMSAFN}}
 & S1 & \ce{.79}{39} & \cL{.60}{4} & \cL{.61}{5} & \ce{.64}{4} & \ce{.64}{6}\\
 & S2 & \cE{.79}{34} & \ce{.55}{5} & \ce{.59}{7} & \cL{.65}{5} & \ce{.64}{7}\\
 & S3 & \ce{.73}{42} & \ce{.52}{2} & \ce{.55}{6} & \ce{.61}{6} & \cL{.65}{16}\\
\cmidrule(l){2-7}
\dset{\textsc{N2C2'08}}
 & S1 & \ce{.55}{4} & \ce{.58}{16} & \ce{.57}{8} & \ce{.57}{13} & \ce{.56}{9}\\
 & S2 & \ce{.60}{11} & \ce{.59}{18} & \cE{.63}{13} & \ce{.61}{12} & \cE{.63}{9}\\
 & S3 & \cE{.64}{14} & \cL{.59}{4} & \ce{.56}{10} & \cE{.61}{9} & \ce{.57}{8}\\
\cmidrule(l){2-7}
\dset{\textsc{EurLex}}
 & S1 & \ce{.80}{52} & \ce{.59}{11} & \ce{.56}{6} & \cL{.58}{9} & \cL{.57}{6}\\
 & S2 & \ce{.83}{60} & \ce{.57}{10} & \ce{.56}{5} & \ce{.58}{9} & \ce{.56}{10}\\
 & S3 & \cE{.83}{61} & \cE{.59}{12} & \cE{.58}{4} & \ce{.55}{9} & \ce{.54}{8}\\
\bottomrule
\end{tabular}
\end{subtable}\hfill
\begin{subtable}[t]{0.32\textwidth}
\centering
\caption{\textsc{Aug-PE} (14\,\textcolor{lexcol}{\textbf{L}} vs.\ 6\,\textcolor{embcol}{\textbf{E}}).}
\label{tab:main_augpe}
\begin{tabular}{@{}ll ccccc@{}}
\toprule
\epshead
\midrule
\dset{\textsc{PsyTAR}}
 & S1 & \ce{.59}{11} & \ce{.59}{14} & \ce{.59}{12} & \ce{.59}{13} & \ce{.61}{12}\\
 & S2 & \cL{.64}{13} & \cL{.62}{12} & \cL{.65}{14} & \cE{.63}{15} & \cL{.63}{12}\\
 & S3 & \ce{.56}{1} & \ce{.56}{8} & \ce{.61}{4} & \ce{.58}{5} & \ce{.59}{7}\\
\cmidrule(l){2-7}
\dset{\textsc{DMSAFN}}
 & S1 & \cL{.61}{4} & \cL{.61}{5} & \ce{.61}{6} & \cL{.61}{5} & \cL{.61}{4}\\
 & S2 & \ce{.60}{5} & \ce{.59}{11} & \ce{.60}{4} & \ce{.60}{4} & \ce{.59}{4}\\
 & S3 & \ce{.58}{11} & \ce{.59}{8} & \cE{.62}{9} & \ce{.58}{8} & \ce{.55}{4}\\
\cmidrule(l){2-7}
\dset{\textsc{N2C2'08}}
 & S1 & \ce{.60}{6} & \ce{.59}{13} & \ce{.56}{16} & \ce{.58}{10} & \ce{.57}{15}\\
 & S2 & \cE{.64}{10} & \cL{.59}{15} & \cE{.62}{13} & \ce{.60}{12} & \cE{.62}{10}\\
 & S3 & \ce{.63}{7} & \ce{.59}{8} & \ce{.60}{7} & \cE{.62}{9} & \ce{.60}{12}\\
\cmidrule(l){2-7}
\dset{\textsc{EurLex}}
 & S1 & \ce{.59}{8} & \cL{.57}{5} & \cL{.59}{7} & \cL{.58}{8} & \ce{.57}{10}\\
 & S2 & \ce{.59}{6} & \ce{.55}{8} & \ce{.59}{10} & \ce{.56}{5} & \ce{.58}{6}\\
 & S3 & \cL{.59}{9} & \ce{.55}{3} & \ce{.56}{9} & \ce{.58}{7} & \cL{.61}{11}\\
\bottomrule
\end{tabular}
\end{subtable}\hfill
\begin{subtable}[t]{0.32\textwidth}
\centering
\caption{\textsc{EPSVec} (16\,\textcolor{lexcol}{\textbf{L}} vs.\ 4\,\textcolor{embcol}{\textbf{E}}).}
\label{tab:main_epsvec}
\begin{tabular}{@{}ll ccccc@{}}
\toprule
\epshead
\midrule
\dset{\textsc{PsyTAR}}
 & S1 & \ce{.59}{12} & \ce{.58}{13} & \cL{.59}{14} & \ce{.59}{10} & \cL{.59}{11}\\
 & S2 & \ce{.58}{9} & \ce{.58}{14} & \ce{.59}{10} & \ce{.59}{8} & \ce{.57}{6}\\
 & S3 & \cL{.60}{19} & \cL{.64}{15} & \ce{.59}{9} & \cL{.60}{6} & \ce{.57}{9}\\
\cmidrule(l){2-7}
\dset{\textsc{DMSAFN}}
 & S1 & \ce{.61}{5} & \ce{.61}{4} & \cL{.61}{5} & \cL{.61}{3} & \ce{.61}{3}\\
 & S2 & \cL{.63}{16} & \ce{.61}{7} & \ce{.60}{3} & \ce{.61}{15} & \cL{.62}{15}\\
 & S3 & \ce{.56}{9} & \cL{.63}{10} & \ce{.56}{11} & \ce{.60}{10} & \ce{.56}{9}\\
\cmidrule(l){2-7}
\dset{\textsc{N2C2'08}}
 & S1 & \ce{.58}{17} & \ce{.55}{13} & \ce{.56}{5} & \ce{.58}{13} & \ce{.58}{6}\\
 & S2 & \cE{.60}{12} & \cE{.60}{6} & \cE{.62}{10} & \ce{.61}{8} & \ce{.60}{12}\\
 & S3 & \ce{.57}{11} & \ce{.55}{9} & \ce{.58}{7} & \cL{.63}{6} & \cL{.63}{9}\\
\cmidrule(l){2-7}
\dset{\textsc{EurLex}}
 & S1 & \ce{.57}{6} & \ce{.57}{11} & \ce{.58}{6} & \ce{.63}{12} & \cL{.59}{7}\\
 & S2 & \cL{.58}{6} & \ce{.58}{9} & \cL{.59}{6} & \cE{.64}{13} & \ce{.56}{6}\\
 & S3 & \ce{.56}{6} & \cL{.59}{8} & \ce{.58}{7} & \ce{.61}{8} & \ce{.58}{6}\\
\bottomrule
\end{tabular}
\end{subtable}
\end{table*}

%% file: tables/table_main_rare_compact.tex
\definecolor{lexcol}{HTML}{1F5FA9}
\definecolor{embcol}{HTML}{A8431F}
\definecolor{best}{HTML}{2A78D6}
\providecommand{\win}{}\renewcommand{\win}[3]{%
  \cellcolor{best!18}\bfseries #1\textsuperscript{\textcolor{#2}{#3}}}
\providecommand{\dset}{}\renewcommand{\dset}[1]{%
  \multirow{3}{*}{\rotatebox[origin=c]{90}{\scriptsize #1}}}
\begin{table}[t]
\caption{Rare-label pool results  for \textsc{DP-Gen} and \textsc{EPSVec}, formatted in the same way as
Table~\ref{tab:main}. 
}
\label{tab:main_rare}
\centering\footnotesize\setlength{\tabcolsep}{2pt}
\begin{threeparttable}
\begin{tabular}{@{}ll ccc ccc@{}}
\toprule
 & & \multicolumn{3}{c}{\textsc{DP-Gen}} & \multicolumn{3}{c}{\textsc{EPSVec}}\\
\cmidrule(lr){3-5}\cmidrule(lr){6-8}
 & & $\bm{\infty}$ & $\bm{4}$ & $\bm{1}$ & $\bm{\infty}$ & $\bm{4}$ & $\bm{1}$\\
\midrule
\dset{\textsc{PsyTAR}}
 & S1 & .82/49 & \win{.59}{embcol}{E}/13 & .57/9 & .57/8 & \win{.60}{embcol}{E}/16 & \win{.58}{embcol}{E}/9\\
 & S2 & .84/46 & .53/11 & .55/4 & .54/6 & .56/7 & .57/8\\
 & S3 & \win{.86}{lexcol}{L}/59 & .56/9 & \win{.58}{lexcol}{L}/11 & \win{.59}{embcol}{E}/7 & .58/13 & .57/8\\
\cmidrule(l){2-8}
\dset{\textsc{N2C2'08}}
 & S1 & .62/14 & .53/5 & .53/7 & .52/3 & .56/16 & \win{.56}{embcol}{E}/6\\
 & S2 & .69/15 & .55/2 & .56/3 & .55/6 & .56/4 & .55/12\\
 & S3 & \win{.77}{lexcol}{L}/28 & \win{.56}{lexcol}{L}/8 & \win{.58}{embcol}{E}/3 & \win{.58}{embcol}{E}/11 & \win{.66}{lexcol}{L}/10 & .55/3\\
\cmidrule(l){2-8}
\dset{\textsc{EurLex}}
 & S1 & .63/13 & \win{.57}{embcol}{E}/9 & .56/6 & \win{.62}{embcol}{E}/9 & \win{.62}{embcol}{E}/10 & \win{.62}{embcol}{E}/9\\
 & S2 & .61/11 & .56/8 & \win{.57}{lexcol}{L}/6 & .58/8 & .62/11 & .58/6\\
 & S3 & \win{.64}{lexcol}{L}/23 & .55/7 & .55/5 & .59/12 & .56/3 & .57/6\\
\bottomrule
\end{tabular}
\end{threeparttable}
\label{tab:main_rare}
\end{table}

%% file: tables/table_baseline_outlier.tex
\definecolor{lexcol}{HTML}{1F5FA9}
\definecolor{embcol}{HTML}{A8431F}
\providecommand{\tbd}{\textcolor{gray}{--}}
\begin{table*}[t]
\centering
\caption{Prior release-based attacks (Top: Outlier pool, bottom: rare label pool) re-instantiated in the subgroup game, across generators. Cells are AUC\,/\,TPR@FPR$\le$0.05. Per $\varepsilon$ block the best AUC of the three methods is in \textbf{bold} and the runner-up is \underline{underlined}. $\Delta$ is \textsc{ours} minus the stronger baseline. Superscripts name the winning variant and are coloured by the channel it uses, on the same scale as the \textcolor{lexcol}{\textbf{L}}/\textcolor{embcol}{\textbf{E}} tags on \textsc{ours}. \textsc{canary} arms: \textcolor{lexcol}{n}\,$n$-gram, \textcolor{lexcol}{nR}\,$n$-gram${+}$RMIA, \textcolor{embcol}{s}\,$\mathrm{SIM}_{\mathrm{emb}}$, \textcolor{embcol}{sR}\,$\mathrm{SIM}_{\mathrm{emb}}$ ratio. \textsc{domias} estimators: \textcolor{embcol}{g}\,Gaussian, \textcolor{embcol}{k}\,KDE, \textcolor{embcol}{pg}\,PCA32${+}$Gaussian, \textcolor{embcol}{pk}\,PCA32${+}$KDE. The superscript on \textsc{ours} marks whether the winning proxy is \textcolor{lexcol}{\textbf{L}}exical or \textcolor{embcol}{\textbf{E}}mbedding.}
\label{tab:baselines-main}
\resizebox{\textwidth}{!}{%
\begin{tabular}{llcccccccc}
\toprule
 & & \multicolumn{4}{c}{$\varepsilon=\infty$} & \multicolumn{4}{c}{$\varepsilon=1$} \\
\cmidrule(lr){3-6}\cmidrule(lr){7-10}
Dataset & Generator & \textsc{canary} & \textsc{domias} & \textsc{ours} & $\Delta$ & \textsc{canary} & \textsc{domias} & \textsc{ours} & $\Delta$ \\
\midrule
PsyTAR & DP-Gen & $0.630$\textsuperscript{\textcolor{lexcol}{nR}}\,/\,21\% & $\underline{0.739}$\textsuperscript{\textcolor{embcol}{g}}\,/\,54\% & $\mathbf{0.804}$\textsuperscript{\textcolor{lexcol}{\textbf{L}}}\,/\,53\% & $+0.065$ & $0.592$\textsuperscript{\textcolor{lexcol}{n}}\,/\,12\% & $\underline{0.597}$\textsuperscript{\textcolor{embcol}{k}}\,/\,9\% & $\mathbf{0.601}$\textsuperscript{\textcolor{embcol}{\textbf{E}}}\,/\,11\% & $+0.004$ \\
PsyTAR & Aug-PE & $\underline{0.587}$\textsuperscript{\textcolor{lexcol}{n}}\,/\,14\% & $0.536$\textsuperscript{\textcolor{embcol}{k}}\,/\,10\% & $\mathbf{0.637}$\textsuperscript{\textcolor{lexcol}{\textbf{L}}}\,/\,13\% & $+0.050$ & $\underline{0.588}$\textsuperscript{\textcolor{lexcol}{n}}\,/\,11\% & $0.563$\textsuperscript{\textcolor{embcol}{k}}\,/\,7\% & $\mathbf{0.631}$\textsuperscript{\textcolor{embcol}{\textbf{E}}}\,/\,15\% & $+0.043$ \\
PsyTAR & EPSVec & $\mathbf{0.595}$\textsuperscript{\textcolor{lexcol}{n}}\,/\,13\% & $0.552$\textsuperscript{\textcolor{embcol}{k}}\,/\,7\% & $\underline{0.595}$\textsuperscript{\textcolor{lexcol}{\textbf{L}}}\,/\,19\% & $+0.000$ & $\underline{0.594}$\textsuperscript{\textcolor{lexcol}{n}}\,/\,11\% & $0.521$\textsuperscript{\textcolor{embcol}{k}}\,/\,9\% & $\mathbf{0.597}$\textsuperscript{\textcolor{lexcol}{\textbf{L}}}\,/\,6\% & $+0.004$ \\
\midrule
DMSAFN & DP-Gen & $0.651$\textsuperscript{\textcolor{lexcol}{nR}}\,/\,22\% & $\underline{0.717}$\textsuperscript{\textcolor{embcol}{g}}\,/\,42\% & $\mathbf{0.789}$\textsuperscript{\textcolor{embcol}{\textbf{E}}}\,/\,34\% & $+0.072$ & $\underline{0.604}$\textsuperscript{\textcolor{lexcol}{n}}\,/\,3\% & $0.575$\textsuperscript{\textcolor{embcol}{k}}\,/\,10\% & $\mathbf{0.647}$\textsuperscript{\textcolor{lexcol}{\textbf{L}}}\,/\,5\% & $+0.043$ \\
DMSAFN & Aug-PE & $\underline{0.603}$\textsuperscript{\textcolor{lexcol}{n}}\,/\,2\% & $0.515$\textsuperscript{\textcolor{embcol}{k}}\,/\,5\% & $\mathbf{0.612}$\textsuperscript{\textcolor{lexcol}{\textbf{L}}}\,/\,4\% & $+0.010$ & $\underline{0.602}$\textsuperscript{\textcolor{lexcol}{n}}\,/\,3\% & $0.494$\textsuperscript{\textcolor{embcol}{k}}\,/\,11\% & $\mathbf{0.613}$\textsuperscript{\textcolor{lexcol}{\textbf{L}}}\,/\,5\% & $+0.012$ \\
DMSAFN & EPSVec & $\underline{0.602}$\textsuperscript{\textcolor{lexcol}{n}}\,/\,3\% & $0.498$\textsuperscript{\textcolor{embcol}{pk}}\,/\,2\% & $\mathbf{0.630}$\textsuperscript{\textcolor{lexcol}{\textbf{L}}}\,/\,16\% & $+0.028$ & $\underline{0.601}$\textsuperscript{\textcolor{lexcol}{n}}\,/\,3\% & $0.513$\textsuperscript{\textcolor{embcol}{k}}\,/\,7\% & $\mathbf{0.611}$\textsuperscript{\textcolor{lexcol}{\textbf{L}}}\,/\,3\% & $+0.010$ \\
\midrule
N2C2'08 & DP-Gen & $\underline{0.607}$\textsuperscript{\textcolor{embcol}{sR}}\,/\,5\% & $0.558$\textsuperscript{\textcolor{embcol}{pk}}\,/\,9\% & $\mathbf{0.636}$\textsuperscript{\textcolor{embcol}{\textbf{E}}}\,/\,14\% & $+0.029$ & $0.504$\textsuperscript{\textcolor{embcol}{s}}\,/\,8\% & $\underline{0.552}$\textsuperscript{\textcolor{embcol}{pk}}\,/\,14\% & $\mathbf{0.615}$\textsuperscript{\textcolor{embcol}{\textbf{E}}}\,/\,9\% & $+0.063$ \\
N2C2'08 & Aug-PE & $\underline{0.570}$\textsuperscript{\textcolor{lexcol}{nR}}\,/\,19\% & $0.555$\textsuperscript{\textcolor{embcol}{pk}}\,/\,7\% & $\mathbf{0.640}$\textsuperscript{\textcolor{embcol}{\textbf{E}}}\,/\,10\% & $+0.070$ & $0.519$\textsuperscript{\textcolor{embcol}{s}}\,/\,2\% & $\underline{0.545}$\textsuperscript{\textcolor{embcol}{k}}\,/\,11\% & $\mathbf{0.623}$\textsuperscript{\textcolor{embcol}{\textbf{E}}}\,/\,9\% & $+0.077$ \\
N2C2'08 & EPSVec & $\underline{0.588}$\textsuperscript{\textcolor{lexcol}{nR}}\,/\,9\% & $0.584$\textsuperscript{\textcolor{embcol}{pk}}\,/\,10\% & $\mathbf{0.605}$\textsuperscript{\textcolor{embcol}{\textbf{E}}}\,/\,12\% & $+0.017$ & $\underline{0.585}$\textsuperscript{\textcolor{embcol}{sR}}\,/\,15\% & $0.570$\textsuperscript{\textcolor{embcol}{pk}}\,/\,9\% & $\mathbf{0.626}$\textsuperscript{\textcolor{lexcol}{\textbf{L}}}\,/\,6\% & $+0.041$ \\
\midrule
EUR-Lex & DP-Gen & $0.709$\textsuperscript{\textcolor{lexcol}{nR}}\,/\,34\% & $\underline{0.809}$\textsuperscript{\textcolor{embcol}{g}}\,/\,61\% & $\mathbf{0.834}$\textsuperscript{\textcolor{embcol}{\textbf{E}}}\,/\,61\% & $+0.025$ & $0.528$\textsuperscript{\textcolor{embcol}{sR}}\,/\,7\% & $\underline{0.540}$\textsuperscript{\textcolor{embcol}{k}}\,/\,7\% & $\mathbf{0.579}$\textsuperscript{\textcolor{lexcol}{\textbf{L}}}\,/\,9\% & $+0.040$ \\
EUR-Lex & Aug-PE & $0.517$\textsuperscript{\textcolor{lexcol}{n}}\,/\,5\% & $\underline{0.536}$\textsuperscript{\textcolor{embcol}{k}}\,/\,10\% & $\mathbf{0.594}$\textsuperscript{\textcolor{lexcol}{\textbf{L}}}\,/\,9\% & $+0.058$ & $0.516$\textsuperscript{\textcolor{lexcol}{n}}\,/\,7\% & $\underline{0.531}$\textsuperscript{\textcolor{embcol}{pk}}\,/\,4\% & $\mathbf{0.584}$\textsuperscript{\textcolor{lexcol}{\textbf{L}}}\,/\,8\% & $+0.054$ \\
EUR-Lex & EPSVec & $0.516$\textsuperscript{\textcolor{lexcol}{n}}\,/\,6\% & $\underline{0.578}$\textsuperscript{\textcolor{embcol}{k}}\,/\,13\% & $\mathbf{0.582}$\textsuperscript{\textcolor{lexcol}{\textbf{L}}}\,/\,6\% & $+0.004$ & $0.535$\textsuperscript{\textcolor{embcol}{sR}}\,/\,8\% & $\underline{0.587}$\textsuperscript{\textcolor{embcol}{g}}\,/\,9\% & $\mathbf{0.644}$\textsuperscript{\textcolor{embcol}{\textbf{E}}}\,/\,13\% & $+0.056$ \\
\toprule
 & & \multicolumn{4}{c}{$\varepsilon=\infty$} & \multicolumn{4}{c}{$\varepsilon=1$} \\
\cmidrule(lr){3-6}\cmidrule(lr){7-10}
Dataset & Generator & \textsc{canary} & \textsc{domias} & \textsc{ours} & $\Delta$ & \textsc{canary} & \textsc{domias} & \textsc{ours} & $\Delta$ \\
\midrule
PsyTAR & DP-Gen & $0.799$\textsuperscript{\textcolor{lexcol}{nR}}\,/\,43\% & $\underline{0.801}$\textsuperscript{\textcolor{embcol}{g}}\,/\,56\% & $\mathbf{0.857}$\textsuperscript{\textcolor{lexcol}{\textbf{L}}}\,/\,59\% & $+0.055$ & $\underline{0.564}$\textsuperscript{\textcolor{embcol}{s}}\,/\,6\% & $0.546$\textsuperscript{\textcolor{embcol}{g}}\,/\,4\% & $\mathbf{0.584}$\textsuperscript{\textcolor{lexcol}{\textbf{L}}}\,/\,11\% & $+0.019$ \\
PsyTAR & EPSVec & $\underline{0.592}$\textsuperscript{\textcolor{embcol}{sR}}\,/\,7\% & $0.575$\textsuperscript{\textcolor{embcol}{pg}}\,/\,7\% & $\mathbf{0.595}$\textsuperscript{\textcolor{embcol}{\textbf{E}}}\,/\,7\% & $+0.003$ & $0.564$\textsuperscript{\textcolor{embcol}{s}}\,/\,8\% & $\mathbf{0.586}$\textsuperscript{\textcolor{embcol}{pg}}\,/\,6\% & $\underline{0.575}$\textsuperscript{\textcolor{embcol}{\textbf{E}}}\,/\,9\% & \textcolor{embcol}{$-0.011$} \\
\midrule
N2C2'08 & DP-Gen & $\underline{0.679}$\textsuperscript{\textcolor{lexcol}{nR}}\,/\,33\% & $0.602$\textsuperscript{\textcolor{embcol}{g}}\,/\,15\% & $\mathbf{0.771}$\textsuperscript{\textcolor{lexcol}{\textbf{L}}}\,/\,28\% & $+0.093$ & $\mathbf{0.591}$\textsuperscript{\textcolor{lexcol}{nR}}\,/\,9\% & $0.550$\textsuperscript{\textcolor{embcol}{pk}}\,/\,4\% & $\underline{0.583}$\textsuperscript{\textcolor{embcol}{\textbf{E}}}\,/\,3\% & \textcolor{embcol}{$-0.008$} \\
N2C2'08 & EPSVec & $\mathbf{0.599}$\textsuperscript{\textcolor{lexcol}{nR}}\,/\,6\% & $0.528$\textsuperscript{\textcolor{embcol}{g}}\,/\,4\% & $\underline{0.584}$\textsuperscript{\textcolor{embcol}{\textbf{E}}}\,/\,11\% & \textcolor{embcol}{$-0.015$} & $0.529$\textsuperscript{\textcolor{embcol}{sR}}\,/\,2\% & $\underline{0.532}$\textsuperscript{\textcolor{embcol}{pg}}\,/\,2\% & $\mathbf{0.557}$\textsuperscript{\textcolor{embcol}{\textbf{E}}}\,/\,6\% & $+0.025$ \\
\midrule
EUR-Lex & DP-Gen & $0.560$\textsuperscript{\textcolor{embcol}{s}}\,/\,9\% & $\underline{0.597}$\textsuperscript{\textcolor{embcol}{pk}}\,/\,10\% & $\mathbf{0.638}$\textsuperscript{\textcolor{lexcol}{\textbf{L}}}\,/\,23\% & $+0.041$ & $\underline{0.570}$\textsuperscript{\textcolor{embcol}{s}}\,/\,8\% & $0.539$\textsuperscript{\textcolor{embcol}{g}}\,/\,12\% & $\mathbf{0.571}$\textsuperscript{\textcolor{lexcol}{\textbf{L}}}\,/\,6\% & $+0.000$ \\
EUR-Lex & EPSVec & $\mathbf{0.621}$\textsuperscript{\textcolor{embcol}{s}}\,/\,11\% & $0.585$\textsuperscript{\textcolor{embcol}{g}}\,/\,9\% & $\underline{0.621}$\textsuperscript{\textcolor{embcol}{\textbf{E}}}\,/\,9\% & \textcolor{embcol}{$-0.001$} & $\underline{0.611}$\textsuperscript{\textcolor{embcol}{s}}\,/\,9\% & $0.563$\textsuperscript{\textcolor{embcol}{pk}}\,/\,7\% & $\mathbf{0.616}$\textsuperscript{\textcolor{embcol}{\textbf{E}}}\,/\,9\% & $+0.005$ \\
\bottomrule
\end{tabular}%
}
\end{table*}

%% file: tables/table_control_dpgen_with_rare.tex
\definecolor{lexcol}{HTML}{1F5FA9}
\definecolor{embcol}{HTML}{A8431F}

\begin{table*}[t]
\centering
\caption{Outlier vs.\ Rare-label subgroups vs.\ the same random-group control for \textsc{DP-Gen} (top) and \textsc{EPSVec} (bottom), evaluated under the identical protocol. Each cell is the best AUC over the candidate proxy set, selected independently per cell so neither group is handicapped by the other's winner; $\pm$ is a 95\% normal interval over the evaluation trials (S~\ref{subsec:eval_protocol}), in units of $10^{-3}$ (i.e. $\pm$9 means $\pm$0.009). The control pools 3 independent draws of the random pool (\texttt{rand0/1/2}), so its interval $1.96\sqrt{\mathrm{se}_{\mathrm{hr}}^2+\mathrm{se}_{\mathrm{ctl}}^2}$ absorbs both the negative-candidate and the pool draw. Gap is defined as subgroup $-$ control.  \textbf{Bold} gaps exclude zero. 
}
\label{tab:control-dpgen-or}
\resizebox{\textwidth}{!}{%
\begin{tabular}{llcccccccccc}
\toprule
 & & \multicolumn{5}{c}{$\varepsilon=\infty$} & \multicolumn{5}{c}{$\varepsilon=4$} \\
\cmidrule(lr){3-7}\cmidrule(lr){8-12}
Dataset & Cal. & Control & Outlier & Gap & Rare-label & Gap & Control & Outlier & Gap & Rare-label & Gap \\
\midrule
PsyTAR & S1 & $0.804$\textsuperscript{\textcolor{lexcol}{\textbf{L}}}\,{\tiny$\pm$9} & $0.791$\textsuperscript{\textcolor{lexcol}{\textbf{L}}}\,{\tiny$\pm$5} & $\mathbf{-0.012}$\,{\tiny$\pm$10} & $0.820$\textsuperscript{\textcolor{lexcol}{\textbf{L}}}\,{\tiny$\pm$5} & $\mathbf{+0.017}$\,{\tiny$\pm$11} & $0.532$\textsuperscript{\textcolor{lexcol}{\textbf{L}}}\,{\tiny$\pm$8} & $0.586$\textsuperscript{\textcolor{embcol}{\textbf{E}}}\,{\tiny$\pm$12} & $\mathbf{+0.054}$\,{\tiny$\pm$14} & $0.588$\textsuperscript{\textcolor{embcol}{\textbf{E}}}\,{\tiny$\pm$9} & $\mathbf{+0.056}$\,{\tiny$\pm$11} \\
 & S2 & $0.810$\textsuperscript{\textcolor{lexcol}{\textbf{L}}}\,{\tiny$\pm$9} & $0.804$\textsuperscript{\textcolor{lexcol}{\textbf{L}}}\,{\tiny$\pm$4} & $-0.006$\,{\tiny$\pm$10}\textsuperscript{ns} & $0.835$\textsuperscript{\textcolor{lexcol}{\textbf{L}}}\,{\tiny$\pm$6} & $\mathbf{+0.025}$\,{\tiny$\pm$11} & $0.525$\textsuperscript{\textcolor{embcol}{\textbf{E}}}\,{\tiny$\pm$7} & $0.587$\textsuperscript{\textcolor{lexcol}{\textbf{L}}}\,{\tiny$\pm$10} & $\mathbf{+0.062}$\,{\tiny$\pm$12} & $0.531$\textsuperscript{\textcolor{embcol}{\textbf{E}}}\,{\tiny$\pm$10} & $+0.006$\,{\tiny$\pm$12}\textsuperscript{ns} \\
\midrule
N2C2'08 & S1 & $0.594$\textsuperscript{\textcolor{lexcol}{\textbf{L}}}\,{\tiny$\pm$7} & $0.551$\textsuperscript{\textcolor{lexcol}{\textbf{L}}}\,{\tiny$\pm$13} & $\mathbf{-0.043}$\,{\tiny$\pm$15} & $0.622$\textsuperscript{\textcolor{lexcol}{\textbf{L}}}\,{\tiny$\pm$10} & $\mathbf{+0.028}$\,{\tiny$\pm$12} & $0.554$\textsuperscript{\textcolor{embcol}{\textbf{E}}}\,{\tiny$\pm$9} & $0.576$\textsuperscript{\textcolor{lexcol}{\textbf{L}}}\,{\tiny$\pm$10} & $\mathbf{+0.022}$\,{\tiny$\pm$13} & $0.525$\textsuperscript{\textcolor{lexcol}{\textbf{L}}}\,{\tiny$\pm$11} & $\mathbf{-0.029}$\,{\tiny$\pm$15} \\
 & S2 & $0.656$\textsuperscript{\textcolor{lexcol}{\textbf{L}}}\,{\tiny$\pm$6} & $0.600$\textsuperscript{\textcolor{lexcol}{\textbf{L}}}\,{\tiny$\pm$9} & $\mathbf{-0.056}$\,{\tiny$\pm$11} & $0.688$\textsuperscript{\textcolor{lexcol}{\textbf{L}}}\,{\tiny$\pm$10} & $\mathbf{+0.032}$\,{\tiny$\pm$12} & $0.550$\textsuperscript{\textcolor{embcol}{\textbf{E}}}\,{\tiny$\pm$9} & $0.590$\textsuperscript{\textcolor{lexcol}{\textbf{L}}}\,{\tiny$\pm$8} & $\mathbf{+0.040}$\,{\tiny$\pm$12} & $0.546$\textsuperscript{\textcolor{embcol}{\textbf{E}}}\,{\tiny$\pm$10} & $-0.004$\,{\tiny$\pm$14}\textsuperscript{ns} \\
\midrule
EUR-Lex & S1 & $0.665$\textsuperscript{\textcolor{embcol}{\textbf{E}}}\,{\tiny$\pm$5} & $0.804$\textsuperscript{\textcolor{embcol}{\textbf{E}}}\,{\tiny$\pm$6} & $\mathbf{+0.139}$\,{\tiny$\pm$8} & $0.625$\textsuperscript{\textcolor{embcol}{\textbf{E}}}\,{\tiny$\pm$10} & $\mathbf{-0.040}$\,{\tiny$\pm$11} & $0.551$\textsuperscript{\textcolor{lexcol}{\textbf{L}}}\,{\tiny$\pm$7} & $0.587$\textsuperscript{\textcolor{embcol}{\textbf{E}}}\,{\tiny$\pm$9} & $\mathbf{+0.036}$\,{\tiny$\pm$11} & $0.575$\textsuperscript{\textcolor{embcol}{\textbf{E}}}\,{\tiny$\pm$12} & $\mathbf{+0.023}$\,{\tiny$\pm$14} \\
 & S2 & $0.655$\textsuperscript{\textcolor{embcol}{\textbf{E}}}\,{\tiny$\pm$7} & $0.831$\textsuperscript{\textcolor{embcol}{\textbf{E}}}\,{\tiny$\pm$4} & $\mathbf{+0.176}$\,{\tiny$\pm$8} & $0.606$\textsuperscript{\textcolor{embcol}{\textbf{E}}}\,{\tiny$\pm$9} & $\mathbf{-0.049}$\,{\tiny$\pm$11} & $0.552$\textsuperscript{\textcolor{lexcol}{\textbf{L}}}\,{\tiny$\pm$12} & $0.570$\textsuperscript{\textcolor{embcol}{\textbf{E}}}\,{\tiny$\pm$9} & $\mathbf{+0.018}$\,{\tiny$\pm$15} & $0.565$\textsuperscript{\textcolor{embcol}{\textbf{E}}}\,{\tiny$\pm$12} & $+0.013$\,{\tiny$\pm$16}\textsuperscript{ns} \\
\bottomrule


 & & \multicolumn{5}{c}{$\varepsilon=\infty$} & \multicolumn{5}{c}{$\varepsilon=4$} \\
\cmidrule(lr){3-7}\cmidrule(lr){8-12}
Dataset & Cal. & Control & Outlier & Gap & Rare-label & Gap & Control & Outlier & Gap & Rare-label & Gap \\
\midrule
PsyTAR & S1 & $0.551$\textsuperscript{\textcolor{lexcol}{\textbf{L}}}\,{\tiny$\pm$7} & $0.594$\textsuperscript{\textcolor{lexcol}{\textbf{L}}}\,{\tiny$\pm$9} & $\mathbf{+0.044}$\,{\tiny$\pm$11} & $0.570$\textsuperscript{\textcolor{embcol}{\textbf{E}}}\,{\tiny$\pm$11} & $\mathbf{+0.019}$\,{\tiny$\pm$13} & $0.539$\textsuperscript{\textcolor{lexcol}{\textbf{L}}}\,{\tiny$\pm$6} & $0.585$\textsuperscript{\textcolor{lexcol}{\textbf{L}}}\,{\tiny$\pm$9} & $\mathbf{+0.046}$\,{\tiny$\pm$11} & $0.597$\textsuperscript{\textcolor{embcol}{\textbf{E}}}\,{\tiny$\pm$11} & $\mathbf{+0.058}$\,{\tiny$\pm$12} \\
 & S2 & $0.550$\textsuperscript{\textcolor{lexcol}{\textbf{L}}}\,{\tiny$\pm$8} & $0.582$\textsuperscript{\textcolor{lexcol}{\textbf{L}}}\,{\tiny$\pm$10} & $\mathbf{+0.032}$\,{\tiny$\pm$13} & $0.543$\textsuperscript{\textcolor{embcol}{\textbf{E}}}\,{\tiny$\pm$12} & $-0.008$\,{\tiny$\pm$14}\textsuperscript{ns} & $0.526$\textsuperscript{\textcolor{lexcol}{\textbf{L}}}\,{\tiny$\pm$5} & $0.578$\textsuperscript{\textcolor{lexcol}{\textbf{L}}}\,{\tiny$\pm$10} & $\mathbf{+0.052}$\,{\tiny$\pm$11} & $0.560$\textsuperscript{\textcolor{embcol}{\textbf{E}}}\,{\tiny$\pm$10} & $\mathbf{+0.033}$\,{\tiny$\pm$12} \\
\midrule
N2C2'08 & S1 & $0.564$\textsuperscript{\textcolor{embcol}{\textbf{E}}}\,{\tiny$\pm$10} & $0.576$\textsuperscript{\textcolor{lexcol}{\textbf{L}}}\,{\tiny$\pm$9} & $+0.012$\,{\tiny$\pm$13}\textsuperscript{ns} & $0.520$\textsuperscript{\textcolor{embcol}{\textbf{E}}}\,{\tiny$\pm$9} & $\mathbf{-0.044}$\,{\tiny$\pm$13} & $0.560$\textsuperscript{\textcolor{embcol}{\textbf{E}}}\,{\tiny$\pm$8} & $0.546$\textsuperscript{\textcolor{lexcol}{\textbf{L}}}\,{\tiny$\pm$10} & $\mathbf{-0.014}$\,{\tiny$\pm$13} & $0.562$\textsuperscript{\textcolor{embcol}{\textbf{E}}}\,{\tiny$\pm$10} & $+0.002$\,{\tiny$\pm$13}\textsuperscript{ns} \\
 & S2 & $0.555$\textsuperscript{\textcolor{embcol}{\textbf{E}}}\,{\tiny$\pm$9} & $0.605$\textsuperscript{\textcolor{embcol}{\textbf{E}}}\,{\tiny$\pm$11} & $\mathbf{+0.049}$\,{\tiny$\pm$14} & $0.549$\textsuperscript{\textcolor{embcol}{\textbf{E}}}\,{\tiny$\pm$11} & $-0.007$\,{\tiny$\pm$14}\textsuperscript{ns} & $0.540$\textsuperscript{\textcolor{embcol}{\textbf{E}}}\,{\tiny$\pm$10} & $0.595$\textsuperscript{\textcolor{embcol}{\textbf{E}}}\,{\tiny$\pm$10} & $\mathbf{+0.055}$\,{\tiny$\pm$14} & $0.563$\textsuperscript{\textcolor{lexcol}{\textbf{L}}}\,{\tiny$\pm$11} & $\mathbf{+0.024}$\,{\tiny$\pm$15} \\
\midrule
EUR-Lex & S1 & $0.534$\textsuperscript{\textcolor{lexcol}{\textbf{L}}}\,{\tiny$\pm$9} & $0.575$\textsuperscript{\textcolor{lexcol}{\textbf{L}}}\,{\tiny$\pm$10} & $\mathbf{+0.041}$\,{\tiny$\pm$13} & $0.621$\textsuperscript{\textcolor{embcol}{\textbf{E}}}\,{\tiny$\pm$12} & $\mathbf{+0.087}$\,{\tiny$\pm$15} & $0.540$\textsuperscript{\textcolor{lexcol}{\textbf{L}}}\,{\tiny$\pm$8} & $0.573$\textsuperscript{\textcolor{lexcol}{\textbf{L}}}\,{\tiny$\pm$8} & $\mathbf{+0.033}$\,{\tiny$\pm$11} & $0.618$\textsuperscript{\textcolor{embcol}{\textbf{E}}}\,{\tiny$\pm$12} & $\mathbf{+0.078}$\,{\tiny$\pm$14} \\
 & S2 & $0.535$\textsuperscript{\textcolor{embcol}{\textbf{E}}}\,{\tiny$\pm$6} & $0.582$\textsuperscript{\textcolor{lexcol}{\textbf{L}}}\,{\tiny$\pm$11} & $\mathbf{+0.048}$\,{\tiny$\pm$12} & $0.583$\textsuperscript{\textcolor{embcol}{\textbf{E}}}\,{\tiny$\pm$12} & $\mathbf{+0.048}$\,{\tiny$\pm$14} & $0.536$\textsuperscript{\textcolor{embcol}{\textbf{E}}}\,{\tiny$\pm$6} & $0.581$\textsuperscript{\textcolor{lexcol}{\textbf{L}}}\,{\tiny$\pm$10} & $\mathbf{+0.045}$\,{\tiny$\pm$12} & $0.617$\textsuperscript{\textcolor{embcol}{\textbf{E}}}\,{\tiny$\pm$11} & $\mathbf{+0.081}$\,{\tiny$\pm$13} \\
\bottomrule
\end{tabular}%
}
\end{table*}

%% file: 6_conclusion_limitations.tex
\section{Conclusion and Limitations}
\label{sec:conclusion}

Auditing synthetic text releases with a game where the membership target pool is an explicit
parameter changes what a membership inference attack experiment can report. Beyond
confirming that subgroup membership leaks and that prior attacks
under-report it, our auditing framework produced findings we
believe are useful to the community: DP protects membership inference signal from high-risk groups less than from random controls. How much residual membership signal a release exposes is a property of the release mechanism, not of the private records, so no release-independent triage of at-risk records is feasible according to our analyses. This implies that average-case MIA performance reports and attempts at attribute-based screening of vulnerable records may both miss the risk to records. Our findings suggest that per-release MIA auditing is the preferred way to identify at-risk records.

Our conclusions have the following limitations: First, the corpora are
small (604--5{,}102 records) and the pools smaller (16--71), with these pool
sizes, no correlation below $|\rho|\approx 0.33$--$0.50$ can reach significance. 
Second, our reported
uncertainty estimates (as reported by confidence intervals) capture negative-sampling variance but not the variance of
drawing the pool itself: re-estimating the latter would mean repeating
the full pipeline over fresh pools, and a single configuration of pool, generator and 
$\varepsilon$ takes about 27-30h to run on a L40S GPU for its 100 releases (\textsc{EurLex}, DP-Gen, one outlier pool)---replicating every combination over multiple pool draws is beyond our budget. Third, our adversary is
strictly release-based and our generators span three mechanism families;
attacks with model access, larger corpora, and further mechanisms may
further refine our conclusions. Our membership inference game and the whole pipeline we release are designed to
make such extensions cheap to run.

%% file: Appendix.tex
\section{Appendix}
\label{appendix}

\subsection{The details of definition of target pool}
\label{app:targetpool}

\begin{table*}[t]
\centering
\small
\setlength{\tabcolsep}{4pt}
\begin{tabular}{llllrlr r}
\toprule
\textbf{Dataset} & \textbf{Embedder} & \multicolumn{2}{c}{\textbf{Outlier}} & & \multicolumn{2}{c}{\textbf{Rare}} & \textbf{Control} \\
\cmidrule(lr){3-4}\cmidrule(lr){6-7}
 & & rule & $|\mathcal{T}|$ & & rule & $|\mathcal{T}|$ & $|\mathcal{T}|$ \\
\midrule
\textsc{N2C2'08} & BioSimCSE-BioLinkBERT & $d_{97}$ + LOF$_{k=3}$  & 19 & & card.\ $\geq 8$ & 63 & 19 \\
\textsc{PsyTAR}  & sentence-t5-base      & $d_{99}$ + LOF$_{k=10}$ & 60 & & freq.\ $\leq 10$ & 54 & 60 \\
\textsc{DMSAFN}  & sentence-t5-base      & $d_{99}$ + LOF$_{k=3}$  & 71 & & --- & --- & 71 \\
\textsc{EurLex}  & bge-m3                & $d_{99}$                & 50 & & freq.\ $\leq 3$ & 55 & 50 \\
\bottomrule
\end{tabular}
\caption{Subgroup construction. $d_{p}$ selects records beyond the $p$-th percentile of distance to the embedding centroid; LOF$_{k}$ is Local Outlier Factor with $k$ neighbours; outlier pools are the union of the two criteria. Rare pools use label statistics: \emph{card.} is label cardinality, \emph{freq.} is the corpus frequency of a record's label combination. Control pools match the outlier pool size and are drawn uniformly at random, with three independent seeds per dataset.}
\label{tab:subgroups}
\end{table*}

For each dataset we construct three target pools (as shown in Table.~\ref{tab:subgroups}). The \emph{outlier} pool captures geometric atypicality in a domain-appropriate embedding space; the \emph{rare} pool captures statistical atypicality in label space; the \emph{control} pool is a uniform random sample, replicated over three seeds, and serves as the null condition against which subgroup effects are measured.

The rarity axis is chosen per dataset because the corpora admit different notions of rarity. In \textsc{N2C2}, 354 of 429 label combinations occur exactly once, so combination frequency is uninformative and we instead define rarity by co-morbidity count. \textsc{DMSAFN} is single-label with three classes and therefore admits no rare-combination subgroup at all; it contributes outlier and control conditions only.

\subsection{Details of Proxies}
\label{app:proxies}

Table~\ref{appd:tab:proxy_families} groups the $m=32$ proxies of
\S\ref{sec:scoring} into 13 families; all families run unchanged in all three
scenarios, with only the observation object $R$ changing
(\S\ref{sec:scenarios}). 
The count expands from 13 to 32 because most
families contribute several concrete features: retrieval-based overlap
statistics are aggregated by both max and mean over the top-$k$ candidates,
and the neighborhood-based embedding families are instantiated at several
neighborhood definitions (nearest neighbor, top-$k$ average, radius count).
The families are chosen to cover distinct signatures of memorization rather
than to be independent: verbatim reuse (character $n$-gram containment,
longest common substring), near-verbatim paraphrase (ROUGE-L, embedding
similarity), distributional support (bigram log-probability, Gaussian
log-likelihood), and rarity-localized copying (rare-$n$-gram and
entity/number overlap), the last targeting the identifier-dense records
typical of clinical and legal text. Within each view the proxies are highly
rank-correlated by construction---they measure the same copying signal at
different granularities---which is why the audit reports the per-proxy
envelope rather than a learned combination, and why the fusion of
Appendix~\ref{app:meta} finds little complementary signal to harvest.

\begin{table*}[!htbp]
\centering
\small
\setlength{\tabcolsep}{6pt}
\renewcommand{\arraystretch}{1.18}
\begin{tabular}{p{3.0cm} p{1.4cm} p{5.8cm} p{2.2cm} p{2.4cm}}
\hline
\textbf{Proxy family} & \textbf{View} & \textbf{What it measures} & \textbf{Probability-like} & \textbf{Used in scenarios} \\
\hline
$n$-gram log-probability & Lexical & Likelihood of the target text under an $n$-gram language model fitted on texts from the observation object $R$ & Yes & S1 / S2 / S3 \\
Gaussian log-likelihood & Embedding & Log-density of the target embedding under a Gaussian fitted on embeddings from $R$ & Yes & S1 / S2 / S3 \\
BM25 relevance & Lexical & Retrieval relevance between the target and texts in $R$ & No & S1 / S2 / S3 \\
Character $n$-gram overlap & Lexical & Local lexical copying measured by containment or Jaccard overlap over character $n$-grams & No & S1 / S2 / S3 \\
Sequence overlap & Lexical & Long-span copying or paraphrase-style overlap measured by ROUGE-L, token LCS, or longest common substring & No & S1 / S2 / S3 \\
Rare phrase overlap & Lexical & Overlap on rare or informative $n$-grams, optionally weighted by IDF & No & S1 / S2 / S3 \\
Entity / number overlap & Lexical & Exact overlap on numbers, identifiers, and entity-like strings & No & S1 / S2 / S3 \\
Cosine similarity & Embedding & Angular similarity to nearest neighbors in the embedding space of $R$ & No & S1 / S2 / S3 \\
Dot-product similarity & Embedding & Raw embedding affinity to nearest neighbors in $R$ & No & S1 / S2 / S3 \\
Euclidean proximity & Embedding & Distance-based closeness and local density in the embedding space of $R$ & No & S1 / S2 / S3 \\
Mahalanobis proximity & Embedding & Covariance-aware closeness to the embedding manifold induced by $R$ & No & S1 / S2 / S3 \\
CSLS similarity & Embedding & Hubness-corrected nearest-neighbor similarity in the embedding space of $R$ & No & S1 / S2 / S3 \\
Anomaly score & Embedding & Normality or atypicality under an embedding-space detector fitted on $R$ & No & S1 / S2 / S3 \\
\hline
\end{tabular}
\caption{Proxy families used in our attacks.
Each proxy is defined as a raw score $\phi(x;R)$ computed for target record $x$ with respect to an observation object $R$.
Depending on the attacker capability scenario, $R$ is the target synthetic release (S1), a raw reference dataset for calibration (S2), or a synthetic reference release for calibration (S3).
Probability-like proxies arise from explicitly fitted probabilistic models; the remaining proxies are empirical similarity, overlap, density, or anomaly signals.
Scenario-specific calibration is handled separately and is not part of the proxy definition itself.}
\label{appd:tab:proxy_families}
\end{table*}

\subsection{Reference Construction: Sharing and Its Caveats}
\label{app:refs}

As in shadow-model
practice~\cite{carlini2022membership,zarifzadeh2023low,meeus2025canary},
reference sets are shared across targets rather than built per target:
several targets are inserted into each set, so each target is IN for half
of the \(M_{\mathrm{ref}}\) references (\(M_{\mathrm{ref}}=4\);
Scenario~3 likewise reuses the generated reference releases across
targets). This introduces two approximations. First, references share one
auxiliary pool rather than being resampled i.i.d., so the baseline
conditions on that pool and misses dataset-level sampling variance
(\textsc{Canary} does the same, as worst-case auditing). Second, OUT
references may contain other records of \(\mathcal G_r\) while the null
world contains none, which can inflate the OUT baseline and make the
audit conservative; the effect is largest for rules selecting mutually
similar records and vanishes for \textsc{Random}.

\subsection{Privacy accounting.}
We evaluate budgets \(\varepsilon \in \{\infty, 4, 2, 1, 0.5\}\), with \(\infty\) the
non-private setting.
\emph{DP-Gen}: accounted at \(\delta_{\mathrm{DP}} = 1/N^2\) via the PRV accountant, with
\(\ell_2\) clip norm \(1.0\), noise multiplier calibrated to the target
\((\varepsilon,\delta_{\mathrm{DP}})\) over 10 epochs at effective batch size 16
(\textsc{PsyTAR}, \textsc{DMSAFN}), 8 (\textsc{EurLex}), 4 (\textsc{N2C2'08}).
\emph{Aug-PE}: 10 refinement rounds, each a Gaussian-noised nearest-neighbor histogram
query on the private data; following~\cite{Xie2024DifferentiallyText} we compose across
rounds at \(\delta_{\mathrm{DP}}=1/(N\log N)\) and calibrate the per-round noise
multiplier.
\emph{EPSVec}: the cost is incurred once at vector release, and is split across two
private steps. Stage~1 selects \(k=2\) few-shot exemplars per class by a DP histogram
over a candidate pool, taking one sixth of the budget
(\(\varepsilon_{\mathrm{hist}}=\varepsilon/6\), \(\delta_{\mathrm{hist}}=10^{-6}\));
the remaining \(5\varepsilon/6\) is spent on the per-layer difference vectors.
At generation the released vectors are injected at layers 8--11 with
scaling factor \(\alpha=1.4\), the value reported by Banayeeanzade et al.~\cite{banayeeanzade2026epsvec};
at \(\varepsilon=\infty\) we use \(\alpha=0.2\), since without privacy noise we find that \(\alpha=1.4\) yields low-quality texts.

\subsection{More Results of Our MIA Auditing}
\label{app:more-results}

Tables~\ref{tab:overall_main_dptransformers}--\ref{tab:overall_main_rare} repeat the main
tables with per-cell 95\% intervals (negative-sampling variability only,
\S\ref{subsec:eval_protocol}).

\input{tables/table_main_per_generator_CI}
\input{tables/table_main_rare_CI}

\subsection{Meta-Classifier: A Negative Result}
\label{app:meta}

\input{tables/table_meta_outlier}
\input{tables/table_meta_rare}

Tables~\ref{tab:meta-outlier}--\ref{tab:meta-rare} evaluate a fusion attack that
combines all 32 proxies through a class-balanced logistic regression with nested
cross-validation (outer 5-fold out-of-fold prediction; inner 5-fold selection of the
feature count and $\ell_2$ penalty), so that no record contributes to the model that
scores it.
The comparison is deliberately unfavorable to the fusion: its reference is the
\emph{oracle} best single proxy, selected per cell after seeing the evaluation
data---the same worst-case selection the main tables report.

The fusion does not close the gap to the oracle.
Across the 234 populated (generator, dataset, scenario, $\varepsilon$) cells it wins
in 4 and loses in 200, with a mean deficit of $\Delta = -0.077$ AUC
(outlier pools: $2/60$ under DP-Gen, $0/60$ under EPSVec, $1/45$ under Aug-PE; rare
pools: $1/54$).
Its absolute performance is nonetheless informative: on the strong non-private
fine-tuning cells the deployable fusion still reaches AUC 0.76--0.79 on the
outlier pools and up to 0.83 on the rare pools, within 0.03--0.05 of the
per-cell oracle, so the headline leakage of \S\ref{sec:results-leakage} does
not depend on oracle proxy selection.
Under DP, where the best single proxy sits at 0.55--0.65, the fusion typically falls
to 0.44--0.57---often \emph{below} the oracle by more than the oracle exceeds
chance.

We attribute the failure to the geometry of the proxy family rather than to the
classifier: the proxies within each view are highly rank-correlated, so the fusion
has few effectively independent signals to combine, while the per-round training sets
(tens of positive records) are small relative to even the reduced feature dimension;
in this regime the cross-validated model's variance costs more than the (near-zero)
complementary signal it could harvest.
Two practical conclusions follow.
First, the oracle best-proxy numbers in the main tables should be read as what they
are---an upper envelope over the family---but the envelope is not an artifact of
selection: a deployable attacker gets within a few points of it wherever leakage is
strong.
Second, there is no evidence that lexical and embedding evidence are complementary at
the record level; the two views agree on which records look like members, which is
consistent with the mechanism-bound picture of \S\ref{sec:attributes}.

\subsection{Extended Baseline Comparisons}
\label{app:baselines}
Tables~\ref{tab:baselines-outlier-rest} and~\ref{tab:baselines-rare-e4} give the full
per-cell results for both baseline families under every privacy budget and variant we
swept (\S\ref{sec:setup-baselines}), complementing the $\varepsilon\in\{\infty,1\}$
cells shown in \S\ref{sec:results-baselines}.
Each method is maximized over its own variants independently in each cell, so all
three columns receive the same worst-case treatment.
Across all 78 populated (dataset, generator, $\varepsilon$, pool) cells---the main-text
and appendix budgets combined---our audit attains the highest AUC in 69, ties in 2,
and is beaten in 7, every loss smaller than $0.02$ AUC and confined to
low-memorization configurations (EPSVec at any budget, or DP-SGD fine-tuning at
$\varepsilon\le 4$).

\input{tables/table_baseline_outlier_appendix}
\input{tables/table_baselines_rare_appendix}

\subsection{Per-Cell Vulnerability Distribution}
\label{app:vdist}

Table~\ref{tab:v-distribution-appendix} reports, for every audited (generator, dataset, $\varepsilon$) cell, the two statistics behind \S\ref{sec:results-concentration}:
the share of qualifying records with $V(x)>0$ and the top-decile leakage
share $S_{10}$ (Eq.~\ref{eq:top-decile-share}), computed with the cell's
best proxy under the same worst-case selection as the main tables. The
medians, ranges, and cell counts quoted in \S\ref{sec:results-concentration} are
computed over this table. Reading across regimes, the two statistics move
in opposite directions: the four non-private fine-tuning cells are the only
ones that combine a large positive share with a small $S_{10}$, while every
low-memorization cell pairs a thinner positive share with a heavier top
decile---the per-cell form of the pattern \S\ref{sec:results-concentration} reports
in aggregate.

\input{tables/table_v_distribution_appendix}

\subsection{Attribute Analysis: Computation and Robustness}
\label{app:attributes}

\begin{figure*}[t]
  \centering
  \includegraphics[width=\textwidth]{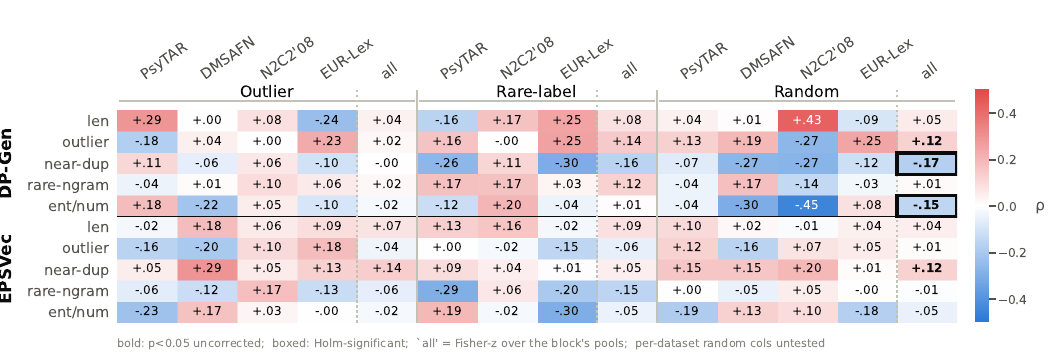}
  \caption{Full attribute--vulnerability matrix: the outlier and rare-label
  pools alongside the random pools of Figure~\ref{fig:attributes-heatmap}.
  Cell format and significance marks as there; within each block, \emph{all}
  is the Fisher-$z$ pool with Holm correction over the block's five attributes.
  None of the 70 per-pool tests in the selected pools reaches nominal
  significance. Outlierness is range-restricted in the outlier pools, which are
  selected on that statistic.}
  \label{fig:attributes-heatmap-full}
\end{figure*}

Figure~\ref{fig:attributes-heatmap-full} extends  Figure~\ref{fig:attributes-heatmap} to the outlier and rare-label pools: no attribute reaches nominal significance in any of them, so the null result of the random pools is not an artifact of how records were selected.

\paragraph{Computing the correlations.}
Every entry of Figure~\ref{fig:attributes-heatmap} and
Table~\ref{tab:attributes} is a Spearman rank correlation computed inside one pool. 
For each pool we take its qualifying records---those scored at least
once as the inserted member and at least once as a non-member candidate
within the 100-instance universe (16--36 per pool)---and their vulnerability $V(x)$ (Eq.~\ref{eq_sample_vul}), pooled over the pool's privacy budgets:
scores are the selected proxy, sign-aligned, $z$-normalized within each
release set, and averaged over all member and all non-member episodes before the difference is taken. 
The five attributes are computed from the record and the fixed pools only, never from a release or an attack proxy: token length
(spaCy tokens); outlierness, the Euclidean distance from the record's
embedding to the corpus centroid under the dataset's encoder, i.e.\ the
selection statistic of \S\ref{sec:subgroup-construction}; near-duplication, the maximum cosine similarity to any other record of that pool's private candidate pool $C$; rare-$n$-gram density, the fraction of the record's token bigrams whose document frequency over the background $B=C\cup R$ falls at or below the occurrence-weighted 10th percentile; and entity/numeral density,
the number of numeral, identifier, all-caps, e-mail, and URL spans (overlaps merged) per token. 
The background is resolved per pool---each pool induces its own $C$ and $R$ (\S\ref{sec:subgroup-construction})---with the record
itself excluded. Per-pool entries report $\rho$ between the attribute and the within-pool rank of $V$, with Holm correction over all attribute--pool tests;
the combined entry pools the twelve random pools of a generator (three
independent draws per dataset, pairwise disjoint in records) by Fisher's $z$, weighting each pool by $n-3$, with Holm correction over the five attributes.
Outlierness is marked $\dagger$ in the outlier columns because those pools were selected on it (\S\ref{sec:attr-worstcase}).
The per-pool detection floor is the significance boundary implied by the pool size: for a two-sided test at $\alpha=0.05$ with
$n$ records, $|\rho|_{\min}=t_{0.975,n-2}/\sqrt{t_{0.975,n-2}^{2}+n-2}$,
i.e.\ $0.33$ at $n=36$ and $0.46$ at $n=19$. A null in a single pool therefore
means ``no correlation as large as $\sim$0.4,'' not ``no correlation'';
detecting effects of the size the random-pool combination suggests
($|\rho|\approx 0.15$) requires pooling, which is why inference is performed
on the blocks' combined columns.

\input{tables/table_attributes.tex}

\paragraph{Selection rules.}
Both rules search the same space---every proxy of \S\ref{sec:proxies}
crossed with every scenario and budget available to the pool---and return
one (scenario, proxy) pair applied to all of the pool's budgets. The
\emph{conservative} rule takes the pair with the best AUC on an 80\%
training partition of instances. This is a pre-commitment, not
independence: $V(x)$ still spans all instances, so selection has seen 80\%
of the episodes behind the tested correlations, but the proxy is fixed by
a deterministic criterion that never sees the remaining fifth and is never
revised afterwards. 
The \emph{worst-case} rule is the main tables':
argmax AUC over all episodes. It is the audit's natural instrument and the more sensitive of the two, but its argmax ranges over the very episodes being tested, so its $p$-values are optimistic---which is why the conservative rule is primary and Table~\ref{tab:selection-robustness} treats the worst-case rule as a sensitivity analysis.

\subsection{Robustness to the proxy-selection rule}
\label{app:selection-robustness}

Because the selected proxy determines $V(x)$, we recompute the full
attribute--pool matrix for every generator and every pool under three
alternative rules: the main table's worst-case rule (argmax AUC over
\emph{all} episodes), and two selection-free rules that fix the same proxy in
every pool---the lexical \texttt{containment\_max} (Scenario~2) and the
embedding \texttt{cos\_max} (Scenario~1).
Tables~\ref{tab:selection-robustness} and~\ref{tab:combined-by-rule}
summarize the outcome. 
Two conclusions are invariant. 
First, no attribute--pool test is significant after Holm correction under any rule
($0$ of $95$, $95$, and $20$ tests for the three generators), so the
cell-level null of \S\ref{sec:attr-null} does not depend on how the proxy was
chosen. 
Second, the direction of every DP-SGD random-pool tendency is stable:
outlierness is positive, and near-duplication and entity density negative,
under all four rules; under activation steering every combined $\rho$ stays
within $\pm0.12$ under every rule. What depends on the rule is significance.
The paper's rule is the most conservative: outlierness is marginal under it
(Holm $p=0.09$) but clears correction under all-episode selection ($+0.20$,
$p=0.002$) and under the fixed embedding proxy ($+0.32$)---the latter being
the natural consequence of selecting and scoring in the same embedding
geometry, not evidence of a stronger effect. Individual pools are noisy:
relative to the paper's rule the median change in a per-pool $\rho$ is at
most $0.12$ but the maximum reaches $0.96$ in the smallest pools
($n=16$--$19$), which is why we interpret pooled directions only. Even under
the most favourable rule the strongest attribute reaches $|\rho|=0.32$, far
below what record-level triage would require.

\input{tables/table_selection_robustness.tex}
\input{tables/table_combined_by_rule.tex}

\subsection{Cross-mechanism consistency of per-record vulnerability}
\label{app:cross-generator}

On identical records scored with the identical proxy (the DP-Gen pool's
selected proxy), the rank correlation of $V(x)$ between DP-SGD fine-tuning and each other mechanism is near zero (Table~\ref{tab:cross-generator}): mean $-0.09$ against activation steering over 16 pools ($-0.12$ over the random pools alone) and $+0.07$ against API-based prompting over the three outlier pools that have both. The per-record channel also differs in magnitude:
without DP, fine-tuning yields mean $|V|=1.99$ on \textsc{PsyTAR}, whereas prompting yields $0.04$--$0.53$ at every noise level and never exhibits the no-DP spike. Both correlations are attenuated by the low reliability of $V$ under the weaker mechanisms (within a mechanism, $V$ at different budgets correlates at $|\rho|\leq0.35$) and are therefore upper bounds on the transferable component rather than point estimates.

\input{tables/table_cross_generator.tex}

%% file: tables/table_main_per_generator_CI.tex
\definecolor{lexcol}{HTML}{1F5FA9}   
\definecolor{embcol}{HTML}{A8431F}   
\definecolor{best}{HTML}{2A78D6}     
\begin{table*}[t]
\centering
\small
\setlength{\tabcolsep}{3pt}
\renewcommand{\arraystretch}{1.05}
\begin{threeparttable}
\begin{tabular}{llccccc}
\toprule
\textbf{Dataset} & \textbf{Scenario} & $\bm{\varepsilon=\infty}$ & $\bm{4}$ & $\bm{2}$ & $\bm{1}$ & $\bm{0.5}$ \\
\midrule
\multirow{3}{*}{\textsc{PsyTAR}} & S1 & \makecell{\small 0.79 (0.79--0.80) \\ \scriptsize 45.7\%} & \makecell{\small 0.59 (0.57--0.60) \\ \scriptsize 12.0\%} & \makecell{\small 0.59 (0.58--0.59) \\ \scriptsize 12.5\%} & \cellcolor{best!18}\makecell{\small \bfseries 0.60\textsuperscript{\textcolor{embcol}{\textbf{E}}} (0.59--0.61) \\ \scriptsize \bfseries 11.2\%} & \makecell{\small 0.61 (0.60--0.63) \\ \scriptsize 4.2\%} \\
                                 & S2 & \cellcolor{best!18}\makecell{\small \bfseries 0.80\textsuperscript{\textcolor{lexcol}{\textbf{L}}} (0.80--0.81) \\ \scriptsize \bfseries 53.2\%} & \cellcolor{best!18}\makecell{\small \bfseries 0.59\textsuperscript{\textcolor{lexcol}{\textbf{L}}} (0.58--0.60) \\ \scriptsize \bfseries 9.8\%} & \makecell{\small 0.57 (0.56--0.59) \\ \scriptsize 5.9\%} & \makecell{\small 0.57 (0.56--0.58) \\ \scriptsize 9.2\%} & \cellcolor{best!18}\makecell{\small \bfseries 0.64\textsuperscript{\textcolor{lexcol}{\textbf{L}}} (0.63--0.65) \\ \scriptsize \bfseries 6.3\%} \\
                                 & S3 & \makecell{\small 0.76 (0.75--0.76) \\ \scriptsize 45.3\%} & \makecell{\small 0.57 (0.56--0.58) \\ \scriptsize 14.7\%} & \cellcolor{best!18}\makecell{\small \bfseries 0.59\textsuperscript{\textcolor{lexcol}{\textbf{L}}} (0.58--0.60) \\ \scriptsize \bfseries 11.4\%} & \makecell{\small 0.57 (0.56--0.59) \\ \scriptsize 7.8\%} & \makecell{\small 0.63 (0.62--0.64) \\ \scriptsize 6.5\%} \\
\cmidrule(lr){1-7}
\multirow{3}{*}{\textsc{DMSAFN}} & S1 & \makecell{\small 0.79 (0.78--0.80) \\ \scriptsize 38.5\%} & \cellcolor{best!18}\makecell{\small \bfseries 0.60\textsuperscript{\textcolor{lexcol}{\textbf{L}}} (0.59--0.62) \\ \scriptsize \bfseries 4.0\%} & \cellcolor{best!18}\makecell{\small \bfseries 0.61\textsuperscript{\textcolor{lexcol}{\textbf{L}}} (0.60--0.62) \\ \scriptsize \bfseries 5.2\%} & \makecell{\small 0.64 (0.63--0.65) \\ \scriptsize 3.5\%} & \makecell{\small 0.64 (0.63--0.65) \\ \scriptsize 6.3\%} \\
                                 & S2 & \cellcolor{best!18}\makecell{\small \bfseries 0.79\textsuperscript{\textcolor{embcol}{\textbf{E}}} (0.78--0.80) \\ \scriptsize \bfseries 33.5\%} & \makecell{\small 0.55 (0.54--0.56) \\ \scriptsize 4.6\%} & \makecell{\small 0.59 (0.58--0.60) \\ \scriptsize 7.4\%} & \cellcolor{best!18}\makecell{\small \bfseries 0.65\textsuperscript{\textcolor{lexcol}{\textbf{L}}} (0.64--0.66) \\ \scriptsize \bfseries 5.2\%} & \makecell{\small 0.64 (0.63--0.65) \\ \scriptsize 7.1\%} \\
                                 & S3 & \makecell{\small 0.73 (0.73--0.74) \\ \scriptsize 41.8\%} & \makecell{\small 0.52 (0.51--0.53) \\ \scriptsize 2.2\%} & \makecell{\small 0.55 (0.54--0.55) \\ \scriptsize 6.1\%} & \makecell{\small 0.61 (0.60--0.62) \\ \scriptsize 6.4\%} & \cellcolor{best!18}\makecell{\small \bfseries 0.65\textsuperscript{\textcolor{lexcol}{\textbf{L}}} (0.64--0.66) \\ \scriptsize \bfseries 16.0\%} \\
\cmidrule(lr){1-7}
\multirow{3}{*}{\textsc{N2C2'08}} & S1 & \makecell{\small 0.55 (0.54--0.56) \\ \scriptsize 3.9\%} & \makecell{\small 0.58 (0.57--0.59) \\ \scriptsize 15.6\%} & \makecell{\small 0.57 (0.56--0.58) \\ \scriptsize 7.9\%} & \makecell{\small 0.57 (0.56--0.58) \\ \scriptsize 12.8\%} & \makecell{\small 0.56 (0.55--0.57) \\ \scriptsize 8.7\%} \\
                                 & S2 & \makecell{\small 0.60 (0.59--0.61) \\ \scriptsize 11.1\%} & \makecell{\small 0.59 (0.58--0.60) \\ \scriptsize 18.4\%} & \cellcolor{best!18}\makecell{\small \bfseries 0.63\textsuperscript{\textcolor{embcol}{\textbf{E}}} (0.61--0.64) \\ \scriptsize \bfseries 12.6\%} & \makecell{\small 0.61 (0.59--0.62) \\ \scriptsize 12.0\%} & \cellcolor{best!18}\makecell{\small \bfseries 0.63\textsuperscript{\textcolor{embcol}{\textbf{E}}} (0.62--0.64) \\ \scriptsize \bfseries 9.2\%} \\
                                 & S3 & \cellcolor{best!18}\makecell{\small \bfseries 0.64\textsuperscript{\textcolor{embcol}{\textbf{E}}} (0.63--0.65) \\ \scriptsize \bfseries 13.9\%} & \cellcolor{best!18}\makecell{\small \bfseries 0.59\textsuperscript{\textcolor{lexcol}{\textbf{L}}} (0.58--0.60) \\ \scriptsize \bfseries 4.0\%} & \makecell{\small 0.56 (0.55--0.57) \\ \scriptsize 9.8\%} & \cellcolor{best!18}\makecell{\small \bfseries 0.61\textsuperscript{\textcolor{embcol}{\textbf{E}}} (0.60--0.63) \\ \scriptsize \bfseries 8.6\%} & \makecell{\small 0.57 (0.56--0.58) \\ \scriptsize 7.9\%} \\
\cmidrule(lr){1-7}
\multirow{3}{*}{\textsc{EurLex}} & S1 & \makecell{\small 0.80 (0.80--0.81) \\ \scriptsize 52.0\%} & \makecell{\small 0.59 (0.58--0.60) \\ \scriptsize 10.7\%} & \makecell{\small 0.56 (0.55--0.57) \\ \scriptsize 5.7\%} & \cellcolor{best!18}\makecell{\small \bfseries 0.58\textsuperscript{\textcolor{lexcol}{\textbf{L}}} (0.57--0.59) \\ \scriptsize \bfseries 8.7\%} & \cellcolor{best!18}\makecell{\small \bfseries 0.57\textsuperscript{\textcolor{lexcol}{\textbf{L}}} (0.56--0.58) \\ \scriptsize \bfseries 6.4\%} \\
                                 & S2 & \makecell{\small 0.83 (0.83--0.84) \\ \scriptsize 60.2\%} & \makecell{\small 0.57 (0.56--0.58) \\ \scriptsize 10.0\%} & \makecell{\small 0.56 (0.55--0.57) \\ \scriptsize 4.7\%} & \makecell{\small 0.58 (0.56--0.59) \\ \scriptsize 9.0\%} & \makecell{\small 0.56 (0.55--0.57) \\ \scriptsize 10.0\%} \\
                                 & S3 & \cellcolor{best!18}\makecell{\small \bfseries 0.83\textsuperscript{\textcolor{embcol}{\textbf{E}}} (0.83--0.84) \\ \scriptsize \bfseries 61.4\%} & \cellcolor{best!18}\makecell{\small \bfseries 0.59\textsuperscript{\textcolor{embcol}{\textbf{E}}} (0.58--0.60) \\ \scriptsize \bfseries 11.9\%} & \cellcolor{best!18}\makecell{\small \bfseries 0.58\textsuperscript{\textcolor{embcol}{\textbf{E}}} (0.57--0.59) \\ \scriptsize \bfseries 3.7\%} & \makecell{\small 0.55 (0.53--0.56) \\ \scriptsize 9.0\%} & \makecell{\small 0.54 (0.53--0.55) \\ \scriptsize 7.5\%} \\
\bottomrule
\end{tabular}
\begin{tablenotes}[flushleft]
\footnotesize
\item Each cell reports the best-feature AUC-ROC with 95\,\% CI via normal approximation over 50 evaluation rounds sharing a fixed set of 100 releases (CI\,=\,mean\,$\pm$\,1.96\,$\sigma$/$\sqrt{50}$; the interval reflects negative-sampling variability only, see Section~\ref{subsec:eval_protocol}) / $\max_{d}\,\mathrm{TPR}(d)$ s.t.\ $\mathrm{FPR}(d)\leq0.05$ for that same feature.
\item \textbf{Shaded} cell = best scenario for that (dataset, $\varepsilon$), ranked by AUC and tie-broken by TPR. The superscript marks whether the winning proxy is \textcolor{lexcol}{\textbf{L}}exical or \textcolor{embcol}{\textbf{E}}mbedding: 11 vs.\ 9 of 20 highlighted cells. Individual feature names are given in the companion CSV rather than here: several proxies are rank-equivalent (e.g.\ Gaussian log-likelihood and the Mahalanobis family), so the view is identified but the particular feature within a tied group is not.
\end{tablenotes}
\end{threeparttable}
\caption{Overall subgroup-targeted membership inference performance across privacy budgets and attacker capability scenarios for \textsc{DP-Gen}. The best-performing feature is selected on the evaluation episodes (empirical worst-case audit; see Section~\ref{subsec:metrics}).}
\label{tab:overall_main_dptransformers}
\end{table*}

\begin{table*}[t]
\centering
\small
\setlength{\tabcolsep}{3pt}
\renewcommand{\arraystretch}{1.05}
\begin{tabular}{llccccc}
\toprule
\textbf{Dataset} & \textbf{Scenario} & $\bm{\varepsilon=\infty}$ & $\bm{4}$ & $\bm{2}$ & $\bm{1}$ & $\bm{0.5}$ \\
\midrule
\multirow{3}{*}{\textsc{PsyTAR}} & S1 & \makecell{\small 0.59 (0.59--0.60) \\ \scriptsize 11.8\%} & \makecell{\small 0.58 (0.58--0.59) \\ \scriptsize 12.7\%} & \cellcolor{best!18}\makecell{\small \bfseries 0.59\textsuperscript{\textcolor{lexcol}{\textbf{L}}} (0.58--0.60) \\ \scriptsize \bfseries 13.8\%} & \makecell{\small 0.59 (0.58--0.60) \\ \scriptsize 10.0\%} & \cellcolor{best!18}\makecell{\small \bfseries 0.59\textsuperscript{\textcolor{lexcol}{\textbf{L}}} (0.58--0.60) \\ \scriptsize \bfseries 10.5\%} \\
                                 & S2 & \makecell{\small 0.58 (0.57--0.59) \\ \scriptsize 8.7\%} & \makecell{\small 0.58 (0.57--0.59) \\ \scriptsize 13.9\%} & \makecell{\small 0.59 (0.58--0.60) \\ \scriptsize 10.3\%} & \makecell{\small 0.59 (0.58--0.60) \\ \scriptsize 7.6\%} & \makecell{\small 0.57 (0.57--0.58) \\ \scriptsize 5.8\%} \\
                                 & S3 & \cellcolor{best!18}\makecell{\small \bfseries 0.60\textsuperscript{\textcolor{lexcol}{\textbf{L}}} (0.58--0.61) \\ \scriptsize \bfseries 19.0\%} & \cellcolor{best!18}\makecell{\small \bfseries 0.64\textsuperscript{\textcolor{lexcol}{\textbf{L}}} (0.63--0.66) \\ \scriptsize \bfseries 14.8\%} & \makecell{\small 0.59 (0.58--0.60) \\ \scriptsize 9.1\%} & \cellcolor{best!18}\makecell{\small \bfseries 0.60\textsuperscript{\textcolor{lexcol}{\textbf{L}}} (0.59--0.61) \\ \scriptsize \bfseries 6.3\%} & \makecell{\small 0.57 (0.56--0.58) \\ \scriptsize 8.5\%} \\
\cmidrule(lr){1-7}
\multirow{3}{*}{\textsc{DMSAFN}} & S1 & \makecell{\small 0.61 (0.60--0.63) \\ \scriptsize 5.3\%} & \makecell{\small 0.61 (0.60--0.62) \\ \scriptsize 3.8\%} & \cellcolor{best!18}\makecell{\small \bfseries 0.61\textsuperscript{\textcolor{lexcol}{\textbf{L}}} (0.60--0.62) \\ \scriptsize \bfseries 4.8\%} & \cellcolor{best!18}\makecell{\small \bfseries 0.61\textsuperscript{\textcolor{lexcol}{\textbf{L}}} (0.60--0.62) \\ \scriptsize \bfseries 3.1\%} & \makecell{\small 0.61 (0.60--0.62) \\ \scriptsize 3.3\%} \\
                                 & S2 & \cellcolor{best!18}\makecell{\small \bfseries 0.63\textsuperscript{\textcolor{lexcol}{\textbf{L}}} (0.62--0.64) \\ \scriptsize \bfseries 16.1\%} & \makecell{\small 0.61 (0.60--0.62) \\ \scriptsize 7.4\%} & \makecell{\small 0.60 (0.58--0.61) \\ \scriptsize 3.0\%} & \makecell{\small 0.61 (0.60--0.61) \\ \scriptsize 14.5\%} & \cellcolor{best!18}\makecell{\small \bfseries 0.62\textsuperscript{\textcolor{lexcol}{\textbf{L}}} (0.61--0.63) \\ \scriptsize \bfseries 15.3\%} \\
                                 & S3 & \makecell{\small 0.56 (0.55--0.57) \\ \scriptsize 8.5\%} & \cellcolor{best!18}\makecell{\small \bfseries 0.63\textsuperscript{\textcolor{lexcol}{\textbf{L}}} (0.62--0.64) \\ \scriptsize \bfseries 9.9\%} & \makecell{\small 0.56 (0.55--0.56) \\ \scriptsize 11.3\%} & \makecell{\small 0.60 (0.59--0.61) \\ \scriptsize 10.4\%} & \makecell{\small 0.56 (0.55--0.57) \\ \scriptsize 8.8\%} \\
\cmidrule(lr){1-7}
\multirow{3}{*}{\textsc{N2C2'08}} & S1 & \makecell{\small 0.58 (0.57--0.59) \\ \scriptsize 16.5\%} & \makecell{\small 0.55 (0.54--0.56) \\ \scriptsize 12.8\%} & \makecell{\small 0.56 (0.55--0.57) \\ \scriptsize 5.3\%} & \makecell{\small 0.58 (0.57--0.59) \\ \scriptsize 13.2\%} & \makecell{\small 0.58 (0.57--0.59) \\ \scriptsize 5.8\%} \\
                                 & S2 & \cellcolor{best!18}\makecell{\small \bfseries 0.60\textsuperscript{\textcolor{embcol}{\textbf{E}}} (0.59--0.62) \\ \scriptsize \bfseries 11.7\%} & \cellcolor{best!18}\makecell{\small \bfseries 0.60\textsuperscript{\textcolor{embcol}{\textbf{E}}} (0.59--0.60) \\ \scriptsize \bfseries 6.2\%} & \cellcolor{best!18}\makecell{\small \bfseries 0.62\textsuperscript{\textcolor{embcol}{\textbf{E}}} (0.61--0.63) \\ \scriptsize \bfseries 9.9\%} & \makecell{\small 0.61 (0.60--0.62) \\ \scriptsize 8.3\%} & \makecell{\small 0.60 (0.59--0.61) \\ \scriptsize 12.4\%} \\
                                 & S3 & \makecell{\small 0.57 (0.56--0.58) \\ \scriptsize 11.3\%} & \makecell{\small 0.55 (0.54--0.56) \\ \scriptsize 8.8\%} & \makecell{\small 0.58 (0.57--0.59) \\ \scriptsize 7.3\%} & \cellcolor{best!18}\makecell{\small \bfseries 0.63\textsuperscript{\textcolor{lexcol}{\textbf{L}}} (0.62--0.63) \\ \scriptsize \bfseries 6.2\%} & \cellcolor{best!18}\makecell{\small \bfseries 0.63\textsuperscript{\textcolor{lexcol}{\textbf{L}}} (0.62--0.64) \\ \scriptsize \bfseries 8.6\%} \\
\cmidrule(lr){1-7}
\multirow{3}{*}{\textsc{EurLex}} & S1 & \makecell{\small 0.57 (0.57--0.58) \\ \scriptsize 5.6\%} & \makecell{\small 0.57 (0.56--0.58) \\ \scriptsize 10.5\%} & \makecell{\small 0.58 (0.57--0.59) \\ \scriptsize 6.1\%} & \makecell{\small 0.63 (0.63--0.64) \\ \scriptsize 11.8\%} & \cellcolor{best!18}\makecell{\small \bfseries 0.59\textsuperscript{\textcolor{lexcol}{\textbf{L}}} (0.58--0.60) \\ \scriptsize \bfseries 7.2\%} \\
                                 & S2 & \cellcolor{best!18}\makecell{\small \bfseries 0.58\textsuperscript{\textcolor{lexcol}{\textbf{L}}} (0.57--0.59) \\ \scriptsize \bfseries 5.5\%} & \makecell{\small 0.58 (0.57--0.59) \\ \scriptsize 9.4\%} & \cellcolor{best!18}\makecell{\small \bfseries 0.59\textsuperscript{\textcolor{lexcol}{\textbf{L}}} (0.58--0.60) \\ \scriptsize \bfseries 6.1\%} & \cellcolor{best!18}\makecell{\small \bfseries 0.64\textsuperscript{\textcolor{embcol}{\textbf{E}}} (0.63--0.65) \\ \scriptsize \bfseries 12.6\%} & \makecell{\small 0.56 (0.55--0.57) \\ \scriptsize 6.2\%} \\
                                 & S3 & \makecell{\small 0.56 (0.55--0.57) \\ \scriptsize 6.3\%} & \cellcolor{best!18}\makecell{\small \bfseries 0.59\textsuperscript{\textcolor{lexcol}{\textbf{L}}} (0.58--0.60) \\ \scriptsize \bfseries 7.8\%} & \makecell{\small 0.58 (0.57--0.59) \\ \scriptsize 7.2\%} & \makecell{\small 0.61 (0.60--0.62) \\ \scriptsize 8.4\%} & \makecell{\small 0.58 (0.57--0.59) \\ \scriptsize 6.0\%} \\
\bottomrule
\end{tabular}
\caption{Overall subgroup-targeted membership inference performance across privacy budgets and attacker capability scenarios for \textsc{EPSVec}. Winning proxy is \textcolor{lexcol}{\textbf{L}}exical or \textcolor{embcol}{\textbf{E}}mbedding: 16 vs.\ 4 of 20 highlighted cells; cell format, shading and notes as in Table~\ref{tab:overall_main_dptransformers}.}
\label{tab:overall_main_EPSVec}
\end{table*}

\begin{table*}[t]
\centering
\small
\setlength{\tabcolsep}{3pt}
\renewcommand{\arraystretch}{1.05}
\begin{tabular}{llccccc}
\toprule
\textbf{Dataset} & \textbf{Scenario} & $\bm{\varepsilon=\infty}$ & $\bm{4}$ & $\bm{2}$ & $\bm{1}$ & $\bm{0.5}$ \\
\midrule
\multirow{3}{*}{\textsc{PsyTAR}} & S1 & \makecell{\small 0.59 (0.58--0.60) \\ \scriptsize 10.9\%} & \makecell{\small 0.59 (0.58--0.60) \\ \scriptsize 13.7\%} & \makecell{\small 0.59 (0.58--0.60) \\ \scriptsize 12.4\%} & \makecell{\small 0.59 (0.59--0.60) \\ \scriptsize 13.3\%} & \makecell{\small 0.61 (0.60--0.62) \\ \scriptsize 11.7\%} \\
                                 & S2 & \cellcolor{best!18}\makecell{\small \bfseries 0.64\textsuperscript{\textcolor{lexcol}{\textbf{L}}} (0.63--0.65) \\ \scriptsize \bfseries 13.3\%} & \cellcolor{best!18}\makecell{\small \bfseries 0.62\textsuperscript{\textcolor{lexcol}{\textbf{L}}} (0.61--0.64) \\ \scriptsize \bfseries 12.4\%} & \cellcolor{best!18}\makecell{\small \bfseries 0.65\textsuperscript{\textcolor{lexcol}{\textbf{L}}} (0.64--0.66) \\ \scriptsize \bfseries 13.5\%} & \cellcolor{best!18}\makecell{\small \bfseries 0.63\textsuperscript{\textcolor{embcol}{\textbf{E}}} (0.62--0.64) \\ \scriptsize \bfseries 15.2\%} & \cellcolor{best!18}\makecell{\small \bfseries 0.63\textsuperscript{\textcolor{lexcol}{\textbf{L}}} (0.62--0.64) \\ \scriptsize \bfseries 12.2\%} \\
                                 & S3 & \makecell{\small 0.56 (0.54--0.57) \\ \scriptsize 1.1\%} & \makecell{\small 0.56 (0.55--0.57) \\ \scriptsize 8.0\%} & \makecell{\small 0.61 (0.60--0.63) \\ \scriptsize 4.4\%} & \makecell{\small 0.58 (0.57--0.60) \\ \scriptsize 4.5\%} & \makecell{\small 0.59 (0.58--0.60) \\ \scriptsize 7.4\%} \\
\cmidrule(lr){1-7}
\multirow{3}{*}{\textsc{DMSAFN}} & S1 & \cellcolor{best!18}\makecell{\small \bfseries 0.61\textsuperscript{\textcolor{lexcol}{\textbf{L}}} (0.60--0.62) \\ \scriptsize \bfseries 4.2\%} & \cellcolor{best!18}\makecell{\small \bfseries 0.61\textsuperscript{\textcolor{lexcol}{\textbf{L}}} (0.60--0.62) \\ \scriptsize \bfseries 5.2\%} & \makecell{\small 0.61 (0.60--0.62) \\ \scriptsize 5.5\%} & \cellcolor{best!18}\makecell{\small \bfseries 0.61\textsuperscript{\textcolor{lexcol}{\textbf{L}}} (0.60--0.62) \\ \scriptsize \bfseries 5.2\%} & \cellcolor{best!18}\makecell{\small \bfseries 0.61\textsuperscript{\textcolor{lexcol}{\textbf{L}}} (0.60--0.62) \\ \scriptsize \bfseries 4.3\%} \\
                                 & S2 & \makecell{\small 0.60 (0.59--0.61) \\ \scriptsize 4.7\%} & \makecell{\small 0.59 (0.58--0.59) \\ \scriptsize 10.5\%} & \makecell{\small 0.60 (0.58--0.61) \\ \scriptsize 3.6\%} & \makecell{\small 0.60 (0.59--0.62) \\ \scriptsize 4.3\%} & \makecell{\small 0.59 (0.58--0.60) \\ \scriptsize 4.1\%} \\
                                 & S3 & \makecell{\small 0.58 (0.57--0.60) \\ \scriptsize 11.3\%} & \makecell{\small 0.59 (0.58--0.61) \\ \scriptsize 7.6\%} & \cellcolor{best!18}\makecell{\small \bfseries 0.62\textsuperscript{\textcolor{embcol}{\textbf{E}}} (0.61--0.63) \\ \scriptsize \bfseries 8.9\%} & \makecell{\small 0.58 (0.57--0.59) \\ \scriptsize 8.4\%} & \makecell{\small 0.55 (0.54--0.55) \\ \scriptsize 3.5\%} \\
\cmidrule(lr){1-7}
\multirow{3}{*}{\textsc{N2C2'08}} & S1 & \makecell{\small 0.60 (0.59--0.61) \\ \scriptsize 5.5\%} & \makecell{\small 0.59 (0.58--0.60) \\ \scriptsize 13.2\%} & \makecell{\small 0.56 (0.55--0.56) \\ \scriptsize 15.7\%} & \makecell{\small 0.58 (0.57--0.59) \\ \scriptsize 10.1\%} & \makecell{\small 0.57 (0.56--0.58) \\ \scriptsize 14.7\%} \\
                                 & S2 & \cellcolor{best!18}\makecell{\small \bfseries 0.64\textsuperscript{\textcolor{embcol}{\textbf{E}}} (0.63--0.65) \\ \scriptsize \bfseries 9.9\%} & \cellcolor{best!18}\makecell{\small \bfseries 0.59\textsuperscript{\textcolor{lexcol}{\textbf{L}}} (0.59--0.60) \\ \scriptsize \bfseries 14.9\%} & \cellcolor{best!18}\makecell{\small \bfseries 0.62\textsuperscript{\textcolor{embcol}{\textbf{E}}} (0.61--0.63) \\ \scriptsize \bfseries 12.5\%} & \makecell{\small 0.60 (0.59--0.61) \\ \scriptsize 12.2\%} & \cellcolor{best!18}\makecell{\small \bfseries 0.62\textsuperscript{\textcolor{embcol}{\textbf{E}}} (0.61--0.63) \\ \scriptsize \bfseries 9.9\%} \\
                                 & S3 & \makecell{\small 0.63 (0.62--0.64) \\ \scriptsize 6.6\%} & \makecell{\small 0.59 (0.58--0.59) \\ \scriptsize 8.4\%} & \makecell{\small 0.60 (0.59--0.60) \\ \scriptsize 6.6\%} & \cellcolor{best!18}\makecell{\small \bfseries 0.62\textsuperscript{\textcolor{embcol}{\textbf{E}}} (0.61--0.63) \\ \scriptsize \bfseries 8.7\%} & \makecell{\small 0.60 (0.60--0.61) \\ \scriptsize 12.3\%} \\
\cmidrule(lr){1-7}
\multirow{3}{*}{\textsc{EurLex}} & S1 & \makecell{\small 0.59 (0.58--0.60) \\ \scriptsize 8.1\%} & \cellcolor{best!18}\makecell{\small \bfseries 0.57\textsuperscript{\textcolor{lexcol}{\textbf{L}}} (0.56--0.58) \\ \scriptsize \bfseries 5.4\%} & \cellcolor{best!18}\makecell{\small \bfseries 0.59\textsuperscript{\textcolor{lexcol}{\textbf{L}}} (0.58--0.60) \\ \scriptsize \bfseries 7.4\%} & \cellcolor{best!18}\makecell{\small \bfseries 0.58\textsuperscript{\textcolor{lexcol}{\textbf{L}}} (0.58--0.59) \\ \scriptsize \bfseries 7.8\%} & \makecell{\small 0.57 (0.56--0.58) \\ \scriptsize 9.7\%} \\
                                 & S2 & \makecell{\small 0.59 (0.58--0.60) \\ \scriptsize 6.3\%} & \makecell{\small 0.55 (0.54--0.56) \\ \scriptsize 8.0\%} & \makecell{\small 0.59 (0.58--0.60) \\ \scriptsize 10.0\%} & \makecell{\small 0.56 (0.55--0.57) \\ \scriptsize 4.9\%} & \makecell{\small 0.58 (0.57--0.59) \\ \scriptsize 5.7\%} \\
                                 & S3 & \cellcolor{best!18}\makecell{\small \bfseries 0.59\textsuperscript{\textcolor{lexcol}{\textbf{L}}} (0.59--0.60) \\ \scriptsize \bfseries 8.8\%} & \makecell{\small 0.55 (0.54--0.56) \\ \scriptsize 2.8\%} & \makecell{\small 0.56 (0.55--0.57) \\ \scriptsize 9.0\%} & \makecell{\small 0.58 (0.57--0.59) \\ \scriptsize 6.5\%} & \cellcolor{best!18}\makecell{\small \bfseries 0.61\textsuperscript{\textcolor{lexcol}{\textbf{L}}} (0.60--0.62) \\ \scriptsize \bfseries 10.8\%} \\
\bottomrule
\end{tabular}
\caption{Overall subgroup-targeted membership inference performance across privacy budgets and attacker capability scenarios for \textsc{Aug-PE}. Winning proxy is \textcolor{lexcol}{\textbf{L}}exical or \textcolor{embcol}{\textbf{E}}mbedding: 14 vs.\ 6 of 20 highlighted cells; cell format, shading and notes as in Table~\ref{tab:overall_main_dptransformers}.}
\label{tab:overall_main_augpe}
\end{table*}

%% file: tables/table_main_rare_CI.tex
\definecolor{lexcol}{HTML}{1F5FA9}   
\definecolor{embcol}{HTML}{A8431F}   
\definecolor{best}{HTML}{2A78D6}     
\begin{table*}[t]
\centering
\small
\setlength{\tabcolsep}{3pt}
\renewcommand{\arraystretch}{1.05}
\begin{threeparttable}
\begin{tabular}{lllccc}
\toprule
\textbf{Generator} & \textbf{Dataset} & \textbf{Scenario} & $\bm{\varepsilon=\infty}$ & $\bm{4}$ & $\bm{1}$ \\
\midrule
\multirow{9}{*}{DP-Gen}
& \multirow{3}{*}{\textsc{PsyTAR}} & S1 & \makecell{\small 0.82 (0.82--0.83) \\ \scriptsize 49.4\%} & \cellcolor{best!18}\makecell{\small \bfseries 0.59\textsuperscript{\textcolor{embcol}{\textbf{E}}} (0.58--0.60) \\ \scriptsize \bfseries 13.1\%} & \makecell{\small 0.57 (0.56--0.58) \\ \scriptsize 9.3\%} \\
&                                  & S2 & \makecell{\small 0.84 (0.83--0.84) \\ \scriptsize 46.4\%} & \makecell{\small 0.53 (0.52--0.54) \\ \scriptsize 11.2\%} & \makecell{\small 0.55 (0.54--0.56) \\ \scriptsize 4.0\%} \\
&                                  & S3 & \cellcolor{best!18}\makecell{\small \bfseries 0.86\textsuperscript{\textcolor{lexcol}{\textbf{L}}} (0.85--0.86) \\ \scriptsize \bfseries 58.8\%} & \makecell{\small 0.56 (0.55--0.58) \\ \scriptsize 9.3\%} & \cellcolor{best!18}\makecell{\small \bfseries 0.58\textsuperscript{\textcolor{lexcol}{\textbf{L}}} (0.57--0.59) \\ \scriptsize \bfseries 10.8\%} \\
\cmidrule(lr){2-6}
& \multirow{3}{*}{\textsc{N2C2'08}} & S1 & \makecell{\small 0.62 (0.61--0.63) \\ \scriptsize 13.7\%} & \makecell{\small 0.53 (0.51--0.54) \\ \scriptsize 5.4\%} & \makecell{\small 0.53 (0.52--0.54) \\ \scriptsize 6.9\%} \\
&                                  & S2 & \makecell{\small 0.69 (0.68--0.70) \\ \scriptsize 15.3\%} & \makecell{\small 0.55 (0.54--0.56) \\ \scriptsize 2.2\%} & \makecell{\small 0.56 (0.56--0.57) \\ \scriptsize 3.2\%} \\
&                                  & S3 & \cellcolor{best!18}\makecell{\small \bfseries 0.77\textsuperscript{\textcolor{lexcol}{\textbf{L}}} (0.76--0.78) \\ \scriptsize \bfseries 28.0\%} & \cellcolor{best!18}\makecell{\small \bfseries 0.56\textsuperscript{\textcolor{lexcol}{\textbf{L}}} (0.56--0.57) \\ \scriptsize \bfseries 8.2\%} & \cellcolor{best!18}\makecell{\small \bfseries 0.58\textsuperscript{\textcolor{embcol}{\textbf{E}}} (0.58--0.59) \\ \scriptsize \bfseries 2.8\%} \\
\cmidrule(lr){2-6}
& \multirow{3}{*}{\textsc{EurLex}} & S1 & \makecell{\small 0.63 (0.62--0.64) \\ \scriptsize 13.2\%} & \cellcolor{best!18}\makecell{\small \bfseries 0.57\textsuperscript{\textcolor{embcol}{\textbf{E}}} (0.56--0.59) \\ \scriptsize \bfseries 8.7\%} & \makecell{\small 0.56 (0.55--0.57) \\ \scriptsize 6.1\%} \\
&                                  & S2 & \makecell{\small 0.61 (0.60--0.61) \\ \scriptsize 10.8\%} & \makecell{\small 0.56 (0.55--0.58) \\ \scriptsize 8.2\%} & \cellcolor{best!18}\makecell{\small \bfseries 0.57\textsuperscript{\textcolor{lexcol}{\textbf{L}}} (0.56--0.58) \\ \scriptsize \bfseries 5.7\%} \\
&                                  & S3 & \cellcolor{best!18}\makecell{\small \bfseries 0.64\textsuperscript{\textcolor{lexcol}{\textbf{L}}} (0.63--0.65) \\ \scriptsize \bfseries 22.9\%} & \makecell{\small 0.55 (0.54--0.56) \\ \scriptsize 7.4\%} & \makecell{\small 0.55 (0.54--0.57) \\ \scriptsize 4.5\%} \\
\midrule
\multirow{9}{*}{EPSVec}
& \multirow{3}{*}{\textsc{PsyTAR}} & S1 & \makecell{\small 0.57 (0.56--0.58) \\ \scriptsize 7.8\%} & \cellcolor{best!18}\makecell{\small \bfseries 0.60\textsuperscript{\textcolor{embcol}{\textbf{E}}} (0.59--0.61) \\ \scriptsize \bfseries 15.8\%} & \cellcolor{best!18}\makecell{\small \bfseries 0.58\textsuperscript{\textcolor{embcol}{\textbf{E}}} (0.56--0.59) \\ \scriptsize \bfseries 8.7\%} \\
&                                  & S2 & \makecell{\small 0.54 (0.53--0.55) \\ \scriptsize 5.5\%} & \makecell{\small 0.56 (0.55--0.57) \\ \scriptsize 6.5\%} & \makecell{\small 0.57 (0.56--0.58) \\ \scriptsize 7.6\%} \\
&                                  & S3 & \cellcolor{best!18}\makecell{\small \bfseries 0.59\textsuperscript{\textcolor{embcol}{\textbf{E}}} (0.58--0.61) \\ \scriptsize \bfseries 7.4\%} & \makecell{\small 0.58 (0.57--0.59) \\ \scriptsize 12.5\%} & \makecell{\small 0.57 (0.56--0.59) \\ \scriptsize 7.7\%} \\
\cmidrule(lr){2-6}
& \multirow{3}{*}{\textsc{N2C2'08}} & S1 & \makecell{\small 0.52 (0.51--0.53) \\ \scriptsize 2.9\%} & \makecell{\small 0.56 (0.55--0.57) \\ \scriptsize 15.5\%} & \cellcolor{best!18}\makecell{\small \bfseries 0.56\textsuperscript{\textcolor{embcol}{\textbf{E}}} (0.55--0.56) \\ \scriptsize \bfseries 5.7\%} \\
&                                  & S2 & \makecell{\small 0.55 (0.54--0.56) \\ \scriptsize 5.8\%} & \makecell{\small 0.56 (0.55--0.57) \\ \scriptsize 3.5\%} & \makecell{\small 0.55 (0.54--0.56) \\ \scriptsize 11.5\%} \\
&                                  & S3 & \cellcolor{best!18}\makecell{\small \bfseries 0.58\textsuperscript{\textcolor{embcol}{\textbf{E}}} (0.58--0.59) \\ \scriptsize \bfseries 11.4\%} & \cellcolor{best!18}\makecell{\small \bfseries 0.66\textsuperscript{\textcolor{lexcol}{\textbf{L}}} (0.65--0.67) \\ \scriptsize \bfseries 9.5\%} & \makecell{\small 0.55 (0.54--0.56) \\ \scriptsize 3.3\%} \\
\cmidrule(lr){2-6}
& \multirow{3}{*}{\textsc{EurLex}} & S1 & \cellcolor{best!18}\makecell{\small \bfseries 0.62\textsuperscript{\textcolor{embcol}{\textbf{E}}} (0.61--0.63) \\ \scriptsize \bfseries 9.2\%} & \cellcolor{best!18}\makecell{\small \bfseries 0.62\textsuperscript{\textcolor{embcol}{\textbf{E}}} (0.61--0.63) \\ \scriptsize \bfseries 9.9\%} & \cellcolor{best!18}\makecell{\small \bfseries 0.62\textsuperscript{\textcolor{embcol}{\textbf{E}}} (0.60--0.63) \\ \scriptsize \bfseries 9.4\%} \\
&                                  & S2 & \makecell{\small 0.58 (0.57--0.59) \\ \scriptsize 8.0\%} & \makecell{\small 0.62 (0.61--0.63) \\ \scriptsize 10.8\%} & \makecell{\small 0.58 (0.57--0.59) \\ \scriptsize 6.4\%} \\
&                                  & S3 & \makecell{\small 0.59 (0.58--0.60) \\ \scriptsize 11.9\%} & \makecell{\small 0.56 (0.55--0.57) \\ \scriptsize 3.1\%} & \makecell{\small 0.57 (0.57--0.58) \\ \scriptsize 6.3\%} \\
\bottomrule
\end{tabular}
\begin{tablenotes}[flushleft]
\footnotesize
\item Each cell reports the best-feature AUC-ROC with 95\,\% CI via normal approximation over 50 evaluation rounds sharing a fixed set of 100 releases (CI\,=\,mean\,$\pm$\,1.96\,$\sigma$/$\sqrt{50}$; the interval reflects negative-sampling variability only, see Section~\ref{subsec:eval_protocol}) / $\max_{d}\,\mathrm{TPR}(d)$ s.t.\ $\mathrm{FPR}(d)\leq0.05$ for that same feature. This sweep covers $\varepsilon\in\{\infty,4,1\}$ only; the remaining budgets were not run for this group.
\item \textbf{Shaded} cell = best scenario for that (dataset, $\varepsilon$), ranked by AUC and tie-broken by TPR. The superscript marks whether the winning proxy is \textcolor{lexcol}{\textbf{L}}exical or \textcolor{embcol}{\textbf{E}}mbedding: 7 vs.\ 11 of 18 highlighted cells. Individual feature names are given in the companion CSV rather than here: several proxies are rank-equivalent (e.g.\ Gaussian log-likelihood and the Mahalanobis family), so the view is identified but the particular feature within a tied group is not.
\end{tablenotes}
\end{threeparttable}
\caption{Overall rare-label subgroup-targeted membership inference performance across generators, privacy budgets and attacker capability scenarios. The best-performing feature is selected on the evaluation episodes (empirical worst-case audit; see Section~\ref{subsec:metrics}).}
\label{tab:overall_main_rare}
\end{table*}

%% file: tables/table_meta_outlier.tex
\definecolor{lexcol}{HTML}{1F5FA9}   
\definecolor{embcol}{HTML}{A8431F}   
\definecolor{best}{HTML}{2A78D6}     
\begin{table*}[t]
\centering
\small
\setlength{\tabcolsep}{3pt}
\renewcommand{\arraystretch}{1.05}
\begin{threeparttable}
\caption{Meta-classifier (fusion of all 32 proxies) on the outlier group, \textsc{DP-Gen} releases: AUC, TPR and the gain over the best single proxy.}
\label{tab:meta-outlier}
\begin{tabular}{lllccccc}
\toprule
\textbf{Generator} & \textbf{Dataset} & \textbf{Scenario} & $\bm{\varepsilon=\infty}$ & $\bm{4}$ & $\bm{2}$ & $\bm{1}$ & $\bm{0.5}$ \\
\midrule
\multirow{12}{*}{DP-Gen}
& \multirow{3}{*}{\textsc{PsyTAR}} & S1 & \makecell{\small 0.76 (0.76--0.77) \\ \scriptsize 51.5\% \quad \scriptsize $-$0.027\,\textcolor{lexcol}{\textbf{L}}} & \makecell{\small 0.52 (0.51--0.54) \\ \scriptsize 7.2\% \quad \scriptsize $-$0.062\,\textcolor{embcol}{\textbf{E}}} & \makecell{\small 0.56 (0.54--0.58) \\ \scriptsize 9.7\% \quad \scriptsize $-$0.025\,\textcolor{lexcol}{\textbf{L}}} & \makecell{\small 0.54 (0.52--0.55) \\ \scriptsize 8.0\% \quad \scriptsize $-$0.066\,\textcolor{embcol}{\textbf{E}}} & \makecell{\small 0.55 (0.53--0.56) \\ \scriptsize 6.8\% \quad \scriptsize $-$0.068\,\textcolor{lexcol}{\textbf{L}}} \\
&                                  & S2 & \makecell{\small 0.76 (0.75--0.77) \\ \scriptsize 46.9\% \quad \scriptsize $-$0.041\,\textcolor{lexcol}{\textbf{L}}} & \makecell{\small 0.53 (0.51--0.55) \\ \scriptsize 8.3\% \quad \scriptsize $-$0.060\,\textcolor{lexcol}{\textbf{L}}} & \makecell{\small 0.56 (0.55--0.58) \\ \scriptsize 10.4\% \quad \scriptsize $-$0.013\,\textcolor{lexcol}{\textbf{L}}} & \makecell{\small 0.51 (0.50--0.53) \\ \scriptsize 8.9\% \quad \scriptsize $-$0.060\,\textcolor{lexcol}{\textbf{L}}} & \makecell{\small 0.54 (0.53--0.56) \\ \scriptsize 7.9\% \quad \scriptsize $-$0.093\,\textcolor{lexcol}{\textbf{L}}} \\
&                                  & S3 & \makecell{\small 0.71 (0.70--0.72) \\ \scriptsize 44.1\% \quad \scriptsize $-$0.049\,\textcolor{lexcol}{\textbf{L}}} & \makecell{\small 0.45 (0.43--0.46) \\ \scriptsize 4.8\% \quad \scriptsize $-$0.117\,\textcolor{embcol}{\textbf{E}}} & \makecell{\small 0.56 (0.55--0.58) \\ \scriptsize 10.3\% \quad \scriptsize $-$0.024\,\textcolor{lexcol}{\textbf{L}}} & \makecell{\small 0.51 (0.50--0.53) \\ \scriptsize 7.9\% \quad \scriptsize $-$0.059\,\textcolor{lexcol}{\textbf{L}}} & \makecell{\small 0.53 (0.51--0.55) \\ \scriptsize 5.5\% \quad \scriptsize $-$0.098\,\textcolor{lexcol}{\textbf{L}}} \\
\cmidrule(lr){2-8}
& \multirow{3}{*}{\textsc{DMSAFN}} & S1 & \makecell{\small 0.70 (0.68--0.71) \\ \scriptsize 28.3\% \quad \scriptsize $-$0.093\,\textcolor{embcol}{\textbf{E}}} & \cellcolor{best!18}\makecell{\small 0.64 (0.62--0.65) \\ \scriptsize 9.3\% \quad \scriptsize \textbf{$+$0.034}\,\textcolor{lexcol}{\textbf{L}}} & \makecell{\small 0.55 (0.54--0.57) \\ \scriptsize 7.1\% \quad \scriptsize $-$0.057\,\textcolor{lexcol}{\textbf{L}}} & \makecell{\small 0.55 (0.54--0.57) \\ \scriptsize 5.9\% \quad \scriptsize $-$0.084\,\textcolor{lexcol}{\textbf{L}}} & \makecell{\small 0.56 (0.54--0.57) \\ \scriptsize 6.3\% \quad \scriptsize $-$0.081\,\textcolor{lexcol}{\textbf{L}}} \\
&                                  & S2 & \makecell{\small 0.69 (0.67--0.70) \\ \scriptsize 29.8\% \quad \scriptsize $-$0.102\,\textcolor{embcol}{\textbf{E}}} & \makecell{\small 0.52 (0.50--0.53) \\ \scriptsize 6.8\% \quad \scriptsize $-$0.036\,\textcolor{lexcol}{\textbf{L}}} & \makecell{\small 0.50 (0.48--0.51) \\ \scriptsize 5.0\% \quad \scriptsize $-$0.094\,\textcolor{lexcol}{\textbf{L}}} & \makecell{\small 0.52 (0.50--0.54) \\ \scriptsize 7.5\% \quad \scriptsize $-$0.123\,\textcolor{lexcol}{\textbf{L}}} & \makecell{\small 0.54 (0.53--0.56) \\ \scriptsize 7.3\% \quad \scriptsize $-$0.100\,\textcolor{lexcol}{\textbf{L}}} \\
&                                  & S3 & \makecell{\small 0.69 (0.68--0.71) \\ \scriptsize 29.4\% \quad \scriptsize $-$0.040\,\textcolor{embcol}{\textbf{E}}} & \cellcolor{best!18}\makecell{\small 0.63 (0.61--0.64) \\ \scriptsize 10.8\% \quad \scriptsize \textbf{$+$0.109}\,\textcolor{embcol}{\textbf{E}}} & \makecell{\small 0.45 (0.43--0.46) \\ \scriptsize 4.8\% \quad \scriptsize $-$0.096\,\textcolor{embcol}{\textbf{E}}} & \makecell{\small 0.54 (0.52--0.56) \\ \scriptsize 6.2\% \quad \scriptsize $-$0.075\,\textcolor{embcol}{\textbf{E}}} & \makecell{\small 0.58 (0.56--0.60) \\ \scriptsize 7.2\% \quad \scriptsize $-$0.062\,\textcolor{lexcol}{\textbf{L}}} \\
\cmidrule(lr){2-8}
& \multirow{3}{*}{\textsc{N2C2'08}} & S1 & \makecell{\small 0.49 (0.47--0.51) \\ \scriptsize 7.5\% \quad \scriptsize $-$0.062\,\textcolor{lexcol}{\textbf{L}}} & \makecell{\small 0.53 (0.51--0.55) \\ \scriptsize 5.5\% \quad \scriptsize $-$0.043\,\textcolor{lexcol}{\textbf{L}}} & \makecell{\small 0.48 (0.46--0.50) \\ \scriptsize 3.8\% \quad \scriptsize $-$0.093\,\textcolor{embcol}{\textbf{E}}} & \makecell{\small 0.53 (0.52--0.55) \\ \scriptsize 7.4\% \quad \scriptsize $-$0.038\,\textcolor{lexcol}{\textbf{L}}} & \makecell{\small 0.46 (0.44--0.48) \\ \scriptsize 3.2\% \quad \scriptsize $-$0.101\,\textcolor{embcol}{\textbf{E}}} \\
&                                  & S2 & \makecell{\small 0.54 (0.52--0.55) \\ \scriptsize 9.2\% \quad \scriptsize $-$0.065\,\textcolor{lexcol}{\textbf{L}}} & \makecell{\small 0.57 (0.55--0.59) \\ \scriptsize 8.2\% \quad \scriptsize $-$0.019\,\textcolor{lexcol}{\textbf{L}}} & \makecell{\small 0.53 (0.51--0.55) \\ \scriptsize 6.2\% \quad \scriptsize $-$0.092\,\textcolor{embcol}{\textbf{E}}} & \makecell{\small 0.56 (0.54--0.57) \\ \scriptsize 8.2\% \quad \scriptsize $-$0.048\,\textcolor{embcol}{\textbf{E}}} & \makecell{\small 0.53 (0.51--0.55) \\ \scriptsize 6.3\% \quad \scriptsize $-$0.097\,\textcolor{embcol}{\textbf{E}}} \\
&                                  & S3 & \makecell{\small 0.55 (0.53--0.57) \\ \scriptsize 5.9\% \quad \scriptsize $-$0.089\,\textcolor{embcol}{\textbf{E}}} & \makecell{\small 0.49 (0.47--0.51) \\ \scriptsize 4.2\% \quad \scriptsize $-$0.102\,\textcolor{lexcol}{\textbf{L}}} & \makecell{\small 0.48 (0.46--0.51) \\ \scriptsize 3.8\% \quad \scriptsize $-$0.079\,\textcolor{embcol}{\textbf{E}}} & \makecell{\small 0.52 (0.50--0.54) \\ \scriptsize 6.9\% \quad \scriptsize $-$0.098\,\textcolor{embcol}{\textbf{E}}} & \makecell{\small 0.44 (0.42--0.46) \\ \scriptsize 4.2\% \quad \scriptsize $-$0.131\,\textcolor{embcol}{\textbf{E}}} \\
\cmidrule(lr){2-8}
& \multirow{3}{*}{\textsc{EurLex}} & S1 & \makecell{\small 0.79 (0.78--0.80) \\ \scriptsize 50.0\% \quad \scriptsize $-$0.013\,\textcolor{embcol}{\textbf{E}}} & \makecell{\small 0.48 (0.46--0.50) \\ \scriptsize 4.2\% \quad \scriptsize $-$0.105\,\textcolor{embcol}{\textbf{E}}} & \makecell{\small 0.47 (0.45--0.49) \\ \scriptsize 5.8\% \quad \scriptsize $-$0.092\,\textcolor{lexcol}{\textbf{L}}} & \makecell{\small 0.44 (0.43--0.46) \\ \scriptsize 4.2\% \quad \scriptsize $-$0.136\,\textcolor{lexcol}{\textbf{L}}} & \makecell{\small 0.48 (0.46--0.50) \\ \scriptsize 4.7\% \quad \scriptsize $-$0.098\,\textcolor{lexcol}{\textbf{L}}} \\
&                                  & S2 & \makecell{\small 0.79 (0.78--0.80) \\ \scriptsize 49.9\% \quad \scriptsize $-$0.037\,\textcolor{embcol}{\textbf{E}}} & \makecell{\small 0.49 (0.47--0.51) \\ \scriptsize 4.0\% \quad \scriptsize $-$0.078\,\textcolor{embcol}{\textbf{E}}} & \makecell{\small 0.45 (0.43--0.47) \\ \scriptsize 4.4\% \quad \scriptsize $-$0.108\,\textcolor{lexcol}{\textbf{L}}} & \makecell{\small 0.43 (0.41--0.44) \\ \scriptsize 3.9\% \quad \scriptsize $-$0.150\,\textcolor{lexcol}{\textbf{L}}} & \makecell{\small 0.44 (0.42--0.46) \\ \scriptsize 5.0\% \quad \scriptsize $-$0.125\,\textcolor{lexcol}{\textbf{L}}} \\
&                                  & S3 & \makecell{\small 0.79 (0.79--0.80) \\ \scriptsize 48.1\% \quad \scriptsize $-$0.041\,\textcolor{embcol}{\textbf{E}}} & \makecell{\small 0.49 (0.47--0.50) \\ \scriptsize 4.2\% \quad \scriptsize $-$0.100\,\textcolor{embcol}{\textbf{E}}} & \makecell{\small 0.49 (0.47--0.50) \\ \scriptsize 5.9\% \quad \scriptsize $-$0.093\,\textcolor{embcol}{\textbf{E}}} & \makecell{\small 0.47 (0.45--0.49) \\ \scriptsize 3.1\% \quad \scriptsize $-$0.078\,\textcolor{lexcol}{\textbf{L}}} & \makecell{\small 0.45 (0.43--0.47) \\ \scriptsize 4.7\% \quad \scriptsize $-$0.089\,\textcolor{lexcol}{\textbf{L}}} \\
\bottomrule
\end{tabular}
\begin{tablenotes}[flushleft]
\footnotesize
\item \textbf{Top line:} AUC-ROC of the meta-classifier (mean $\pm$ 95\,\% CI, normal approximation over 50 evaluation rounds; CI $=$ mean $\pm$ 1.96\,$\sigma/\sqrt{50}$). Every round draws the negative candidates from the same evaluation trial plan the main tables use, so a cell here and the corresponding main-table cell see identical data. Within a round, membership probabilities come from nested cross-validation (outer 5-fold out-of-fold prediction; inner 5-fold grid search over the number of selected features $k\in\{5,8,12,\text{all}\}$ and the $\ell_2$ penalty $C\in\{0.01,0.1,1,10\}$ of a class-balanced logistic regression on standardised, mean-imputed features), so no record contributes to the model that scores it.
\item \textbf{Bottom line:} TPR at FPR $\leq0.05$, then $\Delta$ = meta-classifier AUC $-$ AUC of the best of the 32 individual proxies, recomputed on the same rounds. That reference is an \emph{oracle} --- the proxy is chosen after seeing the evaluation data, which is what the main table reports --- so $\Delta>0$ (shaded) means the fusion beats a single proxy no adversary could have picked in advance. The superscript gives that proxy's view (\textcolor{lexcol}{\textbf{L}}exical / \textcolor{embcol}{\textbf{E}}mbedding); its name, the modal $(k,C)$ and the number of scored records per round are in the companion CSV.
\item Across the 60 populated cells the fusion wins in 2 and loses in 58, mean $\Delta=-0.071$.
\end{tablenotes}
\end{threeparttable}
\end{table*}

\begin{table*}[t]
\centering
\small
\setlength{\tabcolsep}{3pt}
\renewcommand{\arraystretch}{1.05}
\begin{threeparttable}
\caption{Meta-classifier (fusion of all 32 proxies) on the outlier group, \textsc{EPSVec} releases: AUC, TPR and the gain over the best single proxy. Companion to Table~\ref{tab:meta-outlier}.}
\label{tab:meta-outlier-epsvec}
\begin{tabular}{lllccccc}
\toprule
\textbf{Generator} & \textbf{Dataset} & \textbf{Scenario} & $\bm{\varepsilon=\infty}$ & $\bm{4}$ & $\bm{2}$ & $\bm{1}$ & $\bm{0.5}$ \\
\midrule
\multirow{12}{*}{EPSVec}
& \multirow{3}{*}{\textsc{PsyTAR}} & S1 & \makecell{\small 0.51 (0.50--0.53) \\ \scriptsize 6.0\% \quad \scriptsize $-$0.081\,\textcolor{lexcol}{\textbf{L}}} & \makecell{\small 0.55 (0.54--0.57) \\ \scriptsize 8.0\% \quad \scriptsize $-$0.033\,\textcolor{lexcol}{\textbf{L}}} & \makecell{\small 0.54 (0.52--0.56) \\ \scriptsize 5.4\% \quad \scriptsize $-$0.054\,\textcolor{lexcol}{\textbf{L}}} & \makecell{\small 0.55 (0.54--0.57) \\ \scriptsize 9.0\% \quad \scriptsize $-$0.038\,\textcolor{lexcol}{\textbf{L}}} & \makecell{\small 0.54 (0.52--0.55) \\ \scriptsize 6.9\% \quad \scriptsize $-$0.055\,\textcolor{lexcol}{\textbf{L}}} \\
&                                  & S2 & \makecell{\small 0.53 (0.51--0.54) \\ \scriptsize 7.3\% \quad \scriptsize $-$0.055\,\textcolor{lexcol}{\textbf{L}}} & \makecell{\small 0.54 (0.52--0.55) \\ \scriptsize 7.7\% \quad \scriptsize $-$0.042\,\textcolor{lexcol}{\textbf{L}}} & \makecell{\small 0.54 (0.53--0.56) \\ \scriptsize 7.2\% \quad \scriptsize $-$0.042\,\textcolor{lexcol}{\textbf{L}}} & \makecell{\small 0.55 (0.53--0.57) \\ \scriptsize 8.1\% \quad \scriptsize $-$0.040\,\textcolor{lexcol}{\textbf{L}}} & \makecell{\small 0.54 (0.53--0.56) \\ \scriptsize 7.7\% \quad \scriptsize $-$0.030\,\textcolor{lexcol}{\textbf{L}}} \\
&                                  & S3 & \makecell{\small 0.48 (0.46--0.49) \\ \scriptsize 5.9\% \quad \scriptsize $-$0.119\,\textcolor{lexcol}{\textbf{L}}} & \makecell{\small 0.55 (0.54--0.57) \\ \scriptsize 8.1\% \quad \scriptsize $-$0.091\,\textcolor{lexcol}{\textbf{L}}} & \makecell{\small 0.52 (0.51--0.54) \\ \scriptsize 6.8\% \quad \scriptsize $-$0.069\,\textcolor{lexcol}{\textbf{L}}} & \makecell{\small 0.52 (0.50--0.54) \\ \scriptsize 6.6\% \quad \scriptsize $-$0.076\,\textcolor{lexcol}{\textbf{L}}} & \makecell{\small 0.52 (0.51--0.54) \\ \scriptsize 6.1\% \quad \scriptsize $-$0.050\,\textcolor{lexcol}{\textbf{L}}} \\
\cmidrule(lr){2-8}
& \multirow{3}{*}{\textsc{DMSAFN}} & S1 & \makecell{\small 0.56 (0.54--0.58) \\ \scriptsize 6.8\% \quad \scriptsize $-$0.053\,\textcolor{lexcol}{\textbf{L}}} & \makecell{\small 0.57 (0.56--0.59) \\ \scriptsize 5.3\% \quad \scriptsize $-$0.041\,\textcolor{lexcol}{\textbf{L}}} & \makecell{\small 0.56 (0.54--0.57) \\ \scriptsize 6.8\% \quad \scriptsize $-$0.049\,\textcolor{lexcol}{\textbf{L}}} & \makecell{\small 0.60 (0.59--0.62) \\ \scriptsize 8.7\% \quad \scriptsize $-$0.010\,\textcolor{lexcol}{\textbf{L}}} & \makecell{\small 0.55 (0.53--0.56) \\ \scriptsize 5.3\% \quad \scriptsize $-$0.064\,\textcolor{lexcol}{\textbf{L}}} \\
&                                  & S2 & \makecell{\small 0.53 (0.51--0.55) \\ \scriptsize 5.6\% \quad \scriptsize $-$0.100\,\textcolor{lexcol}{\textbf{L}}} & \makecell{\small 0.57 (0.55--0.58) \\ \scriptsize 7.5\% \quad \scriptsize $-$0.044\,\textcolor{lexcol}{\textbf{L}}} & \makecell{\small 0.54 (0.53--0.56) \\ \scriptsize 7.5\% \quad \scriptsize $-$0.052\,\textcolor{lexcol}{\textbf{L}}} & \makecell{\small 0.56 (0.54--0.57) \\ \scriptsize 8.6\% \quad \scriptsize $-$0.048\,\textcolor{lexcol}{\textbf{L}}} & \makecell{\small 0.55 (0.53--0.56) \\ \scriptsize 7.0\% \quad \scriptsize $-$0.075\,\textcolor{lexcol}{\textbf{L}}} \\
&                                  & S3 & \makecell{\small 0.49 (0.48--0.51) \\ \scriptsize 4.3\% \quad \scriptsize $-$0.068\,\textcolor{lexcol}{\textbf{L}}} & \makecell{\small 0.55 (0.53--0.57) \\ \scriptsize 7.4\% \quad \scriptsize $-$0.083\,\textcolor{lexcol}{\textbf{L}}} & \makecell{\small 0.56 (0.54--0.58) \\ \scriptsize 7.7\% \quad \scriptsize $-$0.002\,\textcolor{embcol}{\textbf{E}}} & \makecell{\small 0.51 (0.50--0.53) \\ \scriptsize 5.0\% \quad \scriptsize $-$0.083\,\textcolor{embcol}{\textbf{E}}} & \makecell{\small 0.51 (0.49--0.53) \\ \scriptsize 5.1\% \quad \scriptsize $-$0.044\,\textcolor{lexcol}{\textbf{L}}} \\
\cmidrule(lr){2-8}
& \multirow{3}{*}{\textsc{N2C2'08}} & S1 & \makecell{\small 0.42 (0.40--0.44) \\ \scriptsize 3.5\% \quad \scriptsize $-$0.155\,\textcolor{lexcol}{\textbf{L}}} & \makecell{\small 0.43 (0.41--0.45) \\ \scriptsize 3.4\% \quad \scriptsize $-$0.116\,\textcolor{lexcol}{\textbf{L}}} & \makecell{\small 0.44 (0.42--0.45) \\ \scriptsize 3.0\% \quad \scriptsize $-$0.126\,\textcolor{embcol}{\textbf{E}}} & \makecell{\small 0.56 (0.54--0.57) \\ \scriptsize 7.5\% \quad \scriptsize $-$0.025\,\textcolor{lexcol}{\textbf{L}}} & \makecell{\small 0.46 (0.44--0.48) \\ \scriptsize 3.8\% \quad \scriptsize $-$0.118\,\textcolor{embcol}{\textbf{E}}} \\
&                                  & S2 & \makecell{\small 0.50 (0.48--0.52) \\ \scriptsize 6.2\% \quad \scriptsize $-$0.102\,\textcolor{embcol}{\textbf{E}}} & \makecell{\small 0.48 (0.46--0.49) \\ \scriptsize 4.7\% \quad \scriptsize $-$0.118\,\textcolor{embcol}{\textbf{E}}} & \makecell{\small 0.49 (0.47--0.51) \\ \scriptsize 3.5\% \quad \scriptsize $-$0.125\,\textcolor{embcol}{\textbf{E}}} & \makecell{\small 0.51 (0.49--0.53) \\ \scriptsize 4.9\% \quad \scriptsize $-$0.101\,\textcolor{embcol}{\textbf{E}}} & \makecell{\small 0.51 (0.48--0.53) \\ \scriptsize 7.3\% \quad \scriptsize $-$0.091\,\textcolor{embcol}{\textbf{E}}} \\
&                                  & S3 & \makecell{\small 0.43 (0.41--0.44) \\ \scriptsize 3.4\% \quad \scriptsize $-$0.145\,\textcolor{lexcol}{\textbf{L}}} & \makecell{\small 0.47 (0.45--0.49) \\ \scriptsize 4.2\% \quad \scriptsize $-$0.078\,\textcolor{embcol}{\textbf{E}}} & \makecell{\small 0.44 (0.43--0.46) \\ \scriptsize 3.5\% \quad \scriptsize $-$0.137\,\textcolor{embcol}{\textbf{E}}} & \makecell{\small 0.51 (0.49--0.53) \\ \scriptsize 7.2\% \quad \scriptsize $-$0.115\,\textcolor{lexcol}{\textbf{L}}} & \makecell{\small 0.49 (0.48--0.50) \\ \scriptsize 4.2\% \quad \scriptsize $-$0.138\,\textcolor{lexcol}{\textbf{L}}} \\
\cmidrule(lr){2-8}
& \multirow{3}{*}{\textsc{EurLex}} & S1 & \makecell{\small 0.42 (0.40--0.44) \\ \scriptsize 2.9\% \quad \scriptsize $-$0.155\,\textcolor{lexcol}{\textbf{L}}} & \makecell{\small 0.46 (0.44--0.48) \\ \scriptsize 4.8\% \quad \scriptsize $-$0.113\,\textcolor{lexcol}{\textbf{L}}} & \makecell{\small 0.48 (0.46--0.50) \\ \scriptsize 7.5\% \quad \scriptsize $-$0.105\,\textcolor{lexcol}{\textbf{L}}} & \makecell{\small 0.54 (0.52--0.56) \\ \scriptsize 7.3\% \quad \scriptsize $-$0.093\,\textcolor{embcol}{\textbf{E}}} & \makecell{\small 0.48 (0.46--0.49) \\ \scriptsize 5.6\% \quad \scriptsize $-$0.109\,\textcolor{lexcol}{\textbf{L}}} \\
&                                  & S2 & \makecell{\small 0.41 (0.40--0.43) \\ \scriptsize 4.0\% \quad \scriptsize $-$0.168\,\textcolor{lexcol}{\textbf{L}}} & \makecell{\small 0.45 (0.43--0.47) \\ \scriptsize 5.0\% \quad \scriptsize $-$0.131\,\textcolor{lexcol}{\textbf{L}}} & \makecell{\small 0.45 (0.43--0.47) \\ \scriptsize 5.5\% \quad \scriptsize $-$0.134\,\textcolor{lexcol}{\textbf{L}}} & \makecell{\small 0.55 (0.54--0.57) \\ \scriptsize 9.6\% \quad \scriptsize $-$0.089\,\textcolor{embcol}{\textbf{E}}} & \makecell{\small 0.44 (0.43--0.45) \\ \scriptsize 3.8\% \quad \scriptsize $-$0.119\,\textcolor{lexcol}{\textbf{L}}} \\
&                                  & S3 & \makecell{\small 0.52 (0.51--0.54) \\ \scriptsize 4.4\% \quad \scriptsize $-$0.038\,\textcolor{embcol}{\textbf{E}}} & \makecell{\small 0.49 (0.47--0.50) \\ \scriptsize 5.3\% \quad \scriptsize $-$0.103\,\textcolor{lexcol}{\textbf{L}}} & \makecell{\small 0.46 (0.44--0.48) \\ \scriptsize 4.0\% \quad \scriptsize $-$0.120\,\textcolor{lexcol}{\textbf{L}}} & \makecell{\small 0.50 (0.49--0.52) \\ \scriptsize 5.4\% \quad \scriptsize $-$0.107\,\textcolor{lexcol}{\textbf{L}}} & \makecell{\small 0.48 (0.47--0.50) \\ \scriptsize 5.0\% \quad \scriptsize $-$0.100\,\textcolor{embcol}{\textbf{E}}} \\
\bottomrule
\end{tabular}
\begin{tablenotes}[flushleft]
\footnotesize
\item Cell format, protocol and $\Delta$ as in Table~\ref{tab:meta-outlier}.
\item Across the 60 populated cells the fusion wins in 0 and loses in 60, mean $\Delta=-0.084$.
\end{tablenotes}
\end{threeparttable}
\end{table*}

\begin{table*}[t]
\centering
\small
\setlength{\tabcolsep}{3pt}
\renewcommand{\arraystretch}{1.05}
\begin{threeparttable}
\caption{Meta-classifier (fusion of all 32 proxies) on the outlier group, \textsc{Aug-PE} releases: AUC, TPR and the gain over the best single proxy. Companion to Table~\ref{tab:meta-outlier}.}
\label{tab:meta-outlier-augpe}
\begin{tabular}{lllccccc}
\toprule
\textbf{Generator} & \textbf{Dataset} & \textbf{Scenario} & $\bm{\varepsilon=\infty}$ & $\bm{4}$ & $\bm{2}$ & $\bm{1}$ & $\bm{0.5}$ \\
\midrule
\multirow{12}{*}{Aug-PE}
& \multirow{3}{*}{\textsc{PsyTAR}} & S1 & \makecell{\small 0.54 (0.52--0.55) \\ \scriptsize 8.3\% \quad \scriptsize $-$0.054\,\textcolor{lexcol}{\textbf{L}}} & \makecell{\small 0.53 (0.52--0.55) \\ \scriptsize 7.5\% \quad \scriptsize $-$0.056\,\textcolor{lexcol}{\textbf{L}}} & \makecell{\small 0.55 (0.53--0.56) \\ \scriptsize 7.4\% \quad \scriptsize $-$0.046\,\textcolor{lexcol}{\textbf{L}}} & \makecell{\small 0.56 (0.54--0.57) \\ \scriptsize 9.3\% \quad \scriptsize $-$0.038\,\textcolor{lexcol}{\textbf{L}}} & \makecell{\small 0.53 (0.51--0.55) \\ \scriptsize 6.6\% \quad \scriptsize $-$0.076\,\textcolor{lexcol}{\textbf{L}}} \\
&                                  & S2 & \makecell{\small 0.56 (0.54--0.57) \\ \scriptsize 7.7\% \quad \scriptsize $-$0.079\,\textcolor{lexcol}{\textbf{L}}} & \makecell{\small 0.55 (0.53--0.57) \\ \scriptsize 8.3\% \quad \scriptsize $-$0.076\,\textcolor{lexcol}{\textbf{L}}} & \makecell{\small 0.57 (0.55--0.59) \\ \scriptsize 9.7\% \quad \scriptsize $-$0.073\,\textcolor{lexcol}{\textbf{L}}} & \makecell{\small 0.55 (0.53--0.57) \\ \scriptsize 8.6\% \quad \scriptsize $-$0.082\,\textcolor{embcol}{\textbf{E}}} & \makecell{\small 0.56 (0.54--0.58) \\ \scriptsize 7.9\% \quad \scriptsize $-$0.074\,\textcolor{lexcol}{\textbf{L}}} \\
&                                  & S3 & \makecell{\small 0.52 (0.51--0.54) \\ \scriptsize 7.9\% \quad \scriptsize $-$0.035\,\textcolor{lexcol}{\textbf{L}}} & \makecell{\small 0.53 (0.52--0.54) \\ \scriptsize 6.3\% \quad \scriptsize $-$0.028\,\textcolor{embcol}{\textbf{E}}} & \makecell{\small 0.56 (0.54--0.57) \\ \scriptsize 9.2\% \quad \scriptsize $-$0.056\,\textcolor{lexcol}{\textbf{L}}} & \makecell{\small 0.52 (0.50--0.53) \\ \scriptsize 6.4\% \quad \scriptsize $-$0.069\,\textcolor{lexcol}{\textbf{L}}} & \makecell{\small 0.53 (0.51--0.55) \\ \scriptsize 6.9\% \quad \scriptsize $-$0.055\,\textcolor{lexcol}{\textbf{L}}} \\
\cmidrule(lr){2-8}
& \multirow{3}{*}{\textsc{DMSAFN}} & S1 & \makecell{\small 0.56 (0.54--0.57) \\ \scriptsize 6.4\% \quad \scriptsize $-$0.055\,\textcolor{lexcol}{\textbf{L}}} & \makecell{\small 0.61 (0.59--0.62) \\ \scriptsize 8.1\% \quad \scriptsize $-$0.002\,\textcolor{lexcol}{\textbf{L}}} & \makecell{\small 0.56 (0.54--0.58) \\ \scriptsize 6.7\% \quad \scriptsize $-$0.050\,\textcolor{lexcol}{\textbf{L}}} & \makecell{\small 0.56 (0.54--0.57) \\ \scriptsize 6.2\% \quad \scriptsize $-$0.057\,\textcolor{lexcol}{\textbf{L}}} & \makecell{\small 0.58 (0.57--0.60) \\ \scriptsize 6.2\% \quad \scriptsize $-$0.026\,\textcolor{lexcol}{\textbf{L}}} \\
&                                  & S2 & \makecell{\small 0.51 (0.49--0.53) \\ \scriptsize 4.8\% \quad \scriptsize $-$0.089\,\textcolor{lexcol}{\textbf{L}}} & \makecell{\small 0.58 (0.57--0.60) \\ \scriptsize 8.7\% \quad \scriptsize $-$0.002\,\textcolor{lexcol}{\textbf{L}}} & \makecell{\small 0.53 (0.52--0.55) \\ \scriptsize 6.0\% \quad \scriptsize $-$0.060\,\textcolor{lexcol}{\textbf{L}}} & \makecell{\small 0.52 (0.50--0.53) \\ \scriptsize 5.9\% \quad \scriptsize $-$0.083\,\textcolor{lexcol}{\textbf{L}}} & \makecell{\small 0.53 (0.52--0.55) \\ \scriptsize 5.5\% \quad \scriptsize $-$0.059\,\textcolor{lexcol}{\textbf{L}}} \\
&                                  & S3 & \makecell{\small 0.49 (0.47--0.51) \\ \scriptsize 6.5\% \quad \scriptsize $-$0.093\,\textcolor{embcol}{\textbf{E}}} & \cellcolor{best!18}\makecell{\small 0.64 (0.63--0.66) \\ \scriptsize 14.5\% \quad \scriptsize \textbf{$+$0.048}\,\textcolor{lexcol}{\textbf{L}}} & \makecell{\small 0.48 (0.46--0.50) \\ \scriptsize 5.2\% \quad \scriptsize $-$0.141\,\textcolor{embcol}{\textbf{E}}} & \makecell{\small 0.46 (0.45--0.48) \\ \scriptsize 5.0\% \quad \scriptsize $-$0.113\,\textcolor{lexcol}{\textbf{L}}} & \makecell{\small 0.46 (0.45--0.48) \\ \scriptsize 3.6\% \quad \scriptsize $-$0.083\,\textcolor{embcol}{\textbf{E}}} \\
\cmidrule(lr){2-8}
& \multirow{3}{*}{\textsc{N2C2'08}} & S1 & \makecell{\small 0.46 (0.44--0.49) \\ \scriptsize 3.3\% \quad \scriptsize $-$0.134\,\textcolor{embcol}{\textbf{E}}} & \makecell{\small 0.46 (0.44--0.48) \\ \scriptsize 2.6\% \quad \scriptsize $-$0.130\,\textcolor{lexcol}{\textbf{L}}} & \makecell{\small 0.45 (0.44--0.47) \\ \scriptsize 4.5\% \quad \scriptsize $-$0.106\,\textcolor{lexcol}{\textbf{L}}} & \makecell{\small 0.45 (0.43--0.47) \\ \scriptsize 3.7\% \quad \scriptsize $-$0.126\,\textcolor{lexcol}{\textbf{L}}} & \makecell{\small 0.43 (0.42--0.45) \\ \scriptsize 4.6\% \quad \scriptsize $-$0.138\,\textcolor{lexcol}{\textbf{L}}} \\
&                                  & S2 & \makecell{\small 0.53 (0.52--0.55) \\ \scriptsize 4.5\% \quad \scriptsize $-$0.106\,\textcolor{embcol}{\textbf{E}}} & \makecell{\small 0.50 (0.48--0.52) \\ \scriptsize 4.2\% \quad \scriptsize $-$0.089\,\textcolor{lexcol}{\textbf{L}}} & \makecell{\small 0.51 (0.49--0.52) \\ \scriptsize 7.0\% \quad \scriptsize $-$0.109\,\textcolor{embcol}{\textbf{E}}} & \makecell{\small 0.50 (0.48--0.52) \\ \scriptsize 5.9\% \quad \scriptsize $-$0.098\,\textcolor{embcol}{\textbf{E}}} & \makecell{\small 0.50 (0.48--0.52) \\ \scriptsize 6.2\% \quad \scriptsize $-$0.125\,\textcolor{embcol}{\textbf{E}}} \\
&                                  & S3 & \makecell{\small 0.51 (0.49--0.53) \\ \scriptsize 5.0\% \quad \scriptsize $-$0.122\,\textcolor{lexcol}{\textbf{L}}} & \makecell{\small 0.46 (0.44--0.47) \\ \scriptsize 2.9\% \quad \scriptsize $-$0.129\,\textcolor{lexcol}{\textbf{L}}} & \makecell{\small 0.51 (0.50--0.53) \\ \scriptsize 6.0\% \quad \scriptsize $-$0.085\,\textcolor{lexcol}{\textbf{L}}} & \makecell{\small 0.51 (0.50--0.53) \\ \scriptsize 3.8\% \quad \scriptsize $-$0.109\,\textcolor{embcol}{\textbf{E}}} & \makecell{\small 0.50 (0.48--0.51) \\ \scriptsize 7.3\% \quad \scriptsize $-$0.105\,\textcolor{lexcol}{\textbf{L}}} \\
\cmidrule(lr){2-8}
& \multirow{3}{*}{\textsc{EurLex}} & S1 & \makecell{\small 0.45 (0.43--0.46) \\ \scriptsize 3.8\% \quad \scriptsize $-$0.141\,\textcolor{lexcol}{\textbf{L}}} & \makecell{\small 0.43 (0.41--0.45) \\ \scriptsize 3.7\% \quad \scriptsize $-$0.136\,\textcolor{lexcol}{\textbf{L}}} & \makecell{\small 0.44 (0.42--0.46) \\ \scriptsize 3.1\% \quad \scriptsize $-$0.147\,\textcolor{lexcol}{\textbf{L}}} & \makecell{\small 0.46 (0.44--0.48) \\ \scriptsize 4.6\% \quad \scriptsize $-$0.122\,\textcolor{lexcol}{\textbf{L}}} & \cellcolor{best!18}\makecell{\small 0.58 (0.55--0.60) \\ \scriptsize 8.7\% \quad \scriptsize \textbf{$+$0.006}\,\textcolor{lexcol}{\textbf{L}}} \\
&                                  & S2 & \makecell{\small 0.47 (0.45--0.49) \\ \scriptsize 4.8\% \quad \scriptsize $-$0.123\,\textcolor{lexcol}{\textbf{L}}} & \makecell{\small 0.44 (0.42--0.46) \\ \scriptsize 4.5\% \quad \scriptsize $-$0.115\,\textcolor{embcol}{\textbf{E}}} & \makecell{\small 0.47 (0.45--0.49) \\ \scriptsize 4.5\% \quad \scriptsize $-$0.118\,\textcolor{embcol}{\textbf{E}}} & \makecell{\small 0.43 (0.41--0.45) \\ \scriptsize 4.2\% \quad \scriptsize $-$0.129\,\textcolor{lexcol}{\textbf{L}}} & \makecell{\small 0.47 (0.45--0.49) \\ \scriptsize 4.2\% \quad \scriptsize $-$0.111\,\textcolor{embcol}{\textbf{E}}} \\
&                                  & S3 & \makecell{\small 0.49 (0.47--0.51) \\ \scriptsize 4.7\% \quad \scriptsize $-$0.101\,\textcolor{lexcol}{\textbf{L}}} & \makecell{\small 0.45 (0.43--0.47) \\ \scriptsize 4.2\% \quad \scriptsize $-$0.107\,\textcolor{embcol}{\textbf{E}}} & \makecell{\small 0.50 (0.48--0.52) \\ \scriptsize 6.1\% \quad \scriptsize $-$0.060\,\textcolor{embcol}{\textbf{E}}} & \makecell{\small 0.42 (0.41--0.44) \\ \scriptsize 3.2\% \quad \scriptsize $-$0.159\,\textcolor{lexcol}{\textbf{L}}} & \makecell{\small 0.55 (0.53--0.56) \\ \scriptsize 6.2\% \quad \scriptsize $-$0.059\,\textcolor{lexcol}{\textbf{L}}} \\
\bottomrule
\end{tabular}
\begin{tablenotes}[flushleft]
\footnotesize
\item Cell format, protocol and $\Delta$ as in Table~\ref{tab:meta-outlier}.
\item Across the 60 populated cells the fusion wins in 2 and loses in 58, mean $\Delta=-0.085$.
\end{tablenotes}
\end{threeparttable}
\end{table*}

%% file: tables/table_meta_rare.tex
\definecolor{lexcol}{HTML}{1F5FA9}   
\definecolor{embcol}{HTML}{A8431F}   
\definecolor{best}{HTML}{2A78D6}     
\begin{table*}[t]
\centering
\small
\setlength{\tabcolsep}{3pt}
\renewcommand{\arraystretch}{1.05}
\begin{threeparttable}
\caption{Meta-classifier (fusion of all 32 proxies) on the rare-label group: AUC, TPR and the gain over the best single proxy.}
\label{tab:meta-rare}
\begin{tabular}{lllccc}
\toprule
\textbf{Generator} & \textbf{Dataset} & \textbf{Scenario} & $\bm{\varepsilon=\infty}$ & $\bm{4}$ & $\bm{1}$ \\
\midrule
\multirow{9}{*}{DP-Gen}
& \multirow{3}{*}{\textsc{PsyTAR}} & S1 & \makecell{\small 0.78 (0.77--0.79) \\ \scriptsize 45.7\% \quad \scriptsize $-$0.036\,\textcolor{lexcol}{\textbf{L}}} & \makecell{\small 0.48 (0.46--0.50) \\ \scriptsize 5.6\% \quad \scriptsize $-$0.112\,\textcolor{embcol}{\textbf{E}}} & \makecell{\small 0.48 (0.47--0.50) \\ \scriptsize 5.1\% \quad \scriptsize $-$0.089\,\textcolor{embcol}{\textbf{E}}} \\
&                                  & S2 & \makecell{\small 0.79 (0.78--0.80) \\ \scriptsize 47.4\% \quad \scriptsize $-$0.050\,\textcolor{lexcol}{\textbf{L}}} & \makecell{\small 0.42 (0.40--0.44) \\ \scriptsize 5.1\% \quad \scriptsize $-$0.110\,\textcolor{embcol}{\textbf{E}}} & \makecell{\small 0.46 (0.44--0.48) \\ \scriptsize 4.5\% \quad \scriptsize $-$0.094\,\textcolor{embcol}{\textbf{E}}} \\
&                                  & S3 & \makecell{\small 0.83 (0.82--0.84) \\ \scriptsize 52.6\% \quad \scriptsize $-$0.026\,\textcolor{lexcol}{\textbf{L}}} & \makecell{\small 0.45 (0.43--0.47) \\ \scriptsize 4.4\% \quad \scriptsize $-$0.115\,\textcolor{lexcol}{\textbf{L}}} & \makecell{\small 0.49 (0.47--0.51) \\ \scriptsize 5.7\% \quad \scriptsize $-$0.097\,\textcolor{lexcol}{\textbf{L}}} \\
\cmidrule(lr){2-6}
& \multirow{3}{*}{\textsc{N2C2'08}} & S1 & \makecell{\small 0.56 (0.55--0.58) \\ \scriptsize 7.1\% \quad \scriptsize $-$0.059\,\textcolor{lexcol}{\textbf{L}}} & \makecell{\small 0.46 (0.44--0.48) \\ \scriptsize 2.9\% \quad \scriptsize $-$0.068\,\textcolor{lexcol}{\textbf{L}}} & \makecell{\small 0.43 (0.41--0.45) \\ \scriptsize 3.5\% \quad \scriptsize $-$0.099\,\textcolor{embcol}{\textbf{E}}} \\
&                                  & S2 & \cellcolor{best!18}\makecell{\small 0.69 (0.68--0.71) \\ \scriptsize 11.4\% \quad \scriptsize \textbf{$+$0.005}\,\textcolor{lexcol}{\textbf{L}}} & \makecell{\small 0.47 (0.46--0.49) \\ \scriptsize 3.0\% \quad \scriptsize $-$0.072\,\textcolor{embcol}{\textbf{E}}} & \makecell{\small 0.50 (0.49--0.52) \\ \scriptsize 3.6\% \quad \scriptsize $-$0.059\,\textcolor{embcol}{\textbf{E}}} \\
&                                  & S3 & \makecell{\small 0.72 (0.71--0.73) \\ \scriptsize 18.5\% \quad \scriptsize $-$0.051\,\textcolor{lexcol}{\textbf{L}}} & \makecell{\small 0.44 (0.42--0.46) \\ \scriptsize 2.7\% \quad \scriptsize $-$0.125\,\textcolor{lexcol}{\textbf{L}}} & \makecell{\small 0.50 (0.49--0.51) \\ \scriptsize 6.1\% \quad \scriptsize $-$0.085\,\textcolor{embcol}{\textbf{E}}} \\
\cmidrule(lr){2-6}
& \multirow{3}{*}{\textsc{EurLex}} & S1 & \makecell{\small 0.55 (0.53--0.56) \\ \scriptsize 8.8\% \quad \scriptsize $-$0.078\,\textcolor{embcol}{\textbf{E}}} & \makecell{\small 0.48 (0.46--0.51) \\ \scriptsize 4.3\% \quad \scriptsize $-$0.090\,\textcolor{embcol}{\textbf{E}}} & \makecell{\small 0.47 (0.45--0.49) \\ \scriptsize 4.0\% \quad \scriptsize $-$0.096\,\textcolor{embcol}{\textbf{E}}} \\
&                                  & S2 & \makecell{\small 0.53 (0.51--0.54) \\ \scriptsize 6.5\% \quad \scriptsize $-$0.080\,\textcolor{embcol}{\textbf{E}}} & \makecell{\small 0.47 (0.46--0.49) \\ \scriptsize 4.8\% \quad \scriptsize $-$0.091\,\textcolor{embcol}{\textbf{E}}} & \makecell{\small 0.45 (0.43--0.47) \\ \scriptsize 4.8\% \quad \scriptsize $-$0.118\,\textcolor{lexcol}{\textbf{L}}} \\
&                                  & S3 & \makecell{\small 0.54 (0.53--0.56) \\ \scriptsize 7.1\% \quad \scriptsize $-$0.094\,\textcolor{lexcol}{\textbf{L}}} & \makecell{\small 0.43 (0.42--0.45) \\ \scriptsize 4.0\% \quad \scriptsize $-$0.113\,\textcolor{embcol}{\textbf{E}}} & \makecell{\small 0.42 (0.41--0.44) \\ \scriptsize 3.6\% \quad \scriptsize $-$0.130\,\textcolor{embcol}{\textbf{E}}} \\
\midrule
\multirow{9}{*}{EPSVec}
& \multirow{3}{*}{\textsc{PsyTAR}} & S1 & \makecell{\small 0.46 (0.44--0.48) \\ \scriptsize 3.9\% \quad \scriptsize $-$0.113\,\textcolor{embcol}{\textbf{E}}} & \makecell{\small 0.50 (0.48--0.52) \\ \scriptsize 4.6\% \quad \scriptsize $-$0.096\,\textcolor{embcol}{\textbf{E}}} & \makecell{\small 0.48 (0.46--0.50) \\ \scriptsize 4.7\% \quad \scriptsize $-$0.097\,\textcolor{embcol}{\textbf{E}}} \\
&                                  & S2 & \makecell{\small 0.44 (0.42--0.46) \\ \scriptsize 3.6\% \quad \scriptsize $-$0.102\,\textcolor{embcol}{\textbf{E}}} & \makecell{\small 0.48 (0.45--0.50) \\ \scriptsize 5.8\% \quad \scriptsize $-$0.084\,\textcolor{embcol}{\textbf{E}}} & \makecell{\small 0.52 (0.49--0.54) \\ \scriptsize 6.4\% \quad \scriptsize $-$0.054\,\textcolor{embcol}{\textbf{E}}} \\
&                                  & S3 & \makecell{\small 0.54 (0.52--0.56) \\ \scriptsize 5.2\% \quad \scriptsize $-$0.056\,\textcolor{embcol}{\textbf{E}}} & \makecell{\small 0.50 (0.49--0.52) \\ \scriptsize 7.4\% \quad \scriptsize $-$0.079\,\textcolor{embcol}{\textbf{E}}} & \makecell{\small 0.48 (0.46--0.50) \\ \scriptsize 5.5\% \quad \scriptsize $-$0.092\,\textcolor{embcol}{\textbf{E}}} \\
\cmidrule(lr){2-6}
& \multirow{3}{*}{\textsc{N2C2'08}} & S1 & \makecell{\small 0.45 (0.44--0.47) \\ \scriptsize 3.4\% \quad \scriptsize $-$0.065\,\textcolor{embcol}{\textbf{E}}} & \makecell{\small 0.54 (0.52--0.55) \\ \scriptsize 4.6\% \quad \scriptsize $-$0.027\,\textcolor{embcol}{\textbf{E}}} & \makecell{\small 0.47 (0.45--0.49) \\ \scriptsize 3.4\% \quad \scriptsize $-$0.088\,\textcolor{embcol}{\textbf{E}}} \\
&                                  & S2 & \makecell{\small 0.49 (0.47--0.50) \\ \scriptsize 4.9\% \quad \scriptsize $-$0.061\,\textcolor{embcol}{\textbf{E}}} & \makecell{\small 0.54 (0.52--0.56) \\ \scriptsize 5.4\% \quad \scriptsize $-$0.024\,\textcolor{lexcol}{\textbf{L}}} & \makecell{\small 0.51 (0.50--0.53) \\ \scriptsize 3.7\% \quad \scriptsize $-$0.036\,\textcolor{embcol}{\textbf{E}}} \\
&                                  & S3 & \makecell{\small 0.45 (0.43--0.47) \\ \scriptsize 3.7\% \quad \scriptsize $-$0.133\,\textcolor{embcol}{\textbf{E}}} & \makecell{\small 0.57 (0.55--0.58) \\ \scriptsize 6.0\% \quad \scriptsize $-$0.096\,\textcolor{lexcol}{\textbf{L}}} & \makecell{\small 0.51 (0.50--0.52) \\ \scriptsize 5.1\% \quad \scriptsize $-$0.038\,\textcolor{embcol}{\textbf{E}}} \\
\cmidrule(lr){2-6}
& \multirow{3}{*}{\textsc{EurLex}} & S1 & \makecell{\small 0.51 (0.49--0.54) \\ \scriptsize 6.3\% \quad \scriptsize $-$0.106\,\textcolor{embcol}{\textbf{E}}} & \makecell{\small 0.55 (0.53--0.57) \\ \scriptsize 6.3\% \quad \scriptsize $-$0.071\,\textcolor{embcol}{\textbf{E}}} & \makecell{\small 0.52 (0.49--0.54) \\ \scriptsize 4.4\% \quad \scriptsize $-$0.099\,\textcolor{embcol}{\textbf{E}}} \\
&                                  & S2 & \makecell{\small 0.51 (0.48--0.53) \\ \scriptsize 5.1\% \quad \scriptsize $-$0.077\,\textcolor{embcol}{\textbf{E}}} & \makecell{\small 0.53 (0.51--0.55) \\ \scriptsize 5.2\% \quad \scriptsize $-$0.088\,\textcolor{embcol}{\textbf{E}}} & \makecell{\small 0.45 (0.43--0.47) \\ \scriptsize 4.0\% \quad \scriptsize $-$0.131\,\textcolor{embcol}{\textbf{E}}} \\
&                                  & S3 & \makecell{\small 0.49 (0.47--0.51) \\ \scriptsize 4.5\% \quad \scriptsize $-$0.101\,\textcolor{lexcol}{\textbf{L}}} & \makecell{\small 0.46 (0.44--0.48) \\ \scriptsize 3.5\% \quad \scriptsize $-$0.103\,\textcolor{lexcol}{\textbf{L}}} & \makecell{\small 0.51 (0.49--0.54) \\ \scriptsize 6.4\% \quad \scriptsize $-$0.060\,\textcolor{lexcol}{\textbf{L}}} \\
\bottomrule
\end{tabular}
\begin{tablenotes}[flushleft]
\footnotesize
\item \textbf{Top line:} AUC-ROC of the meta-classifier (mean $\pm$ 95\,\% CI, normal approximation over 50 evaluation rounds; CI $=$ mean $\pm$ 1.96\,$\sigma/\sqrt{50}$). Every round draws the negative candidates from the same evaluation trial plan the main tables use, so a cell here and the corresponding main-table cell see identical data. Within a round, membership probabilities come from nested cross-validation (outer 5-fold out-of-fold prediction; inner 5-fold grid search over the number of selected features $k\in\{5,8,12,\text{all}\}$ and the $\ell_2$ penalty $C\in\{0.01,0.1,1,10\}$ of a class-balanced logistic regression on standardised, mean-imputed features), so no record contributes to the model that scores it.
\item \textbf{Bottom line:} TPR at FPR $\leq0.05$, then $\Delta$ = meta-classifier AUC $-$ AUC of the best of the 32 individual proxies, recomputed on the same rounds. That reference is an \emph{oracle} --- the proxy is chosen after seeing the evaluation data, which is what the main table reports --- so $\Delta>0$ (shaded) means the fusion beats a single proxy no adversary could have picked in advance. The superscript gives that proxy's view (\textcolor{lexcol}{\textbf{L}}exical / \textcolor{embcol}{\textbf{E}}mbedding); its name, the modal $(k,C)$ and the number of scored records per round are in the companion CSV.
\item Across the 54 populated cells the fusion wins in 1 and loses in 53, mean $\Delta=-0.082$.
\end{tablenotes}
\end{threeparttable}
\end{table*}

%% file: tables/table_baseline_outlier_appendix.tex
\definecolor{lexcol}{HTML}{1F5FA9}
\definecolor{embcol}{HTML}{A8431F}
\providecommand{\tbd}{\textcolor{gray}{--}}
\begin{sidewaystable*}[p]
\centering
\caption{Subgroup game: remaining privacy budgets; format, variant codes, shading and notes as in Table~\ref{tab:baselines-main}.}
\label{tab:baselines-outlier-rest}
\resizebox{\textwidth}{!}{%
\begin{tabular}{llcccccccccccc}
\toprule
 & & \multicolumn{4}{c}{$\varepsilon=4$} & \multicolumn{4}{c}{$\varepsilon=2$} & \multicolumn{4}{c}{$\varepsilon=0.5$} \\
\cmidrule(lr){3-6}\cmidrule(lr){7-10}\cmidrule(lr){11-14}
Dataset & Generator & \textsc{canary} & \textsc{domias} & \textsc{ours} & $\Delta$ & \textsc{canary} & \textsc{domias} & \textsc{ours} & $\Delta$ & \textsc{canary} & \textsc{domias} & \textsc{ours} & $\Delta$ \\
\midrule
PsyTAR & DP-Gen & $\underline{0.588}$\textsuperscript{\textcolor{lexcol}{n}}\,/\,14\% & $\mathbf{0.604}$\textsuperscript{\textcolor{embcol}{g}}\,/\,14\% & $0.587$\textsuperscript{\textcolor{lexcol}{\textbf{L}}}\,/\,10\% & \textcolor{embcol}{$-0.017$} & $\mathbf{0.589}$\textsuperscript{\textcolor{lexcol}{n}}\,/\,13\% & $0.571$\textsuperscript{\textcolor{embcol}{k}}\,/\,7\% & $\underline{0.586}$\textsuperscript{\textcolor{lexcol}{\textbf{L}}}\,/\,11\% & \textcolor{embcol}{$-0.003$} & $\underline{0.614}$\textsuperscript{\textcolor{lexcol}{nR}}\,/\,6\% & $0.570$\textsuperscript{\textcolor{embcol}{k}}\,/\,4\% & $\mathbf{0.637}$\textsuperscript{\textcolor{lexcol}{\textbf{L}}}\,/\,6\% & $+0.023$ \\
PsyTAR & Aug-PE & $\underline{0.588}$\textsuperscript{\textcolor{lexcol}{n}}\,/\,14\% & $0.523$\textsuperscript{\textcolor{embcol}{k}}\,/\,8\% & $\mathbf{0.624}$\textsuperscript{\textcolor{lexcol}{\textbf{L}}}\,/\,12\% & $+0.036$ & $\underline{0.589}$\textsuperscript{\textcolor{lexcol}{n}}\,/\,10\% & $0.518$\textsuperscript{\textcolor{embcol}{k}}\,/\,6\% & $\mathbf{0.646}$\textsuperscript{\textcolor{lexcol}{\textbf{L}}}\,/\,14\% & $+0.057$ & $\underline{0.593}$\textsuperscript{\textcolor{lexcol}{n}}\,/\,15\% & $0.558$\textsuperscript{\textcolor{embcol}{k}}\,/\,8\% & $\mathbf{0.631}$\textsuperscript{\textcolor{lexcol}{\textbf{L}}}\,/\,12\% & $+0.037$ \\
PsyTAR & EPSVec & $\underline{0.589}$\textsuperscript{\textcolor{lexcol}{n}}\,/\,10\% & $0.527$\textsuperscript{\textcolor{embcol}{k}}\,/\,5\% & $\mathbf{0.645}$\textsuperscript{\textcolor{lexcol}{\textbf{L}}}\,/\,15\% & $+0.056$ & $\mathbf{0.597}$\textsuperscript{\textcolor{lexcol}{n}}\,/\,15\% & $0.553$\textsuperscript{\textcolor{embcol}{k}}\,/\,7\% & $\underline{0.593}$\textsuperscript{\textcolor{lexcol}{\textbf{L}}}\,/\,14\% & \textcolor{embcol}{$-0.004$} & $\underline{0.590}$\textsuperscript{\textcolor{lexcol}{n}}\,/\,11\% & $0.501$\textsuperscript{\textcolor{embcol}{k}}\,/\,9\% & $\mathbf{0.591}$\textsuperscript{\textcolor{lexcol}{\textbf{L}}}\,/\,10\% & $+0.001$ \\
\midrule
DMSAFN & DP-Gen & $\underline{0.594}$\textsuperscript{\textcolor{lexcol}{n}}\,/\,2\% & $0.549$\textsuperscript{\textcolor{embcol}{pg}}\,/\,12\% & $\mathbf{0.604}$\textsuperscript{\textcolor{lexcol}{\textbf{L}}}\,/\,4\% & $+0.010$ & $\underline{0.599}$\textsuperscript{\textcolor{lexcol}{n}}\,/\,3\% & $0.547$\textsuperscript{\textcolor{embcol}{pg}}\,/\,5\% & $\mathbf{0.610}$\textsuperscript{\textcolor{lexcol}{\textbf{L}}}\,/\,5\% & $+0.011$ & $\underline{0.632}$\textsuperscript{\textcolor{lexcol}{nR}}\,/\,11\% & $0.487$\textsuperscript{\textcolor{embcol}{pk}}\,/\,6\% & $\mathbf{0.647}$\textsuperscript{\textcolor{lexcol}{\textbf{L}}}\,/\,16\% & $+0.014$ \\
DMSAFN & Aug-PE & $\underline{0.604}$\textsuperscript{\textcolor{lexcol}{n}}\,/\,4\% & $0.537$\textsuperscript{\textcolor{embcol}{k}}\,/\,4\% & $\mathbf{0.608}$\textsuperscript{\textcolor{lexcol}{\textbf{L}}}\,/\,5\% & $+0.004$ & $\underline{0.599}$\textsuperscript{\textcolor{lexcol}{n}}\,/\,3\% & $0.454$\textsuperscript{\textcolor{embcol}{g}}\,/\,4\% & $\mathbf{0.621}$\textsuperscript{\textcolor{embcol}{\textbf{E}}}\,/\,9\% & $+0.022$ & $\underline{0.600}$\textsuperscript{\textcolor{lexcol}{n}}\,/\,3\% & $0.537$\textsuperscript{\textcolor{embcol}{pk}}\,/\,5\% & $\mathbf{0.609}$\textsuperscript{\textcolor{lexcol}{\textbf{L}}}\,/\,4\% & $+0.009$ \\
DMSAFN & EPSVec & $\underline{0.599}$\textsuperscript{\textcolor{lexcol}{n}}\,/\,3\% & $0.535$\textsuperscript{\textcolor{embcol}{k}}\,/\,4\% & $\mathbf{0.634}$\textsuperscript{\textcolor{lexcol}{\textbf{L}}}\,/\,10\% & $+0.035$ & $\underline{0.600}$\textsuperscript{\textcolor{lexcol}{n}}\,/\,3\% & $0.499$\textsuperscript{\textcolor{embcol}{k}}\,/\,6\% & $\mathbf{0.608}$\textsuperscript{\textcolor{lexcol}{\textbf{L}}}\,/\,5\% & $+0.007$ & $\underline{0.600}$\textsuperscript{\textcolor{lexcol}{n}}\,/\,3\% & $0.491$\textsuperscript{\textcolor{embcol}{pk}}\,/\,7\% & $\mathbf{0.621}$\textsuperscript{\textcolor{lexcol}{\textbf{L}}}\,/\,15\% & $+0.021$ \\
\midrule
N2C2'08 & DP-Gen & $0.523$\textsuperscript{\textcolor{embcol}{s}}\,/\,6\% & $\underline{0.580}$\textsuperscript{\textcolor{embcol}{pk}}\,/\,8\% & $\mathbf{0.590}$\textsuperscript{\textcolor{lexcol}{\textbf{L}}}\,/\,4\% & $+0.010$ & $0.518$\textsuperscript{\textcolor{embcol}{s}}\,/\,9\% & $\underline{0.568}$\textsuperscript{\textcolor{embcol}{pk}}\,/\,7\% & $\mathbf{0.625}$\textsuperscript{\textcolor{embcol}{\textbf{E}}}\,/\,13\% & $+0.057$ & $0.566$\textsuperscript{\textcolor{embcol}{sR}}\,/\,11\% & $\underline{0.615}$\textsuperscript{\textcolor{embcol}{pk}}\,/\,7\% & $\mathbf{0.629}$\textsuperscript{\textcolor{embcol}{\textbf{E}}}\,/\,9\% & $+0.014$ \\
N2C2'08 & Aug-PE & $0.496$\textsuperscript{\textcolor{lexcol}{nR}}\,/\,8\% & $\underline{0.555}$\textsuperscript{\textcolor{embcol}{pk}}\,/\,9\% & $\mathbf{0.593}$\textsuperscript{\textcolor{lexcol}{\textbf{L}}}\,/\,15\% & $+0.038$ & $0.525$\textsuperscript{\textcolor{lexcol}{nR}}\,/\,6\% & $\underline{0.610}$\textsuperscript{\textcolor{embcol}{pk}}\,/\,9\% & $\mathbf{0.617}$\textsuperscript{\textcolor{embcol}{\textbf{E}}}\,/\,13\% & $+0.007$ & $0.530$\textsuperscript{\textcolor{embcol}{s}}\,/\,2\% & $\underline{0.555}$\textsuperscript{\textcolor{embcol}{k}}\,/\,8\% & $\mathbf{0.622}$\textsuperscript{\textcolor{embcol}{\textbf{E}}}\,/\,10\% & $+0.067$ \\
N2C2'08 & EPSVec & $0.477$\textsuperscript{\textcolor{embcol}{s}}\,/\,5\% & $\underline{0.559}$\textsuperscript{\textcolor{embcol}{pk}}\,/\,6\% & $\mathbf{0.595}$\textsuperscript{\textcolor{embcol}{\textbf{E}}}\,/\,6\% & $+0.036$ & $0.562$\textsuperscript{\textcolor{embcol}{sR}}\,/\,12\% & $\underline{0.595}$\textsuperscript{\textcolor{embcol}{k}}\,/\,12\% & $\mathbf{0.617}$\textsuperscript{\textcolor{embcol}{\textbf{E}}}\,/\,10\% & $+0.022$ & $0.473$\textsuperscript{\textcolor{embcol}{sR}}\,/\,5\% & $\underline{0.525}$\textsuperscript{\textcolor{embcol}{pk}}\,/\,3\% & $\mathbf{0.629}$\textsuperscript{\textcolor{lexcol}{\textbf{L}}}\,/\,9\% & $+0.104$ \\
\midrule
EUR-Lex & DP-Gen & $0.517$\textsuperscript{\textcolor{lexcol}{n}}\,/\,7\% & $\underline{0.565}$\textsuperscript{\textcolor{embcol}{g}}\,/\,9\% & $\mathbf{0.589}$\textsuperscript{\textcolor{embcol}{\textbf{E}}}\,/\,12\% & $+0.024$ & $0.520$\textsuperscript{\textcolor{lexcol}{n}}\,/\,6\% & $\underline{0.545}$\textsuperscript{\textcolor{embcol}{g}}\,/\,6\% & $\mathbf{0.580}$\textsuperscript{\textcolor{embcol}{\textbf{E}}}\,/\,4\% & $+0.035$ & $0.519$\textsuperscript{\textcolor{lexcol}{n}}\,/\,7\% & $\underline{0.537}$\textsuperscript{\textcolor{embcol}{g}}\,/\,5\% & $\mathbf{0.575}$\textsuperscript{\textcolor{lexcol}{\textbf{L}}}\,/\,6\% & $+0.038$ \\
EUR-Lex & Aug-PE & $\underline{0.543}$\textsuperscript{\textcolor{lexcol}{nR}}\,/\,4\% & $0.542$\textsuperscript{\textcolor{embcol}{g}}\,/\,8\% & $\mathbf{0.568}$\textsuperscript{\textcolor{lexcol}{\textbf{L}}}\,/\,5\% & $+0.025$ & $0.513$\textsuperscript{\textcolor{lexcol}{n}}\,/\,6\% & $\underline{0.542}$\textsuperscript{\textcolor{embcol}{k}}\,/\,7\% & $\mathbf{0.589}$\textsuperscript{\textcolor{lexcol}{\textbf{L}}}\,/\,7\% & $+0.048$ & $0.553$\textsuperscript{\textcolor{lexcol}{nR}}\,/\,3\% & $\underline{0.604}$\textsuperscript{\textcolor{embcol}{k}}\,/\,13\% & $\mathbf{0.605}$\textsuperscript{\textcolor{lexcol}{\textbf{L}}}\,/\,11\% & $+0.001$ \\
EUR-Lex & EPSVec & $0.517$\textsuperscript{\textcolor{lexcol}{n}}\,/\,7\% & $\underline{0.559}$\textsuperscript{\textcolor{embcol}{g}}\,/\,8\% & $\mathbf{0.589}$\textsuperscript{\textcolor{lexcol}{\textbf{L}}}\,/\,8\% & $+0.030$ & $0.527$\textsuperscript{\textcolor{lexcol}{nR}}\,/\,7\% & $\underline{0.553}$\textsuperscript{\textcolor{embcol}{k}}\,/\,13\% & $\mathbf{0.586}$\textsuperscript{\textcolor{lexcol}{\textbf{L}}}\,/\,6\% & $+0.032$ & $0.517$\textsuperscript{\textcolor{embcol}{sR}}\,/\,3\% & $\underline{0.583}$\textsuperscript{\textcolor{embcol}{k}}\,/\,9\% & $\mathbf{0.586}$\textsuperscript{\textcolor{lexcol}{\textbf{L}}}\,/\,7\% & $+0.002$ \\
\bottomrule
\end{tabular}%
}
\end{sidewaystable*}

%% file: tables/table_baselines_rare_appendix.tex
\definecolor{lexcol}{HTML}{1F5FA9}
\definecolor{embcol}{HTML}{A8431F}
\providecommand{\tbd}{\textcolor{gray}{--}}
\begin{table}[t]
\centering
\caption{Rare-label pools; format, variant codes, shading and notes as in Table~\ref{tab:baselines-main}.}
\label{tab:baselines-rare-e4}
\resizebox{0.7\columnwidth}{!}{%
\begin{tabular}{llcccc}
\toprule
 & & \multicolumn{4}{c}{$\varepsilon=4$} \\
\cmidrule(lr){3-6}
Dataset & Generator & \textsc{canary} & \textsc{domias} & \textsc{ours} & $\Delta$ \\
\midrule
PsyTAR & DP-Gen & $\underline{0.571}$\textsuperscript{\textcolor{embcol}{s}}\,/\,11\% & $0.531$\textsuperscript{\textcolor{embcol}{pg}}\,/\,7\% & $\mathbf{0.588}$\textsuperscript{\textcolor{embcol}{\textbf{E}}}\,/\,13\% & $+0.018$ \\
PsyTAR & EPSVec & $\underline{0.572}$\textsuperscript{\textcolor{embcol}{s}}\,/\,8\% & $0.570$\textsuperscript{\textcolor{embcol}{pg}}\,/\,6\% & $\mathbf{0.597}$\textsuperscript{\textcolor{embcol}{\textbf{E}}}\,/\,16\% & $+0.025$ \\
\midrule
N2C2'08 & DP-Gen & $0.495$\textsuperscript{\textcolor{lexcol}{n}}\,/\,2\% & $\underline{0.552}$\textsuperscript{\textcolor{embcol}{pk}}\,/\,4\% & $\mathbf{0.565}$\textsuperscript{\textcolor{lexcol}{\textbf{L}}}\,/\,8\% & $+0.013$ \\
N2C2'08 & EPSVec & $0.546$\textsuperscript{\textcolor{lexcol}{nR}}\,/\,11\% & $\underline{0.573}$\textsuperscript{\textcolor{embcol}{pg}}\,/\,3\% & $\mathbf{0.661}$\textsuperscript{\textcolor{lexcol}{\textbf{L}}}\,/\,10\% & $+0.088$ \\
\midrule
EUR-Lex & DP-Gen & $\underline{0.568}$\textsuperscript{\textcolor{embcol}{s}}\,/\,5\% & $0.558$\textsuperscript{\textcolor{embcol}{pk}}\,/\,14\% & $\mathbf{0.575}$\textsuperscript{\textcolor{embcol}{\textbf{E}}}\,/\,9\% & $+0.007$ \\
EUR-Lex & EPSVec & $\underline{0.615}$\textsuperscript{\textcolor{embcol}{s}}\,/\,11\% & $0.569$\textsuperscript{\textcolor{embcol}{g}}\,/\,8\% & $\mathbf{0.618}$\textsuperscript{\textcolor{embcol}{\textbf{E}}}\,/\,10\% & $+0.002$ \\
\bottomrule
\end{tabular}%
}
\end{table}

%% file: tables/table_v_distribution_appendix.tex
\definecolor{s10hi}{HTML}{A8431F}
\begin{table}[t]
\centering
\small
\setlength{\tabcolsep}{4pt}
\caption{Shape of the per-record vulnerability distribution in every audited cell. Each entry is the share of qualifying records with $V(x)>0$ / the top-decile leakage share $S_{10}$ (Eq.~\ref{eq:top-decile-share}); $n$ is the number of qualifying records; even leakage would give $S_{10}\approx10$. $V$ is computed per cell under that cell's best proxy (the main table's worst-case selection). For \textsc{Aug-PE} the columns are its noise levels mapped to the matching budgets, as in Table~\ref{tab:overall_main_augpe}. \textbf{Shaded}: $S_{10}$ above the same row's $\varepsilon=\infty$ value, i.e.\ leakage that DP noise made \emph{more} concentrated; deeper shading marks a larger excess (up to $+40$ points).}
\label{tab:v-distribution-appendix}
\begin{tabular}{llrccccc}
\toprule
Generator & Dataset & $n$ & $\varepsilon=\infty$ & $4$ & $2$ & $1$ & $0.5$ \\
\midrule
\multirow{4}{*}{\textsc{DP-Gen}} & PsyTAR & 35 & 77\,/\,31 & \cellcolor{s10hi!31}46\,/\,58 & \cellcolor{s10hi!29}54\,/\,57 & \cellcolor{s10hi!8}63\,/\,34 & \cellcolor{s10hi!8}63\,/\,38 \\
 & DMSAFN & 31 & 84\,/\,29 & \cellcolor{s10hi!45}29\,/\,69 & \cellcolor{s10hi!23}52\,/\,49 & \cellcolor{s10hi!34}39\,/\,59 & 71\,/\,28 \\
 & N2C2'08 & 19 & 79\,/\,37 & \cellcolor{s10hi!8}79\,/\,38 & \cellcolor{s10hi!8}68\,/\,39 & \cellcolor{s10hi!24}68\,/\,58 & \cellcolor{s10hi!8}68\,/\,38 \\
 & EUR-Lex & 32 & 94\,/\,20 & \cellcolor{s10hi!16}59\,/\,34 & \cellcolor{s10hi!13}69\,/\,31 & \cellcolor{s10hi!25}56\,/\,42 & \cellcolor{s10hi!25}59\,/\,42 \\
\midrule
\multirow{4}{*}{\textsc{EPSVec}} & PsyTAR & 35 & 63\,/\,82 & 49\,/\,69 & 54\,/\,47 & 69\,/\,41 & 51\,/\,51 \\
 & DMSAFN & 31 & 55\,/\,36 & \cellcolor{s10hi!8}52\,/\,41 & \cellcolor{s10hi!21}39\,/\,54 & \cellcolor{s10hi!8}52\,/\,38 & 55\,/\,33 \\
 & N2C2'08 & 19 & 42\,/\,55 & \cellcolor{s10hi!8}47\,/\,59 & 53\,/\,32 & 74\,/\,32 & 74\,/\,41 \\
 & EUR-Lex & 32 & 44\,/\,39 & 56\,/\,35 & 53\,/\,36 & \cellcolor{s10hi!11}81\,/\,49 & 56\,/\,37 \\
\midrule
\multirow{4}{*}{\textsc{Aug-PE}} & PsyTAR & 35 & 60\,/\,34 & \cellcolor{s10hi!9}54\,/\,42 & \cellcolor{s10hi!8}63\,/\,37 & \cellcolor{s10hi!8}66\,/\,35 & \cellcolor{s10hi!8}51\,/\,36 \\
 & DMSAFN & 31 & 58\,/\,49 & \cellcolor{s10hi!10}45\,/\,57 & \cellcolor{s10hi!26}52\,/\,72 & 68\,/\,33 & 68\,/\,37 \\
 & N2C2'08 & 19 & 74\,/\,25 & \cellcolor{s10hi!19}58\,/\,42 & \cellcolor{s10hi!24}58\,/\,46 & 79\,/\,24 & \cellcolor{s10hi!32}42\,/\,54 \\
 & EUR-Lex & 32 & 66\,/\,28 & \cellcolor{s10hi!35}50\,/\,59 & \cellcolor{s10hi!18}53\,/\,44 & \cellcolor{s10hi!13}62\,/\,40 & \cellcolor{s10hi!14}62\,/\,41 \\
\bottomrule
\end{tabular}
\end{table}

%% file: tables/table_attributes.tex
\begin{table}[t]
\centering
\footnotesize
\setlength{\tabcolsep}{3pt}
\caption{Spearman correlation between five release-independent record attributes and per-record vulnerability rank, per generator, pooled over budgets. Left: the outlier pools; middle: the rare-label pools; right: Fisher-$z$ combination over the 12 disjoint random pools (337 records). No cell is significant after Holm correction; per-pool detection floors are $|\rho|\geq 0.33$--$0.46$ (raw). $\dagger$: range-restricted in the outlier pools. $\ast$: the only directionally consistent effect (9/12 pools positive, Holm $p=0.09$).}
\label{tab:attributes}
\resizebox{0.7\columnwidth}{!}{%
\begin{tabular}{@{}lcccccccc@{}}
\toprule
 & \multicolumn{4}{c}{Outlier pools} & \multicolumn{3}{c}{Rare-label pools} & Random pools \\
\cmidrule(lr){2-5}\cmidrule(lr){6-8}\cmidrule(lr){9-9}
Attribute & \rotatebox[origin=l]{90}{PsyTAR} & \rotatebox[origin=l]{90}{DMSAFN} & \rotatebox[origin=l]{90}{N2C2'08} & \rotatebox[origin=l]{90}{EUR-Lex} & \rotatebox[origin=l]{90}{PsyTAR} & \rotatebox[origin=l]{90}{N2C2'08} & \rotatebox[origin=l]{90}{EUR-Lex} & \rotatebox[origin=l]{90}{combined $\rho$} \\
\midrule
\multicolumn{9}{@{}l}{\itshape DP-Gen} \\
length & $+.29$ & $+.00$ & $+.08$ & $-.24$ & $-.16$ & $+.17$ & $+.25$ & $+.05$ \\
outlierness$^\dagger$ & $-.18$ & $+.04$ & $+.00$ & $+.23$ & $+.16$ & $-.00$ & $+.25$ & $+.12$$^\ast$ \\
near-duplication & $+.11$ & $-.06$ & $+.06$ & $-.10$ & $-.26$ & $+.11$ & $-.30$ & $-.17$ \\
rare $n$-gram density & $-.04$ & $+.01$ & $+.10$ & $+.06$ & $+.17$ & $+.17$ & $+.03$ & $+.01$ \\
entity/numeral density & $+.18$ & $-.22$ & $+.05$ & $-.10$ & $-.12$ & $+.20$ & $-.04$ & $-.15$ \\
\midrule
\multicolumn{9}{@{}l}{\itshape EPSVec} \\
length & $-.02$ & $+.18$ & $+.06$ & $+.09$ & $+.13$ & $+.16$ & $-.02$ & $+.04$ \\
outlierness$^\dagger$ & $-.16$ & $-.20$ & $+.10$ & $+.18$ & $+.00$ & $-.02$ & $-.15$ & $+.01$ \\
near-duplication & $+.05$ & $+.29$ & $+.05$ & $+.13$ & $+.09$ & $+.04$ & $+.01$ & $+.12$ \\
rare $n$-gram density & $-.06$ & $-.12$ & $+.17$ & $-.13$ & $-.29$ & $+.06$ & $-.20$ & $-.01$ \\
entity/numeral density & $-.23$ & $+.17$ & $+.03$ & $-.00$ & $+.19$ & $-.02$ & $-.30$ & $-.05$ \\
\midrule
\multicolumn{9}{@{}l}{\itshape Aug-PE} \\
length & $-.30$ & $-.17$ & $+.25$ & $+.15$ & -- & -- & -- & -- \\
outlierness$^\dagger$ & $-.03$ & $-.24$ & $+.32$ & $+.05$ & -- & -- & -- & -- \\
near-duplication & $+.06$ & $+.03$ & $+.39$ & $+.00$ & -- & -- & -- & -- \\
rare $n$-gram density & $+.26$ & $-.14$ & $-.08$ & $+.11$ & -- & -- & -- & -- \\
entity/numeral density & $-.08$ & $+.09$ & $+.01$ & $-.12$ & -- & -- & -- & -- \\
\bottomrule
\end{tabular}%
}
\end{table}

%% file: tables/table_selection_robustness.tex
\begin{table*}[!htbp]
\centering
\small
\setlength{\tabcolsep}{4pt}
\caption{Table~\ref{tab:attributes} under four proxy-selection rules, per generator, over every pool the table covers (high-risk, rare-label, and random). Columns: number of attribute--pool tests significant after Holm correction (of $N$); the Fisher-$z$ combined $\rho$ over that generator's random pools per attribute; and the largest change in any per-pool $\rho$ relative to the paper's rule.}
\label{tab:selection-robustness}
\begin{tabular}{lccccccc}
\toprule
Selection rule & sig.$/N$ & length & outlierness & near-dup. & rare $n$-gram & entity/num. & max $|\Delta\rho|$ \\
\midrule
\multicolumn{8}{@{}l}{\itshape DP-Gen} \\
80\% split (paper) & 0/90 & $+0.05$ & $+0.12$ & $-0.17$ & $+0.01$ & $-0.15$ & -- \\
all episodes & 0/90 & $+0.01$ & $+0.20$ & $-0.20$ & $+0.04$ & $-0.15$ & 0.40 \\
fixed \texttt{containment\_max} & 0/90 & $-0.05$ & $+0.18$ & $-0.19$ & $+0.02$ & $-0.13$ & 0.70 \\
fixed \texttt{cos\_max} & 0/90 & $-0.17$ & $+0.32$ & $-0.24$ & $+0.13$ & $-0.11$ & 0.96 \\
\midrule
\multicolumn{8}{@{}l}{\itshape EPSVec} \\
80\% split (paper) & 0/95 & $+0.04$ & $+0.01$ & $+0.12$ & $-0.01$ & $-0.05$ & -- \\
all episodes & 0/95 & $+0.10$ & $+0.02$ & $-0.03$ & $+0.01$ & $+0.01$ & 0.64 \\
fixed \texttt{containment\_max} & 0/95 & $+0.08$ & $-0.08$ & $+0.06$ & $+0.03$ & $-0.04$ & 0.64 \\
fixed \texttt{cos\_max} & 0/95 & $+0.04$ & $+0.01$ & $+0.00$ & $+0.08$ & $-0.04$ & 0.65 \\
\midrule
\bottomrule
\end{tabular}
\end{table*}

%% file: tables/table_combined_by_rule.tex
\begin{table*}[!htbp]
\centering
\small
\setlength{\tabcolsep}{5pt}
\caption{Random-pool combined correlation between each attribute and per-record vulnerability under four proxy-selection rules (Fisher-$z$ over 12 pools for \textsc{DP-Gen} and 12 pools for \textsc{EPSVec}). Each entry is $\rho$ with the Holm-corrected $p$ over the five attributes in parentheses; \textbf{bold} marks $p<0.05$. The paper's rule is the most conservative: the same tendencies that are marginal under it reach significance under all-episode (worst-case) selection and under fixed proxies.}
\label{tab:combined-by-rule}
\begin{tabular}{lcccc}
\toprule
Attribute & \makecell{80\% split\\(paper)} & \makecell{all\\episodes} & \makecell{fixed\\\texttt{containment\_max}} & \makecell{fixed\\\texttt{cos\_max}} \\
\midrule
\multicolumn{5}{@{}l}{\itshape DP-Gen} \\
length & $+0.05$ (0.725) & $+0.01$ (0.895) & $-0.05$ (0.707) & \textbf{$-0.17$ (0.009)} \\
outlierness & $+0.12$ (0.088) & \textbf{$+0.20$ (0.002)} & \textbf{$+0.18$ (0.006)} & \textbf{$+0.32$ ($<$0.001)} \\
near-duplication & \textbf{$-0.17$ (0.013)} & \textbf{$-0.20$ (0.002)} & \textbf{$-0.19$ (0.004)} & \textbf{$-0.24$ ($<$0.001)} \\
rare $n$-gram density & $+0.01$ (0.826) & $+0.04$ (0.895) & $+0.02$ (0.761) & \textbf{$+0.13$ (0.048)} \\
entity/numeral density & \textbf{$-0.15$ (0.033)} & \textbf{$-0.15$ (0.023)} & $-0.13$ (0.067) & \textbf{$-0.11$ (0.048)} \\
\midrule
\multicolumn{5}{@{}l}{\itshape EPSVec} \\
length & $+0.04$ (1.000) & $+0.10$ (0.372) & $+0.08$ (0.742) & $+0.04$ (1.000) \\
outlierness & $+0.01$ (1.000) & $+0.02$ (1.000) & $-0.08$ (0.725) & $+0.01$ (1.000) \\
near-duplication & $+0.12$ (0.172) & $-0.03$ (1.000) & $+0.06$ (0.824) & $+0.00$ (1.000) \\
rare $n$-gram density & $-0.01$ (1.000) & $+0.01$ (1.000) & $+0.03$ (0.954) & $+0.08$ (0.782) \\
entity/numeral density & $-0.05$ (1.000) & $+0.01$ (1.000) & $-0.04$ (0.954) & $-0.04$ (1.000) \\
\bottomrule
\end{tabular}
\end{table*}

%% file: tables/table_cross_generator.tex
\begin{table}[t]
\centering
\small
\caption{Cross-mechanism consistency of per-record vulnerability. Each entry is the Spearman correlation between $V(x)$ under DP-Gen and $V(x)$ under the named generator, on the same records scored with the same proxy (the DP-Gen pool's selected proxy); parentheses give the number of records qualifying under both. \textsc{Aug-PE} has no random-pool releases.}
\label{tab:cross-generator}
\begin{tabular}{lrcc}
\toprule
Pool & $n$ & vs.\ \textsc{EPSVec} & vs.\ \textsc{Aug-PE} \\
\midrule
PsyTAR outlier & 35 & $+0.01$ (35) & $+0.34$ (35) \\
DMSAFN outlier & 31 & $-0.05$ (31) & $-0.05$ (31) \\
N2C2'08 outlier & 19 & $+0.28$ (19) & $-0.08$ (19) \\
EUR-Lex outlier & 32 & $-0.13$ (32) & $-0.14$ (32) \\
PsyTAR rand0 & 32 & $+0.08$ (32) & -- \\
PsyTAR rand1 & 30 & $+0.07$ (30) & -- \\
PsyTAR rand2 & 36 & $-0.06$ (36) & -- \\
DMSAFN rand0 & 34 & $-0.34$ (34) & -- \\
DMSAFN rand1 & 31 & $-0.16$ (31) & -- \\
DMSAFN rand2 & 34 & $+0.04$ (34) & -- \\
N2C2'08 rand0 & 18 & $-0.03$ (18) & -- \\
N2C2'08 rand1 & 17 & $-0.41$ (17) & -- \\
N2C2'08 rand2 & 16 & $-0.55$ (16) & -- \\
EUR-Lex rand0 & 32 & $+0.05$ (32) & -- \\
EUR-Lex rand1 & 26 & $-0.13$ (26) & -- \\
EUR-Lex rand2 & 31 & $-0.02$ (31) & -- \\
\midrule
mean & & $-0.09$ & $+0.02$ \\
\bottomrule
\end{tabular}
\end{table}